\documentclass[conference]{IEEEtran}

\usepackage[utf8]{luainputenc}  

\IEEEoverridecommandlockouts
\usepackage{cite}
\usepackage{amsmath,amssymb,amsfonts}
\usepackage{algorithmic}
\usepackage{graphicx}
\usepackage{textcomp}
\usepackage{xcolor}
\usepackage{booktabs}
\usepackage{tabularx}
\usepackage{multirow}
\usepackage{physics}
\usepackage{tikz}
\usepackage{tikzit}
\usepackage{hyperref}
\usepackage{longtable}
\usepackage{array}
\usepackage{footnote}
\tikzstyle{black}=[draw=black, fill=darkgray, tikzit draw=black, shape=circle, tikzit shape=circle, tikzit fill=darkgray]
\tikzstyle{white}=[draw=black, fill=white, tikzit draw=black, tikzit fill=white, shape=circle]
\tikzstyle{small white}=[inner sep=0mm, minimum size=2.5mm, shape=circle, draw=black, fill=white]
\tikzstyle{small black}=[inner sep=0mm, minimum size=2.5mm, shape=circle, draw=black, fill=darkgray]
\tikzstyle{red}=[draw=black, fill=red, shape=circle, tikzit shape=circle]
\tikzstyle{rededge}=[-, draw=red, shape=circle]
\tikzstyle{bluefill}=[-, draw=black, fill={rgb,255: red,182; green,185; blue,236}]
\tikzstyle{fibre}=[-, double, draw=black]
\tikzstyle{arrow}=[->, draw=black]
\tikzstyle{snake arrow}=[->, tikzit draw=black, decorate, decoration={snake,amplitude=1mm, pre length=1mm, post length=1mm}]
\tikzstyle{crosshatched edge}=[-, draw=black, fill=white, pattern=crosshatch, tikzit fill=blue]
\tikzstyle{hatched edge}=[-, draw=black, fill=white, pattern=vertical lines]
\tikzstyle{new style 1}=[-]
\tikzstyle{grey fill}=[-, draw=black, fill=gray]

\usetikzlibrary{quantikz2}
\def\BibTeX{{\rm B\kern-.05em{\sc i\kern-.025em b}\kern-.08em
    T\kern-.1667em\lower.7ex\hbox{E}\kern-.125emX}}

\begin{document}
\bstctlcite{IEEEexample:BSTcontrol}

\title{Quantum Networking Fundamentals: From Physical Protocols to Network Engineering
\\

\thanks{The authors acknowledge funding from the Engineering and Physical Sciences Research Council (EPSRC) funded Distributed Quantum Computing project, grant number EP/W032643/1. Athanasios Gkelias and Kin K. Leung were also supported by the EPSRC grant EP/Y037243/1.}
}
\author{\IEEEauthorblockN{Athanasios Gkelias}
\IEEEauthorblockA{\textit{EEE Department} \\
\textit{Imperial College}\\
London, UK \\
a.gkelias@imperial.ac.uk}

\and
\IEEEauthorblockN{Felix T. A. Burt}
\IEEEauthorblockA{\textit{EEE Department} \\
\textit{Imperial College}\\
London, UK\\
f.burt23@imperial.ac.uk}

\and
\IEEEauthorblockN{Kin K. Leung}
\IEEEauthorblockA{\textit{EEE Department} \\
\textit{Imperial College}\\
London, UK\\
kin.leung@imperial.ac.uk}

}

\maketitle

\begin{abstract}
Abstract—The realization of the Quantum Internet promises transformative capabilities in unconditionally secure communication, distributed quantum computing, and high-precision quantum metrology. However, transitioning from isolated laboratory experiments to a scalable, multi-tenant network utility introduces deep orchestration challenges. Current development is largely siloed within the physics and optics communities, prioritizing hardware fidelities and photon sources, while the classical networking community lacks the architectural models required to dynamically manage these fragile quantum resources. This tutorial bridges this disciplinary divide by providing a comprehensive, network-centric view of quantum networking. We systematically dismantle the idealized assumptions prevalent in current network simulators to directly address the ``simulation–reality gap," and we recast them as explicit control-plane constraints. To bridge this gap, we establish \textit{Software-Defined Quantum Networking (SDQN)} not merely as an evolutionary management tool, but as a mandatory prerequisite for scale, and we prioritize the orchestration of a symbiotic, dual-plane architecture in which classical control dictates quantum data flow. Specifically, we synthesize reference models for SDQN and the \textit{Quantum Network Operating System (QNOS)} for hardware abstraction, and we adapt a \textit{Quantum Network Utility Maximization (Q-NUM)} framework as a unifying mathematical lens to help network engineers reason about the inherent trade-offs between entanglement routing, scheduling, and fidelity targets. Furthermore, we analyze \textit{Distributed Quantum AI (DQAI)} over imperfect networks as a case study, illustrating how physical constraints such as probabilistic stragglers and decoherence fundamentally dictate application-layer viability. Ultimately, this tutorial equips network engineers with the operational mindset and architectural tools required to transition quantum networking from a bespoke physics experiment into a programmable, multi-tenant global infrastructure.

\end{abstract}

\begin{IEEEkeywords}
Quantum Internet, Software-Defined Quantum Networking (SDQN), Quantum Network Utility Maximization (Q-NUM), Entanglement Routing, Protocol Stack, Distributed Quantum AI. 
\end{IEEEkeywords}

\section{Introduction}

The concept of a global Quantum Internet is no longer confined to the realm of theoretical physics; it is rapidly transitioning into a highly complex network engineering challenge. While the classical Internet has revolutionized global communication by transmitting deterministic bits of information (0s and 1s) over robust channels, it is fundamentally bound by the limits of classical physics. The Quantum Internet seeks to transcend these limits by leveraging the unique phenomena of quantum mechanics---namely, superposition and entanglement. By transmitting quantum bits (qubits), this future network promises unconditionally secure communication, distributed quantum computing (DQC), and high-precision quantum metrology.

To understand the current urgency of quantum network design, it is necessary to view it through the lens of generational hardware evolution. In quantum computing, the industry is currently navigating the Noisy Intermediate-Scale Quantum (NISQ) era \cite{preskillQuantumComputingNISQ2018}, where processors possess dozens to hundreds of qubits but lack the resources for full error correction. The ultimate goal is the transition to Fault-Tolerant Quantum Computing (FTQC) \cite{preskillFaulttolerantQuantumComputation1997}, which will require millions of physical qubits to create stable, logical qubits. 

Quantum networking is undergoing a parallel evolution, generally categorized into three generations:
\begin{enumerate}
    \item \textbf{Generation 1 (Prepare-and-Measure):} The current state-of-the-art, primarily focused on Quantum Key Distribution (QKD) using single photons. These networks are limited by distance and rely on vulnerable ``trusted repeater'' nodes.
    \item \textbf{Generation 2 (Entanglement Distribution):} The active frontier. This generation introduces quantum memory and entanglement swapping, allowing entangled states to be distributed across multi-hop networks without relying on trusted classical nodes.
    
    \item \textbf{Generation 3 (Fault-Tolerant Quantum Communication):} The ultimate vision of the Quantum Internet, supporting universal DQC and blind quantum computing through robust quantum error correction over global distances.
\end{enumerate}

Transitioning to Generation 2 and beyond requires a radical departure from classical networking paradigms. Classical networks operate on the principle of replication. If a signal degrades over a long fiber-optic cable, classical repeaters simply read the signal, copy it, amplify it, and retransmit it. Furthermore, if a classical packet is lost due to congestion, the network simply requests a retransmission (e.g., via ARQ protocols).

Quantum networks strip away these foundational luxuries. The fundamental laws of quantum mechanics dictate that an arbitrary unknown quantum state cannot be copied (the \textit{No-Cloning Theorem} \cite{woottersSingleQuantumCannot1982}), meaning the classical principles of amplification cannot directly be applied in the quantum domain. Furthermore, quantum states are incredibly fragile; interactions with the environment or classical measurements can irreversibly collapse the quantum state. Consequently, quantum networks cannot easily buffer, copy, store or retransmit data. Instead, they must rely on the continuous, probabilistic generation of entanglement and the complex orchestration of ``quantum teleportation'' \cite{bennettTeleportingUnknownQuantum1993} to move information, alongside entanglement purification~\cite{deutsch1996quantum} and quantum error correction~\cite{shorSchemeReducingDecoherence1995} to combat noise. Achieving the core primitives of networking in the quantum domain is not impossible, but requires new operational paradigms that circumvent the restrictions and no-go theorems posed by quantum mechanics.

A common misconception is that the Quantum Internet is an ``Internet 2.0'' destined to replace existing classical infrastructure. In reality, the Quantum Internet is a specialized, symbiotic overlay. It is designed to provide entirely distinct services---such as entanglement distribution and DQC---that cannot be simulated classically. It will never be used to stream high-definition video or browse web pages. Critically, quantum network operations are fundamentally reliant upon a parallel classical communication infrastructure to function. Operations such as entanglement swapping and teleportation mandate continuous, strict-latency classical communication to share measurement outcomes. Therefore, classical networks must shoulder the heavy burden of quantum telemetry, forging an inseparable dual-plane architecture where classical control dictates quantum data flow.

Despite these profound networking challenges, the majority of quantum communication development has historically been driven by the physics and optics communities, focusing naturally on hardware fidelities and photon sources. However, as quantum networks scale into multi-tenant infrastructures, a critical disciplinary divide has emerged. There is a significant technical language barrier between physicists and the network engineering community. Where a physicist sees ``decoherence,'' a network engineer must see a strict ``Time-to-Live (TTL)'' constraint. Where a physicist performs ``entanglement purification,'' a network engineer must account for a ``throughput-versus-reliability'' resource trade-off. If the Quantum Internet is to become a reality, it must be architected by the network engineering community. We must move beyond treating the quantum network as a bespoke physics experiment and begin applying established network engineering principles to orchestrate fragile quantum resources dynamically.

Motivated by this imperative, the primary objective of this tutorial is to serve as a definitive, network-centric translation guide. We aim to bridge the persistent knowledge gap by systematically translating fundamental quantum mechanical phenomena into actionable network engineering primitives. Rather than proposing isolated algorithms, this paper seeks to provide a unified operational mindset for the network engineer. Ultimately, this tutorial is designed to equip the classical networking community with the pragmatic understanding required to transition quantum networking from isolated, bespoke physical experiments to a globally orchestratable, multi-tenant Internet.

\subsection{Related Work and Prior Literature}

While the rapid development of quantum technologies has generated a vast body of literature, existing surveys tend to isolate specific physical layers, applications, or theoretical bounds. These works generally fall into five distinct thematic categories. As detailed below, each group exhibits specific limitations when addressing the practical, network-centric orchestration of a multi-tenant Quantum Internet.

\begin{table*}[t]
\centering
\caption{Summary of Existing Surveys vs. This Tutorial}
\label{tab:literature_comparison}
\renewcommand{\arraystretch}{1.3} 
\begin{tabular}{|p{0.2\linewidth}|p{0.15\linewidth}|p{0.25\linewidth}|p{0.3\linewidth}|}
\hline
\textbf{Reference / Author} & \textbf{Primary Focus Area} & \textbf{Main Contribution} & \textbf{Key Difference from Our Tutorial} \\
\hline
Singh~\cite{singh2021} / Yang~\cite{yang2023} / Kimble~\cite{kimbleQuantumInternet2008a} / Wehner~\cite{wehnerQuantumInternetVision2018a} & 
Broad Overviews & 
Catalogs basic quantum internet concepts, high-level visions, and parallel tech domains. & 
Lacks concrete SDQN/QNOS orchestration and economic resource allocation models for a multi-tenant network. \\\hline

Li \cite{li2024} / Zhang \cite{zhang202x} & Layered Architectures & Maps quantum protocols to OSI layers / details quantum stack structures. & We reject rigid OSI mapping; propose a dual-plane cross-layer SDQN model. \\\hline

Gyongyosi \cite{gyongyosi2018} / Koudia \cite{koudia2022} & Information Theory & Calculates Shannon limits, channel capacities, and superadditivity. & Shifts focus from offline theoretical capacities to real-time SDQN metrics. \\\hline

Lukens~\cite{lukens2025} / Li~\cite{li2023} / Amiri~\cite{amiri2024} / Cacciapuoti~\cite{cacciapuotiWhenEntanglementMeets2020} & 
Physical Integration & 
Details physical coexistence (WDM), teleportation-based links, and noise/decoherence models over hybrid channels. & 
Abstracts link-level physics to focus on multi-hop, logical control-plane routing and embeds noise into SDQN metrics. \\
\hline

Khalighi \cite{khalighi2014} / Hosseinidehaj \cite{hosseinidehaj2019} / Kumar \cite{kumar2021} & Modality \& Topology & Analyzes FSO channels, satellite links, and drone network topologies. & Topology and hardware-agnostic; focuses on universal QNOS abstraction APIs. \\
\hline
Pan \cite{pan2024} / Birhanu \cite{birhanu2025} / Sharma \cite{sharma2021} / Cao \cite{cao2022} / Huang \cite{huang2013} & Single-Application (QKD) & Details evolution, architecture, and deployment of cryptographic QKD networks. & Proposes an application-agnostic, multi-tenant resource sharing architecture. \\
\hline
Huang \cite{huang2025} / Babar \cite{babar2019} & Algorithms \& Coding & Mathematical theory of quantum circuits and Error Correction Codes (QECC). & Translates QECC math into network-level telemetry burdens and latency constraints. \\
\hline
Abane \cite{abane2025survey} / Glisic \cite{glisic2024} & Routing \& Allocation & Surveys entanglement routing algorithms and resource allocation tools. & Integrates routing into a joint SDQN triad governed by non-additive Q-NUM models. \\
\hline
Mahmud \cite{mahmud2025} & Classical AI & Uses classical AI to optimize quantum/RF networks. & Explores the reverse paradigm: Distributed Quantum AI over physical quantum networks. \\\hline

Zhao \cite{zhao2025} / Botsinis \cite{botsinis2019} / Butt \cite{butt2025} & Quantum for Classical & Applies quantum computing/search to optimize classical 5G/6G wireless. & Focuses entirely on orchestrating native quantum data across a Quantum Internet. \\\hline

Cacciapuoti~\cite{cacciapuotiQuantumInternetNetworking2020} / Caleffi~\cite{caleffiDistributedQuantumComputing2024} / Barral~\cite{barralReviewDistributedQuantum2025} / Larasati~\cite{larasatiFaulttolerantDistributedQuantum2025} /
Vanmeter~\cite{vanmeterLocalDistributedQuantum2016} /
Knorzer~\cite{knorzerDistributedQuantumInformation2025}& 
Distributed Quantum Computing (DQC) & 
Surveys the Quantum Internet as an enabler for DQC, reviewing FT-DQC architectures, protocols, and compilers. & 
Treats DQC as a flagship workload within a multi-tenant SDQN/QNOS framework, using Q-NUM to resolve real-time resource conflicts. \\\hline

\hline
\textbf{This Tutorial} & \textbf{Network-Centric Orchestration \& Physics-to-Engineering Translation} & \textbf{Dual-Plane SDQN Architecture, Application-Agnostic QNOS, Q-NUM optimization framework, and DQAI case study.} & \textbf{Explicitly bridges the physics-to-engineering gap by translating fundamental quantum constraints into a unified, hardware-aware control framework for multi-tenant applications.} \\
\hline
\end{tabular}
\end{table*}

\subsubsection{Broad Overviews and Rigid Layered Architectures}
A foundational segment of the literature provides broad, holistic overviews of quantum internet concepts. For example, Singh et al. \cite{singh2021} provide a wide-ranging catalog of enabling hardware technologies and potential applications (such as teleportation and key distribution), while Yang et al. \cite{yang2023} categorize the quantum landscape into four distinct pillars: computing, networks, cryptography, and machine learning. Foundational vision papers such as Kimble's ``The Quantum Internet''~\cite{kimbleQuantumInternet2008a} and the roadmap by Wehner et al.~\cite{wehnerQuantumInternetVision2018a} likewise paint a high-level picture of globally interconnected quantum networks, emphasizing repeaters, quantum memories, and entanglement-enabled services without prescribing how a classical control plane should expose and manage these capabilities. To formalize these broad concepts, recent surveys have attempted to structure quantum operations using layered network models. Li et al. \cite{li2024} present a comprehensive review that explicitly maps quantum protocols onto strict classical layered architectures, mimicking the vertical OSI model. Similarly, Zhang et al. \cite{zhang202x} survey the structural development of quantum network stacks, categorizing them into distinct node stacks and programmable control stacks.
\begin{itemize}
    \item  While these foundational overviews are excellent for introducing the technology, they treat the quantum network either as a collection of isolated parallel domains or force it into rigid classical hierarchies. Attempting to fit quantum protocols into a strict vertical OSI stack fails to account for the physical reality of the hardware. For instance, operations like entanglement purification inherently demand cross-layer interactions: they rely on physical-layer photon generation but are triggered by network-layer routing thresholds. Furthermore, while existing literature catalogues stack architectures, it frequently omits the economic and mathematical resource allocation models required to drive them. This tutorial bridges this gap by synthesizing an explicitly \textit{dual-plane} (quantum and classical) reference model, providing a framework that acknowledges and operationalizes these mandatory cross-layer interactions. 
    
\end{itemize}

\subsubsection{Physical Layer, Channel Capacities, and Hardware Modalities}
A highly significant portion of the literature remains deeply rooted in pure information theory and physical-layer mechanics. Surveys by Gyongyosi et al. \cite{gyongyosi2018} and Koudia et al. \cite{koudia2022} dive deeply into the mathematical calculation of quantum channel capacities, exploring theoretical phenomena like superadditivity and superactivation. Li et al. \cite{li2023} detail the physical mechanics and node architectures of entanglement-assisted links. Other works focus heavily on the physical coexistence and integration of signals. Lukens et al. \cite{lukens2025} and Amiri et al. \cite{amiri2024} explore the physical-layer challenges of running hybrid classical-quantum communications over shared optical fibers using wavelength-division multiplexing. Expanding on the physical transmission paradigm, Cacciapuoti et al.~\cite{cacciapuotiWhenEntanglementMeets2020} further recast quantum teleportation itself as the fundamental communication primitive, developing a teleportation-centered system model and analyzing decoherence and noise from a communications-engineering perspective. Furthermore, other surveys focus entirely on specific hardware modalities, analyzing free-space optical (FSO) channels \cite{khalighi2014}, continuous-variable satellite links \cite{hosseinidehaj2019}, or the specific line-of-sight constraints of quantum drone swarms \cite{kumar2021}.
\begin{itemize}
    \item These works successfully solve vital physical-layer transmission or pure link-capacity problems. However, establishing a high-capacity link or a theoretical Shannon limit does not equate to a functional Internet. This tutorial takes a different explanatory path by conceptually abstracting away these physical and topological variances. Rather than focusing on offline theoretical capacity bounds or physical-layer fiber multiplexing, we focus on \textit{logical control-plane orchestration}. We frame the Quantum Network Operating System (QNOS) APIs as a conceptual method to abstract hardware heterogeneity, shifting the focus to real-time Software Defined Quantum Networking (SDQN) operational metrics (like Quantum Delay and Timing Jitter) that dictate multi-hop network viability. 
\end{itemize}

\subsubsection{Cryptographic Focus and Single-Application Networks}
Many surveys are highly specialized, treating the quantum network strictly as a cryptographic tool. For instance, Pan et al. \cite{pan2024} focus extensively on the theoretical and experimental verification of Quantum Secure Direct Communication (QSDC). Birhanu et al. \cite{birhanu2025} provide a deep dive restricted to Continuous Variable QKD (CV-QKD), while Sharma et al. \cite{sharma2021} and Cao et al. \cite{cao2022} survey the evolution, architecture, and deployment of networks designed exclusively for QKD. Additionally, Huang et al. \cite{huang2013} survey Device-Independent (DI) protocols focused strictly on endpoint cryptographic trust models. Other works in this silo focus purely on the theoretical mathematics of the payloads, such as Huang et al.'s \cite{huang2025} review of specific quantum communication circuit algorithms, or Babar et al.'s \cite{babar2019} exploration of the mathematical duality between classical and Quantum Error Correction (QEC) codes.
\begin{itemize}
    \item These studies typically treat the quantum network as a bespoke, static pipeline serving a single cryptographic purpose. While they ask ``how do we securely transmit a key?'', this tutorial asks ``how do we route, schedule, and allocate resources in a multi-tenant environment to support \textit{any} application?'' We synthesize an application-agnostic architectural perspective. Furthermore, rather than focusing on the mathematical topology of QEC codes, we conceptually translate those codes into concrete network engineering burdens—specifically, visualizing the massive classical telemetry overhead (the syndrome cycle) they impose on the network's latency-coherence budget. 
\end{itemize}

\subsubsection{DQC and FT-DQC Surveys}

A distinct line of work surveys DQC as the primary vehicle for scaling quantum processing. Van Meter and Devitt~\cite{vanmeterLocalDistributedQuantum2016} lay out some of the criteria for DQC, analyzing a variety of hardware platforms as potential hosts of DQC. Cacciapuoti et al.~\cite{cacciapuotiQuantumInternetNetworking2020} review Quantum Internet networking challenges with a particular emphasis on DQC, highlighting how teleportation, entanglement distribution, decoherence, and no-cloning fundamentally reshape link, routing, and transport primitives. Caleffi et al.~\cite{caleffiDistributedQuantumComputing2024} survey a wider variety of emerging problems in DQC, including software level problems such as distributed algorithms, compilers and simulation tools. Barral et al.~\cite{barralReviewDistributedQuantum2025} further extend the scope, providing a comprehensive bottom-up survey ``from single QPU to high-performance quantum computing,'' covering physical interconnects, DQC network architectures, distributed compilers, partitioning techniques, and application classes such as \textit{Quantum Machine Learning (QML)} and \textit{Variational Quantum Eigensolvers (VQE)}. Moving toward long-term reliability, Larasati and Choi~\cite{larasatiFaulttolerantDistributedQuantum2025} extend this perspective to fault-tolerant DQC, proposing a layered taxonomy of FT-DQC challenges and identifying key reliability issues such as modular QEC schemes, qubit allocation, remote-gate scheduling, and entanglement resource management across noisy interconnects. Finaly, Knorzer et al.~\cite{knorzerDistributedQuantumInformation2025} highlights recent developments though focusing on the theoretical foundations of distributed quantum information processing.

\begin{itemize}
    \item While these surveys establish DQC and FT-DQC as central scaling paradigms and map out their design space across physical, compilation, and application layers, they primarily serve as mechanisms-based taxonomies. They describe hardware and protocol challenges rather than prescribing a network-oriented orchestration framework for managing these heterogeneous workloads. In contrast, this tutorial treats DQC---together with DQAI and other services---as first-class, concurrent tenants of a shared Quantum Internet. We introduce SDQN/QNOS and the Q-NUM optimization model as reusable control-plane tools to dynamically orchestrate joint routing, scheduling, and resource allocation under realistic noise and decoherence constraints.
\end{itemize}

\subsubsection{Routing, AI, and Quantum-for-Classical Optimization}
The final category explores the intersection of quantum communications with routing and artificial intelligence. However, the majority of this literature focuses on the ``Quantum for Classical'' paradigm—using quantum computing or quantum-inspired mathematics to optimize classical 5G/6G wireless networks \cite{zhao2025, botsinis2019, butt2025}. Conversely, surveys like Mahmud et al. \cite{mahmud2025} explore using classical AI (like Deep Reinforcement Learning) to optimize quantum network parameters. Even among surveys dedicated to native quantum network operations, such as Abane et al.'s \cite{abane2025survey} comprehensive review of entanglement routing or Glisic \& Lorenzo's \cite{glisic2024} survey of resource allocation, the physical hardware constraints are frequently abstracted away, allowing for the use of classical additive-cost algorithms.
\begin{itemize}
    \item This tutorial addresses these gaps by exploring the ``additive cost fallacy'' as a pedagogical teaching point. Because quantum path costs are fundamentally non-additive and probabilistic, we reframe routing so it is not treated as an isolated graph-theory problem. Instead, we present it conceptually as one-third of an inseparable SDQN control triad (Routing, Scheduling, and Allocation). Furthermore, rather than exploring ``Quantum for Classical'' networks or ``Classical AI for Quantum,'' this tutorial guides the reader through the emerging frontier of Distributed Quantum AI (DQAI) over imperfect quantum networks. We analyze how physical constraints—such as the ``quantum straggler'' problem and decoherence-induced gradient noise—must dictate the architectural design of future application-layer learning models. 
\end{itemize}

Ultimately, while the existing literature provides invaluable deep dives into quantum mechanics, information theory, isolated cryptographic applications, and even dedicated DQC taxonomies, it collectively leaves a structural void for the network engineer. To fulfill our objective of bridging this disciplinary divide, this tutorial adopts a fundamentally different, network-centric approach. Rather than relying on classical additive-cost algorithms or abstracting away hardware limitations, we explicitly target the ``simulation-reality gap'' by turning idealized assumptions into concrete architectural constraints for the control plane. We do so through physically grounded, tutorial frameworks that emphasize a dual-plane view of quantum data and classical control, and that equip network engineers and classical communications researchers to reason about rate-fidelity trade-offs and probabilistic QoS guarantees that prior surveys largely overlook. To encapsulate this analysis, Table~\ref{tab:literature_comparison} provides a comprehensive comparison of the existing survey literature against the pedagogical focus of this tutorial.

\subsection{Bridging the Physics-to-Engineering Gap: The Case for SDQN}

While the physics community has made extraordinary strides in physical-layer hardware, and computer scientists continue to advance quantum compilers and algorithms, the intermediate networking layer remains critically underdeveloped. Physical links and isolated repeater chains do not intrinsically constitute an Internet; transitioning from bespoke, laboratory-scale experiments to a global, scalable network utility requires solving a fundamental orchestration problem. Unlike classical bits, quantum entanglement is fragile, probabilistic, and short-lived due to decoherence, making static or fully distributed routing too slow to match entanglement generation with tenant demand. At the same time, the future Quantum Internet will be physically heterogeneous, comprising varied qubit modalities (e.g., superconducting circuits, trapped ions, photonic platforms) and memory technologies. To manage this complexity, a logically centralized SDQN control plane, supported by a QNOS, becomes an architectural necessity. SDQN provides dual-plane synchronization between classical control signals and quantum data flow, while QNOS abstracts hardware diversity into device-agnostic metrics and interfaces, enabling concurrent, multi-tenant workloads such as DQC and DQAI to be orchestrated using familiar network engineering principles.

\subsection{Contributions and Organization of this Tutorial}

Existing surveys and roadmaps predominantly adopt either a physics-driven, information-theoretic, or single-application perspective on the Quantum Internet, often cataloguing technologies, protocols, and use cases without prescribing how a network engineer should orchestrate these resources as a programmable utility. In contrast, this tutorial is explicitly written as a network-centric, expository framework that translates fundamental quantum phenomena into the language, abstractions, and control concepts familiar to the networking community. Rather than proposing a new algorithm or protocol, the novelty of this work lies in the synthesis and structuring of existing knowledge into cohesive operational frameworks that close the simulation--reality gap and enable multi-tenant quantum network design.

The main contributions of this tutorial are as follows:
\begin{itemize}
    \item \textbf{A physics-to-network engineering translation guide:} We systematically reinterpret key quantum concepts---such as superposition, entanglement, decoherence, and the no-cloning theorem---as concrete network engineering primitives (e.g., TTL-like lifetimes, non-additive path costs, probabilistic service guarantees). This creates a common vocabulary that allows network practitioners to reason about quantum constraints without requiring deep expertise in quantum physics.
    
    \item \textbf{A dual-plane reference architecture, instead of rigid layering:} Building on, but deliberately departing from, OSI-style quantum stacks and purely hardware-centric taxonomies, we introduce a dual-plane architectural model that separates a vertical quantum service stack from a horizontal classical control plane. This model is presented as a teaching tool to highlight mandatory cross-layer interactions (e.g., purification triggers, teleportation signaling) rather than as a fixed standard, emphasizing how entanglement services and classical coordination must co-evolve.
    
    \item \textbf{SDQN and QNOS as conceptual scaffolding:} Whereas prior works often focus on individual layers (physical, routing, or DQC protocols), we frame SDQN and the QNOS as unifying abstractions that hide hardware heterogeneity and expose device-agnostic metrics such as quantum delay, timing jitter, and memory fidelity. The emphasis is pedagogical: we articulate why logical centralization, dual-plane synchronization, and hardware abstraction are architecturally necessary for scale, and how they relate to familiar SDN paradigms in classical networks.
    
    \item \textbf{The SDQN control triad as a pedagogical lens:} Rather than surveying routing or resource allocation in isolation, we introduce the SDQN control triad---joint routing, scheduling, and resource allocation---as an intuitive framework for understanding why classical shortest-path and additive-cost intuitions break down in quantum settings. This triad is used throughout the tutorial to organize discussion of probabilistic entanglement generation, memory decoherence, and multi-tenant contention, guiding readers away from common modeling pitfalls in simulators.
    
    \item \textbf{A unifying Quantum Network Utility Maximization (Q-NUM) formulation for navigating trade-offs:} While optimization-oriented works exist, they typically treat rate, fidelity, or topology in isolation. Here, Q-NUM is presented as a streamlined, reference formulation that network engineers can use as a ``mental model'' for encoding rate-fidelity trade-offs, probabilistic QoS constraints, and slice-level service objectives within a single framework. The focus is not on solving a specific numerical problem, but on demonstrating how utility-based thinking can replace classical bandwidth-centric intuition when designing control policies for quantum networks.
    
    \item \textbf{Tutorial case studies for multi-tenant workloads, including Distributed Quantum AI:} Finally, this tutorial treats applications such as QKD, DQC, and DQAI as concurrent tenants of a shared infrastructure, rather than as isolated end-goals. Using DQAI over imperfect networks as a forward-looking case study, we illustrate how physical constraints (e.g., quantum stragglers, decoherence-induced gradient noise) propagate all the way to application-layer design choices, thereby demonstrating how the proposed SDQN/QNOS/Q-NUM framework can be applied in practice.
\end{itemize}

The remainder of the tutorial is structured to gradually build this operational mindset: from fundamental quantum states and entanglement, through memory and error mechanisms, to the dual‑plane protocol stack, SDQN control triad, Q‑NUM utility modeling, and finally to multi-tenant applications such as DQC and DQAI.

Section~\ref{sec:Fundamentals} begins by introducing the fundamental rules of single quantum states, translating concepts like qubits, superposition, and the no-cloning theorem into actionable classical communications constraints so network engineers can understand the fragility of the payload. Building upon these mechanics, Section~\ref{sec:entanglement} explains how multi-qubit entanglement is generated, swapped, and utilized to move information via quantum teleportation without violating physical laws. This naturally leads into Section~\ref{sec:quantum_networks}, which introduces the hybrid control and data plane architecture required to support advanced applications like QKD, DQC, and Distributed Quantum Sensing (DQS). To ground these applications in physical reality, Section~\ref{sec:core_challenges} details the core networking challenges that disrupt them, analyzing quantum noise channels, exponential photon loss, and the non-additivity of quantum channel capacity. Section~\ref{sec:quantum_memory} further isolates the critical hardware bottleneck—the quantum memory—by dismantling optimistic assumptions regarding its coherence time, fidelity limits, and multimode capacity. These localized hardware limits are then scaled in Section~\ref{sec:performance_metrics} into a comprehensive taxonomy of real-time end-to-end performance metrics, such as Quantum Delay and Secret Key Rate, that the network must actively monitor. To combat this continuous degradation, Section~\ref{sec:Error_management} explores the necessary error management protocols, comparing link-layer purification, application-layer Quantum Error Mitigation (QEM), and fault-tolerant Quantum Error Correction (QEC).

Transitioning from physical constraints to network orchestration, Section~\ref{sec:protocol-stack} synthesizes all the preceding physical and error-correction constraints into a dual-plane Quantum Protocol Stack, explicitly separating the quantum data payload from classical control functions. Section~\ref{sec:sdqn} operationalizes this stack by introducing SDQN and the QNOS to abstract the severe hardware heterogeneity of the physical layer. With this abstraction in place, Section~\ref{sec:control-functions} breaks down the SDQN control triad, detailing how routing, scheduling, and resource allocation must be jointly managed to respect the strict latency-coherence budget. To scale these control functions for multi-tenant environments, Section~\ref{sec:partitioning} explores Quantum Network Virtualization and joint circuit partitioning, while Section~\ref{sec:q-num} formalizes the Q-NUM framework as a universal cost model to resolve the NP-hard optimization conflicts between rate and fidelity across network slices. Section~\ref{sec:dqai} then investigates the deployment of DQAI over these imperfect networks as a practical case study, highlighting how probabilism and decoherence fundamentally dictate model training and viability. Finally, Section~\ref{sec:future-directions} outlines remaining open challenges, such as multi-domain federation and the critical need for standardization of the quantum-classical interface, before Section~\ref{sec:conclusion} concludes the tutorial by summarizing the paradigm shift required to transition quantum networking into a programmable, multi-tenant utility.

\section{The Fundamentals}
\label{sec:Fundamentals}

To understand the architecture and operational constraints of a quantum network, it is first necessary to define the physical and mathematical rules governing single quantum states. This section introduces the foundational concepts of quantum information, providing the required terminology for the network-level protocols discussed later in this paper.

\subsection{The Qubit}

A \textbf{qubit} is the fundamental unit of quantum information, analogous to the classical bit. While a classical bit occupies a definite state of $0$ or $1$, a qubit exists as a complex vector defined over two basis states, corresponding to $0$ and $1$. This means that a qubit can represent not only $0$ and $1$, but also weighted linear combinations of the two, known as \textit{superpositions}.

\textit{Classical Analogy and Physical Difference:} From a classical networking perspective, a classical bit is akin to a coin lying flat on a table, showing either heads (0) or tails (1). An initial intuitive picture for a qubit in superposition might be a coin actively spinning in the air. While it spins, it is not strictly heads or tails, but a dynamic, probabilistic combination of both. However, it is a critical misunderstanding to view a system in superposition as occupying two readable states simultaneously -- the qubit still only contains a single measurable `bit' of information. This information is simply dependent on a continuous rotation of the measurement axis, rather than a discrete binary value.

\begin{figure}
    \centering
    \usetikzlibrary{3d}
\begin{tikzpicture}[z={(10:10mm)},x={(-45:5mm)},font=\small]
  \def\waveamp{0.45}
  \def\wave{
    \draw[fill,thick,fill opacity=.2,samples=180,smooth,domain=0:2]
      plot (\x,{\waveamp*sin(360*\x)}) -- (2,0) -- (0,0) -- cycle;
    \foreach \u in {0.1,0.2,...,2.00}
    {
      \pgfmathsetmacro{\amp}{\waveamp*sin(360*\u)}
      \draw[thin] (\u,0) -- (\u,\amp);
    }
  }
  \def\axes{
    \draw[->,thin] (0,0,0) -- (1.0,0,0);
    \draw[->,thin] (0,0,0) -- (0,1.0,0);
    \draw[->,thin] (0,0,0) -- (0,0,2.50);
  }

  \begin{scope}
    \axes
    \node at (1.35,0,0) {$E$};
    \node at (0,1.35,0) {$B$};
    \begin{scope}[canvas is zy plane at x=0,fill=black]
      \wave
    \end{scope}
    \begin{scope}[canvas is zx plane at y=0,fill=white]
      \wave
    \end{scope}
    \node[align=center] at (0,-1.25,1.1) {\textbf{Horizontal polarization:}\\$\ket{H} = \ket{0}$};
  \end{scope}

  \begin{scope}[xshift=5cm]
    \axes
    \node at (1.35,0,0) {$B$};
    \node at (0,1.35,0) {$E$};
    \begin{scope}[canvas is zy plane at x=0,fill=white]
      \wave
    \end{scope}
    \begin{scope}[canvas is zx plane at y=0,fill=black]
      \wave
    \end{scope}
    \node[align=center] at (0,-1.25,1.1) {\textbf{Vertical polarization:}\\$\ket{V} = \ket{1}$};
  \end{scope}
\end{tikzpicture}
    \caption{\textbf{Single-photon qubit.} An example of a single photon qubit. The qubit is encoded in the plane of oscillation of the electric field  distinguishing horizontally and vertically polarized photons.}
    \label{fig:photon}
\end{figure}
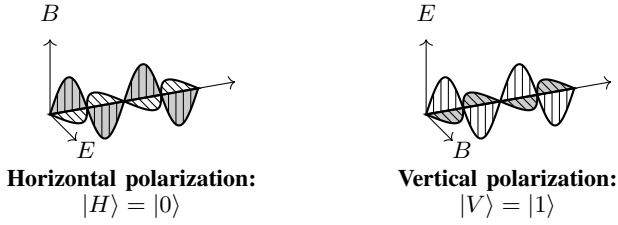

Like a bit, a qubit is a mathematical object -- some physical system is needed to implement the qubit. There are many different systems that can encode a qubit -- electrons, photons and atoms are some of the most obvious examples. Moreover, the qubit can be encoded into various different degrees of freedom of the system~\cite{laddQuantumComputers2010}. Most relevant to communication systems are \textbf{photonic qubits}, in which the qubit is encoded into degrees of freedom of light particles. A common example is to encode the qubit in the \textbf{polarization}, i.e., the plane of oscillation a single photon, such that horizontal polarization corresponds to $0$ and vertical to $1$ (see Fig. \ref{fig:photon}). Alternatively, in \textbf{matter-based qubits} quantum information is encoded in degrees of freedom of atoms or sub-atomic particles, or in bulk properties of particles in solid-state systems such as charge or current direction. In all cases, the abstract qubit state is the same, even though its physical representation is different. Different qubits, as we see in the following, have different strengths and weaknesses, making them more or less suitable for particular tasks. Photonic qubits are examples of  ``flying qubits'' and thus serve as carriers of quantum information. In contrast, \textbf{matter-based} qubits are typically `static', and are thus more appropriate for storing or processing quantum information.

\subsubsection{Quantum Bases and States}
To truly understand a qubit, we must first define its \textbf{basis}. A basis is a set of mutually orthogonal quantum states that can be used to describe any other state in a given quantum system. Think of a basis as a coordinate system for quantum states. Just as any point in a 3D Euclidean space is described by its coordinates along the $x$, $y$, and $z$ axes, any quantum state is described as a linear combination (a superposition) of its basis states. This description exists within a complex vector space known as a \textbf{Hilbert space}. 

\begin{equation}
| \psi \rangle = c_1 | \phi_1 \rangle + c_2 | \phi_2 \rangle = \sum_{i=1}^{2} c_i | \phi_i \rangle
\end{equation}

Here, the $| \phi_i \rangle$ are the \textbf{basis vectors} (in Dirac notation), and the $c_i$ are \textbf{complex coefficients} (also known as probability amplitudes). The basis must be \textbf{orthonormal}. This implies that the basis vectors are orthogonal, $\braket{\phi_i}{\phi_j} = 0$ for $i \neq j$ and normalized, so $\braket{\phi_i}{\phi_j} = 1$ for $i = j$. The orthogonality ensures that the two states are perfectly distinguishable. The coefficients correspond to the probability of measuring the qubit in the corresponding state. However, the coefficients are complex numbers, so the actual probabilities come from their magnitude $P(\phi_1) = |c_1|^2$ and $P(\phi_2) = |c_2|^2$. The coefficients must satisfy the \textbf{normalization condition}, $\sum_{i=1}^{2} |c_i|^2 = 1$, which ensures that the total probability of finding the system in any of the basis states is 1. Relative to a measurement basis, a qubit can be in a definite state (when either $c_1$ or $c_2$ is zero or one) or in a superposition state. Within the context of quantum information, the vector $\vert \psi \rangle$ is typically known as the \textbf{state vector}.

\begin{figure}
    \centering
    \usetikzlibrary{angles, quotes}

\begin{tikzpicture}[scale=0.9]

  \def\r{3}

  \draw[thick, ->] (0, 0) node[circle, fill, inner sep=1] (orig) {} -- (\r/3, \r/2) node[label=above:$\ket{\psi}$  ] (a) {};
  \draw[dashed] (orig) -- (\r/3, -\r/5) node (phi) {} -- (a);

  \draw (orig) circle (\r);
  \draw[dashed] (orig) ellipse (\r{} and \r/3);

  \draw[->] (orig) -- ++(-\r/5, -\r/3) node[below] (x1) {$\ket{+}$};
  \draw[->] (orig) -- ++(\r, 0) node[right] (x2) {$\ket{+i}$};
  \draw[->] (orig) -- ++(0, \r) node[above] (x3) {$\ket{0}$};

  \pic [draw=black, text=black, ->, "$\phi$"] {angle = x1--orig--phi};
  \pic [draw=black, text=black, <-, "$\theta$", angle eccentricity=1.4] {angle = a--orig--x3};

\end{tikzpicture}
    \caption{\textbf{Bloch sphere.} A mathematical construction for representing single qubit transformation and states, where $\ket{\psi} = \cos{(\theta/2)}\ket{0} + e^{i\phi}\sin{(\theta/2)}\ket{1}$.}
    \label{fig:bloch}
\end{figure}
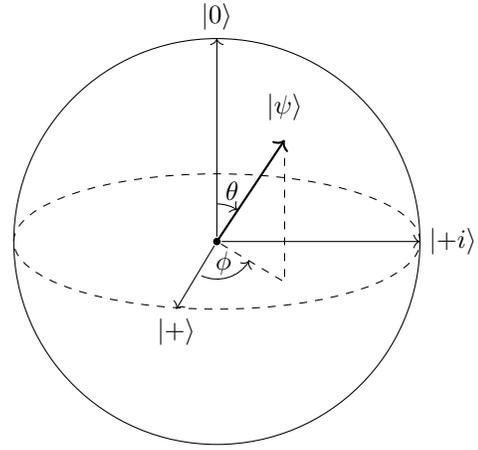

When a projective measurement is performed on a qubit, its superposition collapses, and it is found in one of the definite states, either $|\phi_1\rangle$ or $|\phi_2\rangle$, with the probability determined by the amplitudes. The rule for determining measurement probabilities as the squared modulus of the state amplitudes is known as the \textbf{Born rule}~\cite{wildeNoiselessQuantumTheory2013}. Our freedom in the choice of basis, corresponds to a freedom in the choice of projective measurement
This collapse is a key aspect of quantum mechanics. Using the reduced notation ${|\phi_1\rangle \rightarrow \lvert0\rangle,}$ ${|\phi_2\rangle \rightarrow \lvert1\rangle}$ by convention, a common way to visualize a qubit's state is with the \textbf{Bloch sphere} (see Fig. \ref{fig:bloch}), where the north pole represents the $|0\rangle$ state, the south pole represents the $|1\rangle$ state, and the coefficients $c_1$ and $c_2$ correspond to $\cos (\theta/2)$ and $e^{i\phi}\sin{(\theta/2)}$, respectively. Any point on the surface of the sphere is a valid single-qubit state.

The three most common bases for a single qubit are the \textbf{Pauli bases}, which correspond to measurements along the $x$, $y$, and $z$ axes of the Bloch sphere.

\begin{itemize}
\item \textbf{Z-basis (Computational Basis)}
Its two basis states are $|0\rangle$ and $|1\rangle$.
These states are typically used for representing classical information within a quantum computer, and a common assumption in quantum circuits is the ability to perform state initialization and measurement in the computational basis.
$$|0\rangle = \begin{pmatrix} 1 \\ 0 \end{pmatrix}$$
$$|1\rangle = \begin{pmatrix} 0 \\ 1 \end{pmatrix}$$

Qubits are most commonly written in the computational basis, often as $\ket{\psi} = \alpha \ket{0} + \beta \ket{1}$.

\item \textbf{X-basis}
This basis is often referred to as the \textbf{Hadamard basis}. Its two basis states are $|+\rangle$ and $|-\rangle$, which are superpositions of the computational basis states,
\begin{align*}
    |+\rangle &= \frac{1}{\sqrt{2}}(|0\rangle + |1\rangle) = \frac{1}{\sqrt{2}}\begin{pmatrix} 1 \\ 1 \end{pmatrix} , \\
    |-\rangle &= \frac{1}{\sqrt{2}}(|0\rangle - |1\rangle) = \frac{1}{\sqrt{2}}\begin{pmatrix} 1 \\ -1 \end{pmatrix} .
\end{align*}
For example, a qubit in either state $|+\rangle$ or $|-\rangle$ has a 50\% chance of being measured as a 0 and a 50\% chance of being measured as a 1. These states are essential for many quantum algorithms. Corresponding relations hold for $Z$-basis states in terms of $X$-basis states.
$$|0\rangle = \frac{1}{\sqrt{2}}(|+\rangle + |-\rangle)$$
$$|1\rangle = \frac{1}{\sqrt{2}}(|+\rangle - |-\rangle)$$
A substitution shows the conversion between the $Z$ and $X$ bases:

$$\ket{\psi} = \alpha \ket{0} + \beta \ket{1} = \frac{(\alpha + \beta)}{\sqrt{2}}\ket{+} + \frac{(\alpha - \beta)}{\sqrt{2}}\ket{-}.$$

\item \textbf{Y-basis}
This basis uses basis states that also involve a complex number, $i$.
$$|+i\rangle = \frac{1}{\sqrt{2}}(|0\rangle + i|1\rangle) = \frac{1}{\sqrt{2}}\begin{pmatrix} 1 \\ i \end{pmatrix}$$
$$|-i\rangle = \frac{1}{\sqrt{2}}(|0\rangle - i|1\rangle) = \frac{1}{\sqrt{2}}\begin{pmatrix} 1 \\ -i \end{pmatrix}$$
\end{itemize}

\textbf{NOTE:}
We briefly note two important differences between these descriptions of quantum states and the corresponding behaviour of classical coordinates.
First, in the Bloch sphere representation, orthogonal vectors do not correspond to orthogonal coordinates, because the 2D complex vector space has been mapped to a sphere in a 3D real space.
Second, in a classical scenario, we might think about orienting an antenna to measure the amplitude of a signal along some axis. This would provide a continuous-valued outcome and would not have any effect on the source of the signal.
In the quantum case, the projective effect of measurement collapses the state to one of two discrete points on the Bloch sphere.
Therefore, it is not possible to make simultaneous measurements along any two Pauli axes.
For example, a measurement in the Z-basis is like asking, ``Is the coin pointing heads-up or heads-down?", while a measurement in the X-basis asks, ``Is the coin pointing heads-left or heads-right?".
A qubit that is definitely `up' (Z-basis) is in a superposition of `left' and `right' states (X-basis).
The act of measurement itself collapses the qubit's superposition into one of its definite states, erasing information along other axes.
Where correlations exist between qubits (between the measured qubit and a source, for instance)
this creates an important `back-action' effect, as we discuss in Section~\ref{sec:entanglement} below.

\subsubsection{Multiple Qubits}

The independent states of multiple qubits can be represented by the \textit{tensor product} of the individual state vectors. For two qubits
in states
$\ket{\psi}$ and 
$\ket{\psi'}$, their \textbf{joint state} is given by:
\begin{equation}\label{eq:two-qubit-product}
   \begin{aligned}
   \ket{\Psi} &= \ket{\psi} \otimes \ket{\psi'} \\
           &= \sum_{i,j \in \{0,1\}} c_i c_j' \ket{\phi_i} \otimes \ket{\phi_j '}.
   \end{aligned}
\end{equation}
Each new qubit doubles the dimension of the vector space, and a complete basis for a two-qubit state now comprises four vectors, corresponding to $\ket{\phi_0} \otimes  \ket{\phi_0'}$, $\ket{\phi_0} \otimes  \ket{\phi_1'}$, $\ket{\phi_1} \otimes  \ket{\phi_0'}$ and $\ket{\phi_1} \otimes  \ket{\phi_1'}$, which are also orthonormal. For convenience, the tensor product is often left implicit by writing $\ket{\phi_i}\ket{\phi_j'}$ or even $\ket{\phi_i\phi_j'}$.
A fully general two-qubit state is written as
\begin{equation}\label{eq:two-qubit}
   \begin{aligned}
   \ket{\Psi} &= \sum_{i,j \in \{0,1\}} c_{ij} \ket{\phi_i} \otimes \ket{\phi_j '}.
   \end{aligned}
\end{equation}
Within the larger two-qubit space there also exist states which cannot be
decomposed into independent vectors ${\bar{c}}$ and ${\bar{c}^{\prime} }$ (are not `separable'). 
These states are important and are known as \textbf{entangled} states.

Extending to $n$-qubits, an arbitrary quantum state can be written in the simplified form:
\begin{equation}
   \ket{\Psi} = \sum_{\mathbf{x} \in \{0,1\}^{n}} c_{\mathbf{x}} \ket{\phi_{\mathbf{x}}},
\end{equation}
which is a superposition of basis vectors corresponding to all possible $n$-bit strings. The normalization condition must continue to hold $\sum_{\mathbf{x} \in \{0,1\}^{n}}\vert c_{\mathbf{x}} \vert ^{2} = 1$, since $P(\phi_{\mathbf{x}}) = \vert c_{\mathbf{x}} \vert ^{2}$ and all measurement probabilities must sum to one.

\subsubsection{Quantum Logic Gates}
In classical computing, operations are performed using logic gates (e.g., AND, OR, NOT) which irreversibly manipulate electrical voltages. In quantum mechanics, operations are performed using \textbf{Quantum Logic Gates}, which mathematically correspond to reversible, unitary rotations of the state vector.

Understanding these single-qubit gates is a prerequisite for network engineers, as they form the ``alphabet'' for the quantum error models and correction protocols discussed later in this paper. If we identify classical information with the computational basis, then we have:
\begin{itemize}
    \item \textbf{Pauli-$X$ Gate:} Analogous to a classical NOT gate. It flips $|0\rangle$ to $|1\rangle$ and vice versa, representing a $180^\circ$ rotation around the X-axis of the Bloch sphere. This is the source of a \textit{bit-flip error}.
    \item \textbf{Pauli-$Z$ Gate:} This gate has no classical equivalent. It leaves the probability of 0 and 1 unchanged but flips the sign of the complex phase. This is the source of a \textit{phase-flip error}, which actively destroys quantum interference.
    \item \textbf{Hadamard ($H$) Gate:} Also lacking a classical equivalent, this gate transforms a deterministic computational basis state (like $|0\rangle$) into a perfectly balanced superposition state (like $|+\rangle$), rotating the vector from the pole to the equator of the Bloch sphere.
\end{itemize}

Though the $Z$-gate does not directly have a classical equivalent, notice that in the $X$-basis (spanned by $\ket{+}$ and $\ket{-}$) the $Z$ gate behaves like a classical NOT gate, flipping $\ket{+}$ to $\ket{-}$ and $\ket{-}$ to $\ket{+}$. This illustrates that the choice of basis is in fact arbitrary. When defining a qubit we must choose \emph{some} pair of orthogonal states to represent the computational basis states $0$ and $1$.
We could just as well have identified classical information with the $X$-basis states.
This would be equivalent to rotating the axes of the Bloch sphere shown in Fig. \ref{fig:bloch} (performing a Hadamard gate).
Some common single-qubit and two-qubit gates are shown in Tab. \ref{tab:quantum_gates}.

\begin{table}
\centering
\caption{Common single-qubit and two-qubit quantum gates and their matrix representations.}
\label{tab:quantum_gates}
    
\begin{tabular}{llll}
\hline
\textbf{Gate} & \textbf{Symbol} & \textbf{Circuit} & \textbf{Matrix Representation} \\
\hline
Identity & $I$ & \centering \begin{quantikz} \qw & \qw & \qw & \qw \end{quantikz} & $\begin{pmatrix} 1 & 0 \\ 0 & 1 \end{pmatrix}$ \\[0.8em]
Pauli-X & $X$ & \centering \begin{quantikz} \qw & \gate{X} & \qw \end{quantikz} & $\begin{pmatrix} 0 & 1 \\ 1 & 0 \end{pmatrix}$ \\[0.8em]
Pauli-Y & $Y$ & \centering \begin{quantikz} \qw & \gate{Y} & \qw \end{quantikz} & $\begin{pmatrix} 0 & -i \\ i & 0 \end{pmatrix}$ \\[0.8em]
Pauli-Z & $Z$ & \centering \begin{quantikz} \qw & \gate{Z} & \qw \end{quantikz} & $\begin{pmatrix} 1 & 0 \\ 0 & -1 \end{pmatrix}$ \\[0.8em]
$\frac{\pi}{8}$-gate & $T$ & \centering \begin{quantikz} \qw & \gate{T} & \qw \end{quantikz} & $\begin{pmatrix} e^{-i\pi/8} & 0 \\ 0 & e^{i \pi/8} \end{pmatrix}$ \\[0.8em]
Hadamard & $H$ & \centering \begin{quantikz} \qw & \gate{H} & \qw \end{quantikz} & $\frac{1}{\sqrt{2}}\begin{pmatrix} 1 & 1 \\ 1 & -1 \end{pmatrix}$ \\[0.8em]
CNOT & $CX$ & \centering \begin{quantikz} \qw & \ctrl{1} & \qw \\ \qw & \targ{} & \qw \end{quantikz} & $\begin{pmatrix} 1 & 0 & 0 & 0 \\ 0 & 1 & 0 & 0 \\ 0 & 0 & 0 & 1 \\ 0 & 0 & 1 & 0 \end{pmatrix}$ \\[0.8em]
SWAP & $SWAP$ & \centering \begin{quantikz} \qw & \swap{1} & \qw \\ \qw & \targX{} & \qw \end{quantikz} & $\begin{pmatrix} 1 & 0 & 0 & 0 \\ 0 & 0 & 1 & 0 \\ 0 & 1 & 0 & 0 \\ 0 & 0 & 0 & 1 \end{pmatrix}$ \\[0.8em]
\hline
\end{tabular}
\end{table}

In principle, any $2^{n} \times 2^{n}$ dimensional unitary matrix $U$ corresponds to a valid ${n}$-qubit quantum gate.
In practice, however, quantum devices use only a small subset of quantum gates to generate more complex unitary transformations. A set of gates that can approximate all unitary transformations to arbitrary precision is known as a universal gate set \cite{deutschUniversalityQuantumComputation1995}. A universal gate-set might only contain a small number of single-qubit gates (e.g. ${H}$ and ${T}$) and one two-qubit entangling gate (e.g. ${\mathrm{CNOT}}$)
\cite{barencoElementaryGatesQuantum1995}.

Certain network protocols do not require a universal gate set, and may not even require two-qubit operations (see e.g. Section~\ref{sec:qkd} on QKD).

\subsubsection{Qubit Preparation and Measurement}

The process of extracting information from a qubit is called measurement, which forces it to ``choose" a definitive state. 

\textbf{Classical Analogy and Physical Difference:} In classical networks, a router can passively read a packet header in a memory buffer without altering the payload. The data can be copied and inspected infinitely. In quantum mechanics, observing a state actively and irreversibly alters it. Returning to the spinning coin analogy, measuring the qubit might correspond to the physical act of flattening the coin onto the surface of the table.
This is not a unitary operation, and is not reversible.
Once collapsed, the complex phase information is permanently lost, and the qubit becomes a simple classical bit.
It is important to emphasize that this is not due to the use of excessive force or imprecision in the measurement, it is a fundamental property of quantum mechanics. The measurement basis must be chosen to correspond to the information that is desired to be extracted, making it a critical step in any quantum algorithm.

The measurement implementation varies, depending on physical medium used:

In single-photons, measurement is often implemented using a polarizing beam splitter and single-photon detectors. The beam splitter acts as a filter, directing photons with one polarization (e.g., vertical) to one detector and those with the orthogonal polarization (e.g., horizontal) to another. The detection of a photon by a specific detector constitutes the measurement and determines the qubit's final state. Matter-based systems use different techniques. For example, trapped ion qubits can be measured using lasers. A laser is tuned to a frequency that will only cause a specific ground state of the atom (e.g., $|0\rangle$) to fluoresce (emit light), while the other state ($|1\rangle$) remains dark. The presence or absence of fluorescence, detected by a camera, indicates the qubit's state~\cite{winelandExperimentalIssuesCoherent1998}. Most measurement techniques follow a similar principle -- perform some operation on the qubit that has a state dependent outcome that is measurable, and infer the state from this outcome.

\subsection{The No-Cloning Theorem}

One of the most profound differences between classical and quantum information is dictated by the \textbf{No-Cloning Theorem}~\cite{woottersSingleQuantumCannot1982}. This fundamental theorem of quantum mechanics states that it is physically impossible to create an identical copy from a single instance of an arbitrary, unknown quantum state.

\textbf{Classical Analogy and Physical Difference:} In a classical networking system, the foundation of data transmission and reliability is the ``CTRL+C / CTRL+V'' copy-paste mechanism. Bits can be perfectly duplicated to amplify signals, buffer packets, and implement Automatic Repeat reQuest (ARQ) retransmissions. In a quantum network, if a node possesses a qubit in an unknown state $|\Psi\rangle$, there is no physical operation or quantum gate that can duplicate it. Attempting to measure the qubit to learn its state will irrevocably collapse and destroy the original superposition.
Consequently, classical approaches to amplification, broadcasting and buffering for retransmission are not directly applicable to quantum information.

Quantum networks require entirely new operational paradigms to move data across distances.

\section{ENTANGLEMENT AND COMMUNICATIONS}
\label{sec:entanglement}

Building on the mechanics of single qubits, the foundation of quantum networking relies on multi-qubit systems. The primary resource that enables quantum communication is not the transmission of payload data itself, but the distribution of a unique quantum mechanical correlation. For the network engineer, understanding these specific multi-qubit mechanics is not just a physics prerequisite; it is the blueprint for the entire network control plane. Unlike classical networks where the physical layer simply dictates bandwidth, in a quantum network, the physical mechanics of entanglement dictate the routing logic, the scheduling deadlines, and the topology itself. The concepts introduced in this section—teleportation, swapping, and heralding—are the direct physical causes of the severe Quantum Delay, Timing Jitter, and Fidelity bottlenecks that the SDQN frameworks (detailed in Sections \ref{sec:performance_metrics} and \ref{sec:sdqn}) must actively manage.

\subsection{The Phenomenon of Entanglement}

Entanglement is often loosely described as ``quantum correlations'' between two subsystems. Classical statistics is of course capable of accounting for correlations between random variables. The aspect of entanglement that makes it useful in the computational setting is that these correlations exist simultaneously across \emph{incompatible} measurement bases.

Mathematically, the fundamental unit of entanglement in a quantum network is a maximally entangled pair of qubits, often called an EPR pair, an e-bit, or a \textbf{Bell state}. These states act as the ``blank copper wire'' of the Quantum Internet. There are four distinct Bell states:

\begin{align}
|\Phi^+\rangle &= \frac{1}{\sqrt{2}}(|00\rangle + |11\rangle) \\
|\Phi^-\rangle &= \frac{1}{\sqrt{2}}(|00\rangle - |11\rangle) \\
|\Psi^+\rangle &= \frac{1}{\sqrt{2}}(|01\rangle + |10\rangle) \\
|\Psi^-\rangle &= \frac{1}{\sqrt{2}}(|01\rangle - |10\rangle)
\end{align}

Consider the Bell state ${\lvert \Phi^{+} \rangle }$, where Alice and Bob each hold one of the two qubits. This state is perfectly correlated in the computational 0/1 basis.
If the first qubit is measured to be $|0\rangle$, the second qubit is known to also be $|0\rangle$. If the first qubit collapses to $|1\rangle$, the second will also be $|1\rangle$. To Alice and Bob, however, the measurement results still appear completely random, since they cannot know what the other's result was without communicating it classically.

Interestingly, if we convert ${\lvert \Phi^{+} \rangle}$ into the ${X}$ basis, we get the following state \begin{equation}
\lvert \Phi^{+} \rangle = \frac{\lvert ++ \rangle + \lvert -- \rangle}{\sqrt{2}}.
\end{equation}
This shows that a perfect correlation also exists in the $X$-basis. The existence of strong correlations across multiple incompatible measurement bases is what fundamentally distinguishes entanglement from classical correlations, and is formalized by \textbf{Bell's theorem}~\cite{bellEinsteinPodolskyRosen1964}, which shows that no classical model can reproduce these statistics.

While entanglement alone is not sufficient for communication, when complemented by classical communication it becomes a powerful networking tool that can be used to teleport qubits, secure communication channels and perform remote unitary operations.

\textbf{Classical Analogy and Physical Difference:} From a traditional networking perspective, entanglement is often misunderstood as a ``telephone line'' for transmitting data. A better classical analogy is a pair of perfectly synchronized, tamper-proof random number generators. If Alice and Bob each hold one, they will both generate the exact same sequence of random numbers simultaneously. However, neither Alice nor Bob can \textit{force} the generator to output a specific number (like a `1' instead of a `0'). Because they cannot control the output, they cannot use it to send a deterministic message. It is a shared correlation, not a communication channel.

Entanglement can be created in different ways depending partly on the physical medium. In many matter-based systems, entanglement can be deterministically generated using entangling gates. For example, trapped ions can be entangled by using a laser pulse to couple their internal states to their collective motion. Entanglement generation in photons is more challenging. One method uses a nonlinear crystal to split a single high-energy photon into two lower-energy entangled photons. This process, called spontaneous parametric down-conversion (SPDC), creates entangled polarizations, though the spontaneous nature of the process means the entanglement cannot be generated deterministically~\cite{kwiatNewHighintensitySource1995}. Alternatively, two-qubit gates can be probabilistically implemented using only linear optics, though boosting the overall success probabilities toward near-deterministic operation requires large numbers of additional photons and measurements~\cite{obrienOpticalQuantumComputing2007}.

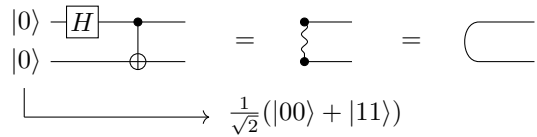
\begin{figure}[h]
    \centering
\usetikzlibrary{decorations.pathmorphing,decorations.markings}
\begin{tikzpicture}[scale=1.000000,x=1pt,y=1pt]
\draw[color=black] (0.000000,15.000000) -- (51.000000,15.000000);
\draw[color=white] (51.000000,15.000000) -- (96.000000,15.000000);
\draw[color=black] (96.000000,15.000000) -- (114.000000,15.000000);
\draw[color=white] (114.000000,15.000000) -- (162.000000,15.000000);
\draw[color=black] (162.000000,15.000000) -- (183.000000,15.000000);
\draw[color=black] (0.000000,15.000000) node[left] {$\ket{0}$};
\draw[color=black] (0.000000,0.000000) -- (51.000000,0.000000);
\draw[color=white] (51.000000,0.000000) -- (96.000000,0.000000);
\draw[color=black] (96.000000,0.000000) -- (114.000000,0.000000);
\draw[color=white] (114.000000,0.000000) -- (162.000000,0.000000);
\draw[color=black] (162.000000,0.000000) -- (183.000000,0.000000);
\draw[color=black] (0.000000,0.000000) node[left] {$\ket{0}$};
\begin{scope}
\draw[fill=white] (12.000000, 15.000000) +(-45.000000:8.485281pt and 8.485281pt) -- +(45.000000:8.485281pt and 8.485281pt) -- +(135.000000:8.485281pt and 8.485281pt) -- +(225.000000:8.485281pt and 8.485281pt) -- cycle;
\clip (12.000000, 15.000000) +(-45.000000:8.485281pt and 8.485281pt) -- +(45.000000:8.485281pt and 8.485281pt) -- +(135.000000:8.485281pt and 8.485281pt) -- +(225.000000:8.485281pt and 8.485281pt) -- cycle;
\draw (12.000000, 15.000000) node {$H$};
\end{scope}
\draw (33.000000,15.000000) -- (33.000000,0.000000);
\filldraw (33.000000, 15.000000) circle(1.500000pt);
\begin{scope}
\draw[fill=white] (33.000000, 0.000000) circle(3.000000pt);
\clip (33.000000, 0.000000) circle(3.000000pt);
\draw (30.000000, 0.000000) -- (36.000000, 0.000000);
\draw (33.000000, -3.000000) -- (33.000000, 3.000000);
\end{scope}
\filldraw (51.000000, 15.000000) circle(0.000000pt);
\filldraw (51.000000, 0.000000) circle(0.000000pt);
\draw[fill=white,color=white] (66.000000, -6.000000) rectangle (81.000000, 21.000000);
\draw (73.500000, 7.500000) node {$=$};
\draw[decorate,decoration={snake,amplitude=.4mm,segment length=2mm,post length=1mm}] (96.000000,15.000000) -- (96.000000,0.000000);
\filldraw (96.000000, 15.000000) circle(1.500000pt);
\filldraw (96.000000, 0.000000) circle(1.500000pt);
\filldraw (114.000000, 15.000000) circle(0.000000pt);
\filldraw (114.000000, 0.000000) circle(0.000000pt);
\draw[fill=white,color=white] (129.000000, -6.000000) rectangle (144.000000, 21.000000);
\draw (136.500000, 7.500000) node {$=$};
\draw[color=black] (162.000000, 15.000000) .. controls (154.800000, 15.000000) and (154.800000, 0.000000) .. (162.000000, 0.000000);
\filldraw (183.000000, 15.000000) circle(0.000000pt);
\filldraw (183.000000, 0.000000) circle(0.000000pt);

\draw (-10,-10) to (-10, -21);
\draw[->] (-10, -21) -- (60,-21);

\draw (100.000000, -20.500000) node {$\frac{1}{\sqrt{2}}(\ket{00}+\ket{11})$};

\end{tikzpicture}
    \caption{\textbf{Bell pair generation}. Different representations for the generation of a $\ket{\Phi^{+}}$ state. At the circuit level, the state is generated by converting one of the qubits to a $\ket{+}$ superposition, followed by a $\mathrm{CNOT}$, which flips the target qubit in one branch of the superposition. This state can be prepared in different ways, and alternative diagrammatic notation may be used.}
    \label{fig:bell-pair}
\end{figure}

Though the specific mechanisms differ, we can view the process of generating entanglement from separable qubit states using a quantum circuit. For example, Fig. \ref{fig:bell-pair} shows the generation of $\ket{\Phi^{+}}$ from $\ket{00}$.

It is also possible to perform joint entangling measurements, in which the measurement projects the qubits into an entangled state. One important example is a Bell-state measurement (BSM). An ideal BSM can distinguish all four two-qubit Bell states. In linear-optical photonic encodings with two photons and passive elements, however, at most two of the four Bell states can be unambiguously discriminated~\cite{calsamigliaMaximumEfficiencyLinearoptical2001}, so the success probability is bounded by 50\%. Moreover, standard single-photon detection is destructive, so the photons are not available after the measurement for subsequent processing. On the other hand, if each photon is entangled with a stationary qubit (for example, an ion), a successful BSM on the photons probabilistically projects the stationary qubits into an entangled state~\cite{barrettEfficientHighfidelityQuantum2005, moehringEntanglementSingleatomQuantum2007a}. Crucially, this can be done in a heralded fashion, where specific detection outcomes signal (herald) that entanglement generation has succeeded~\cite{forbesHeraldedGenerationEntanglement2025}.

\subsection{The No-Communication Theorem}

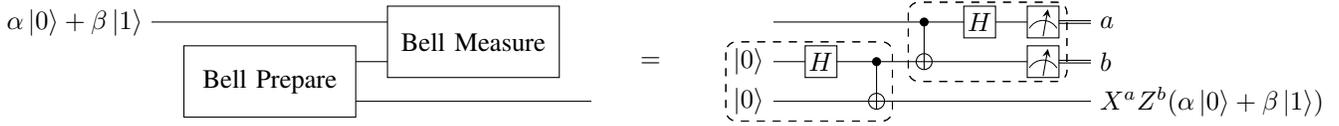
\begin{figure*}[h!]
    \centering
    \usetikzlibrary{decorations.pathmorphing,decorations.markings}
\providecommand{\ket}[1]{\left|#1\right\rangle}
\begin{tikzpicture}[scale=1.000000,x=1pt,y=1pt]
\filldraw[color=white] (0.000000, -7.500000) rectangle (397.000000, 37.500000);
\draw[color=black] (13.500000,30.000000) -- (142.500000,30.000000);
\draw[color=white] (142.500000,30.000000) -- (194.500000,30.000000);
\draw[color=black] (248.500000,30.000000) -- (358.000000,30.000000);
\draw[color=black] (358.000000,29.500000) -- (383.500000,29.500000);
\draw[color=black] (358.000000,30.500000) -- (383.500000,30.500000);
\draw[color=white] (13.500000,15.000000) -- (65.500000,15.000000);
\draw[color=black] (65.500000,15.000000) -- (142.500000,15.000000);
\draw[color=white] (142.500000,15.000000) -- (194.500000,15.000000);
\draw[color=black] (248.500000,15.000000) -- (358.000000,15.000000);
\draw[color=black] (358.000000,14.500000) -- (383.500000,14.500000);
\draw[color=black] (358.000000,15.500000) -- (383.500000,15.500000);
\draw[color=white] (13.500000,0.000000) -- (65.500000,0.000000);
\draw[color=black] (65.500000,0.000000) -- (194.500000,0.000000);
\draw[color=black] (248.500000,0.000000) -- (383.500000,0.000000);
\draw[color=black] (21.000000,30.000000) node[fill=white,left,minimum height=15.000000pt,minimum width=15.000000pt,inner sep=0pt] {\phantom{$\alpha\ket{0}+\beta\ket{1}$}};
\draw[color=black] (21.000000,30.000000) node[left] {$\alpha\ket{0}+\beta\ket{1}$};
\draw[color=white] (21.000000,15.000000) node[fill=white,left,minimum height=15.000000pt,minimum width=15.000000pt,inner sep=0pt] {\phantom{$\ket{0}$}};
\draw[color=white] (21.000000,15.000000) node[left] {$\ket{0}$};
\draw[color=white] (21.000000,0.000000) node[fill=white,left,minimum height=15.000000pt,minimum width=15.000000pt,inner sep=0pt] {\phantom{$\ket{0}$}};
\draw[color=white] (21.000000,0.000000) node[left] {$\ket{0}$};
\draw (65.500000,15.000000) -- (65.500000,0.000000);
\begin{scope}
\draw[fill=white] (65.500000, 7.500000) +(-45.000000:45.961941pt and 19.091883pt) -- +(45.000000:45.961941pt and 19.091883pt) -- +(135.000000:45.961941pt and 19.091883pt) -- +(225.000000:45.961941pt and 19.091883pt) -- cycle;
\clip (65.500000, 7.500000) +(-45.000000:45.961941pt and 19.091883pt) -- +(45.000000:45.961941pt and 19.091883pt) -- +(135.000000:45.961941pt and 19.091883pt) -- +(225.000000:45.961941pt and 19.091883pt) -- cycle;
\draw (65.500000, 7.500000) node {{Bell Prepare}};
\end{scope}
\draw (142.500000,30.000000) -- (142.500000,15.000000);
\begin{scope}
\draw[fill=white] (142.500000, 22.500000) +(-45.000000:45.961941pt and 19.091883pt) -- +(45.000000:45.961941pt and 19.091883pt) -- +(135.000000:45.961941pt and 19.091883pt) -- +(225.000000:45.961941pt and 19.091883pt) -- cycle;
\clip (142.500000, 22.500000) +(-45.000000:45.961941pt and 19.091883pt) -- +(45.000000:45.961941pt and 19.091883pt) -- +(135.000000:45.961941pt and 19.091883pt) -- +(225.000000:45.961941pt and 19.091883pt) -- cycle;
\draw (142.500000, 22.500000) node {{Bell Measure}};
\end{scope}
\draw[color=white] (187.000000,30.000000) node[fill=white,right,minimum height=15.000000pt,minimum width=15.000000pt,inner sep=0pt] {\phantom{${}$}};
\draw[color=white] (187.000000,30.000000) node[right] {${}$};
\draw[color=white] (187.000000,15.000000) node[fill=white,right,minimum height=15.000000pt,minimum width=15.000000pt,inner sep=0pt] {\phantom{${}$}};
\draw[color=white] (187.000000,15.000000) node[right] {${}$};
\draw[color=black] (187.000000,0.000000) node[fill=white,right,minimum height=15.000000pt,minimum width=15.000000pt,inner sep=0pt] {\phantom{${}$}};
\draw[color=black] (187.000000,0.000000) node[right] {${}$};
\draw[fill=white,color=white] (214.000000, -6.000000) rectangle (229.000000, 36.000000);
\draw (209.500000, 15.000000) node {$=$};
\draw[color=black] (256.000000,30.000000) node[fill=white,left,minimum height=15.000000pt,minimum width=15.000000pt,inner sep=0pt] {\phantom{${}$}};
\draw[color=black] (256.000000,30.000000) node[left] {${}$};
\draw[color=black] (256.000000,15.000000) node[fill=white,left,minimum height=15.000000pt,minimum width=15.000000pt,inner sep=0pt] {\phantom{${\ket{0}}$}};
\draw[color=black] (256.000000,15.000000) node[left] {${\ket{0}}$};
\draw[color=black] (256.000000,0.000000) node[fill=white,left,minimum height=15.000000pt,minimum width=15.000000pt,inner sep=0pt] {\phantom{${\ket{0}}$}};
\draw[color=black] (256.000000,0.000000) node[left] {${\ket{0}}$};
\begin{scope}
\draw[fill=white] (274.000000, 15.000000) +(-45.000000:8.485281pt and 8.485281pt) -- +(45.000000:8.485281pt and 8.485281pt) -- +(135.000000:8.485281pt and 8.485281pt) -- +(225.000000:8.485281pt and 8.485281pt) -- cycle;
\clip (274.000000, 15.000000) +(-45.000000:8.485281pt and 8.485281pt) -- +(45.000000:8.485281pt and 8.485281pt) -- +(135.000000:8.485281pt and 8.485281pt) -- +(225.000000:8.485281pt and 8.485281pt) -- cycle;
\draw (274.000000, 15.000000) node {$H$};
\end{scope}
\draw (295.000000,15.000000) -- (295.000000,0.000000);
\filldraw (295.000000, 15.000000) circle(1.500000pt);
\begin{scope}
\draw[fill=white] (295.000000, 0.000000) circle(3.000000pt);
\clip (295.000000, 0.000000) circle(3.000000pt);
\draw (292.000000, 0.000000) -- (298.000000, 0.000000);
\draw (295.000000, -3.000000) -- (295.000000, 3.000000);
\end{scope}
\draw (313.000000,30.000000) -- (313.000000,15.000000);
\begin{scope}
\draw[fill=white] (313.000000, 15.000000) circle(3.000000pt);
\clip (313.000000, 15.000000) circle(3.000000pt);
\draw (310.000000, 15.000000) -- (316.000000, 15.000000);
\draw (313.000000, 12.000000) -- (313.000000, 18.000000);
\end{scope}
\filldraw (313.000000, 30.000000) circle(1.500000pt);
\begin{scope}
\draw[fill=white] (334.000000, 30.000000) +(-45.000000:8.485281pt and 8.485281pt) -- +(45.000000:8.485281pt and 8.485281pt) -- +(135.000000:8.485281pt and 8.485281pt) -- +(225.000000:8.485281pt and 8.485281pt) -- cycle;
\clip (334.000000, 30.000000) +(-45.000000:8.485281pt and 8.485281pt) -- +(45.000000:8.485281pt and 8.485281pt) -- +(135.000000:8.485281pt and 8.485281pt) -- +(225.000000:8.485281pt and 8.485281pt) -- cycle;
\draw (334.000000, 30.000000) node {$H$};
\end{scope}
\draw[fill=white] (352.000000, 24.000000) rectangle (364.000000, 36.000000);
\draw[very thin] (358.000000, 30.600000) arc (90:150:6.000000pt);
\draw[very thin] (358.000000, 30.600000) arc (90:30:6.000000pt);
\draw[->,>=stealth] (358.000000, 24.600000) -- +(80:10.392305pt);
\draw[fill=white] (352.000000, 9.000000) rectangle (364.000000, 21.000000);
\draw[very thin] (358.000000, 15.600000) arc (90:150:6.000000pt);
\draw[very thin] (358.000000, 15.600000) arc (90:30:6.000000pt);
\draw[->,>=stealth] (358.000000, 9.600000) -- +(80:10.392305pt);
\draw[color=black] (376.000000,30.000000) node[fill=white,right,minimum height=15.000000pt,minimum width=15.000000pt,inner sep=0pt] {\phantom{${a}$}};
\draw[color=black] (376.000000,30.000000) node[right] {${a}$};
\draw[color=black] (376.000000,15.000000) node[fill=white,right,minimum height=15.000000pt,minimum width=15.000000pt,inner sep=0pt] {\phantom{${b}$}};
\draw[color=black] (376.000000,15.000000) node[right] {${b}$};
\draw[color=black] (376.000000,0.000000) node[fill=white,right,minimum height=15.000000pt,minimum width=15.000000pt,inner sep=0pt] {\phantom{$X^{b}Z^{a}(\alpha\ket{0}+\beta\ket{1})$}};
\draw[color=black] (376.000000,0.000000) node[right] {$X^{a}Z^{b}(\alpha\ket{0}+\beta\ket{1})$};
\draw[draw opacity=1.000000,fill opacity=0.200000,color=black,dashed,rounded corners=4pt] (238.000000,22.500000) rectangle (301.000000,-7.500000);
\draw[draw opacity=1.000000,fill opacity=0.200000,color=black,dashed,rounded corners=4pt] (238.000000,22.500000) rectangle (301.000000,-7.500000);
\draw[draw opacity=1.000000,fill opacity=0.200000,color=black,dashed,rounded corners=4pt] (307.000000,37.500000) rectangle (367.000000,7.500000);
\draw[draw opacity=1.000000,fill opacity=0.200000,color=black,dashed,rounded corners=4pt] (307.000000,37.500000) rectangle (367.000000,7.500000);
\end{tikzpicture}
    \caption{Quantum state teleportation. A qubit can be reconstructed at distant location using a pre-shared Bell state and local Bell-state measurement (BSM). The BSM projects the unmeasured qubit into the original qubit's state up to two possible correction operations. The bits $a$ and $b$ are classical measurement results which determine whether or not correction operations must be applied.}
    \label{fig:state-telep}
\end{figure*}

Entanglement allows two parties, Alice and Bob, to share a quantum link by each holding one of a pair of entangled qubits. Given that entanglement establishes a non-local correlation between distant qubits, can this shared state be leveraged as a direct, Faster-Than-Light (FTL) communication channel? If Alice could locally manipulate her qubit in such a way that it induces a locally detectable change on Bob's qubit, it would represent a new signaling mechanism. This is the most common misconception we must address.

The answer to this question is a definitive no. Any such hypothesis is invalidated by a fundamental principle known as the \textbf{no-communication theorem} (also known as no-signaling therem) . The explanation is that for a ``signal” to be received, the manipulation must cause a \textbf{locally detectable change} for the other person (Bob). The only thing Bob can locally detect is the probability of measuring 0 or 1 on his qubit. Any manipulation Alice performs (like a Pauli gate flip) is a \textbf{unitary operation}, which, by its very nature, conserves the norms of the superposition coefficients—the values that determine probability via the Born rule. 

For example, if Alice and Bob share the state $\frac{1}{\sqrt{2}}(|00\rangle + |11\rangle)$, Bob's local probability of measuring 0 is 50\% and 1 is 50\%. If Alice ``flips" her qubit using an X-gate, the global state becomes $\frac{1}{\sqrt{2}}(|10\rangle + |01\rangle)$. From Bob's local perspective, his probability of measuring 0 (now from the $|10\rangle$ term) is still 50\%, and his probability of measuring 1 (from the $|01\rangle$ term) is also 50\%. Since Bob's local measurement statistics remain unchanged, he has no way of knowing Alice did anything, and thus no information can be sent; Alice's action only changes the \textbf{global correlation pattern} which is undetectable without cross-referencing via classical communication.
From an architectural standpoint, the no-communication theorem mandates that a quantum network can never operate in isolation. It proves that the Quantum Internet must be designed as a strictly dual-plane architecture (as synthesized later in Section \ref{sec:protocol-stack}), where a classical network data-plane operates in tandem to carry the essential measurement results required to unlock the quantum correlations.

\subsection{Quantum Teleportation: Moving State Information}

If the no-communication theorem proves you cannot send information just by manipulating entanglement, how does a ``quantum network" actually move quantum payload data? The answer is the \textbf{Quantum Teleportation} procedure, introduced by Bennett \cite{bennettTeleportingUnknownQuantum1993}. 

\textbf{Classical Analogy and Physical Difference:} In classical networks, moving a payload involves copying the file and transmitting it over a link. In quantum networks, the No-Cloning theorem forbids copying. Teleportation is conceptually similar to a "Live Migration" of a Virtual Machine (VM) between servers, where the VM state is recreated on the destination server and explicitly deleted from the source server. However, teleportation does not physically move matter or energy; it transfers the \textit{quantum information} (the exact superposition coefficients $\alpha$ and $\beta$) of a data qubit onto a distant target qubit.

To teleport a data qubit, Alice and Bob must first share a pre-established entangled Bell state. Alice performs a joint Bell-state measurement (BSM) on her data qubit and her half of the entangled pair. This measurement completely destroys the original data qubit (satisfying the No-Cloning rule) and forces Bob's distant qubit into a state correlated with the original data. However, Bob's qubit is scrambled by a random Pauli rotation. To fix this, Alice must use a \textit{classical network} to transmit the 2-bit result of her measurement to Bob. Bob uses these two classical bits to apply the correct inverse Pauli gate (X, Z, both, or neither) to his qubit, perfectly recovering the original state. Therefore, teleporting one qubit requires the consumption of one entangled pair \textit{and} the transmission of two classical bits, strictly bounding quantum network speeds by classical network latency.

Since different forms of teleportation within quantum information theory, this is often disambiguated as \textbf{quantum state teleportation}. The procedure is illustrated in Fig. \ref{fig:state-telep}.

\subsection{Entanglement Distribution and Swapping}

\begin{figure*}[htpb!]
    \input{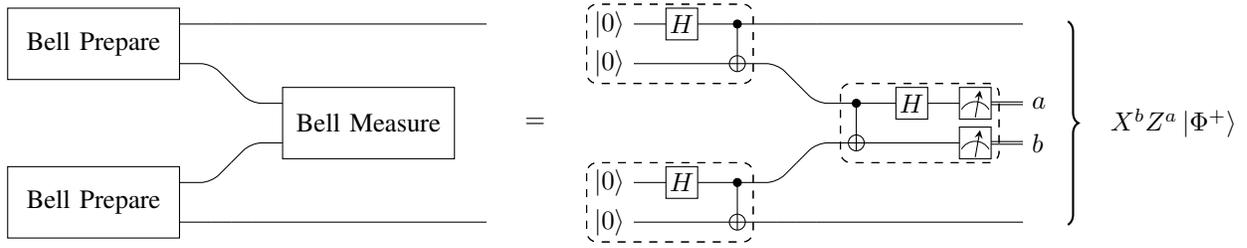}
    \caption{The entanglement swap circuit. The entanglement swap uses pre-shared entanglement and a local BSM to transfer the entanglement onto the unmeasured qubits. The procedure is identical to quantum state teleportation, but where the teleported qubit is already entangled with another qubit.}
    \label{fig:ent-swap-circuit}
\end{figure*}

\begin{figure*}
\centering
    \begin{tikzpicture}[scale=0.8]
	\begin{pgfonlayer}{nodelayer}
		\node [style=white, label=above:{A}] (203) at (2.5, 1.5) {};
		\node [style=white, label=above:{B}] (211) at (15.5, 1.5) {};
		\node [style=none] (212) at (4, 2) {};
		\node [style=none] (213) at (4, 1) {};
		\node [style=none] (214) at (7, 2) {};
		\node [style=none] (215) at (7, 1) {};
		\node [style=none] (216) at (5.5, 1.5) {Photon source};
		\node [style=none] (217) at (4, 1.5) {};
		\node [style=none] (218) at (7, 1.5) {};
		\node [style=none] (221) at (11, 2) {};
		\node [style=none] (222) at (11, 1) {};
		\node [style=none] (223) at (14, 2) {};
		\node [style=none] (224) at (14, 1) {};
		\node [style=none] (225) at (12.5, 1.5) {Photon source};
		\node [style=none] (226) at (11, 1.5) {};
		\node [style=none] (227) at (14, 1.5) {};
		\node [style=white] (232) at (8.5, 1.5) {};
		\node [style=white] (233) at (9.5, 1.5) {};
		\node [style=none] (234) at (9, 2.25) {Repeater};

		\node [style=white, label=above:{A}] (275) at (2.5, -3.25) {};
		\node [style=white, label=above:{B}] (276) at (15.5, -3.25) {};
		\node [style=none] (277) at (4, -2.75) {};
		\node [style=none] (278) at (4, -3.75) {};
		\node [style=none] (279) at (7, -2.75) {};
		\node [style=none] (280) at (7, -3.75) {};
		\node [style=none] (281) at (5.5, -3.25) {Photon source};
		\node [style=none] (282) at (4, -3.25) {};
		\node [style=none] (283) at (7, -3.25) {};
		\node [style=none] (284) at (11, -2.75) {};
		\node [style=none] (285) at (11, -3.75) {};
		\node [style=none] (286) at (14, -2.75) {};
		\node [style=none] (287) at (14, -3.75) {};
		\node [style=none] (288) at (12.5, -3.25) {Photon source};
		\node [style=none] (289) at (11, -3.25) {};
		\node [style=none] (290) at (14, -3.25) {};
		\node [style=white] (295) at (8.5, -3.25) {};
		\node [style=white] (296) at (9.5, -3.25) {};
		\node [style=black] (298) at (9, 1.5) {};
		\node [style=black] (299) at (9, -3.25) {};
		\node [style=none] (301) at (5.5, -0.25) {$\ket{\Phi^+}$};
		\node [style=none] (306) at (12.5, -0.25) {$\ket{\Phi^+}$};;
		\node [style=none] (307) at (9, -5.75) {$X^{b}Z^{a}\ket{\Phi^+}$};
		\node [style=none] (308) at (9, -0.25) {BSM};
		\node [style=none] (309) at (9, -0.75) {};
		\node [style=none] (310) at (9, -1.75) {};
		\node [style=none] (311) at (9, -2.5) {Bits $a,b$};
		\node [style=none] (324) at (5.5, 0.25) {};
		\node [style=none] (404) at (2.5, 1) {};
		\node [style=none] (405) at (8.5, 1) {};
		\node [style=none] (406) at (12.5, 0.25) {};
		\node [style=none] (407) at (9.5, 1) {};
		\node [style=none] (408) at (15.5, 1) {};
		\node [style=none] (409) at (9, -5.25) {};
		\node [style=none] (410) at (2.5, -3.75) {};
		\node [style=none] (411) at (15.5, -3.75) {};
	\end{pgfonlayer}
	\begin{pgfonlayer}{edgelayer}
		\draw [style=fibre] (212.center) to (212.center);
		\draw [style=bluefill] (212.center) to (213.center);
		\draw [style=bluefill] (213.center) to (215.center);
		\draw [style=bluefill] (215.center) to (214.center);
		\draw [style=bluefill] (214.center) to (212.center);
		\draw [style=snake arrow] (217.center) to (203);
		\draw [style=fibre] (221.center) to (221.center);
		\draw [style=bluefill] (221.center) to (222.center);
		\draw [style=bluefill] (222.center) to (224.center);
		\draw [style=bluefill] (224.center) to (223.center);
		\draw [style=bluefill] (223.center) to (221.center);
		\draw [style=snake arrow, out=0, in=180] (218.center) to (232);
		\draw [style=snake arrow] (226.center) to (233);
		\draw [style=snake arrow] (227.center) to (211);
		\draw [style=fibre] (277.center) to (277.center);
		\draw [style=bluefill] (277.center) to (278.center);
		\draw [style=bluefill] (278.center) to (280.center);
		\draw [style=bluefill] (280.center) to (279.center);
		\draw [style=bluefill] (279.center) to (277.center);
		\draw [style=fibre] (284.center) to (284.center);
		\draw [style=bluefill] (284.center) to (285.center);
		\draw [style=bluefill] (285.center) to (287.center);
		\draw [style=bluefill] (287.center) to (286.center);
		\draw [style=bluefill] (286.center) to (284.center);
		\draw (232) to (233);
		\draw (295) to (299);
		\draw (299) to (296);
		\draw [style=arrow] (309.center) to (310.center);
		\draw [out=-30, in=135] (404.center) to (324.center);
		\draw [out=45, in=-150] (324.center) to (405.center);
		\draw [out=-30, in=135] (407.center) to (406.center);
		\draw [out=45, in=-150] (406.center) to (408.center);
		\draw [out=-30, in=135] (410.center) to (409.center);
		\draw [out=45, in=-150] (409.center) to (411.center);
	\end{pgfonlayer}

    \draw[<-, double, thick] (2.6,-2.5) -- (8,-2.5);
    \draw[->, double, thick] (10,-2.5) -- (15.4,-2.5);
\end{tikzpicture}
    \caption{Entanglement swap in a network. Two photon sources first share Bell pairs along the intermediate links. The central node, the \textit{repeater}, performs a BSM and communicates the measurement results to the end nodes. The end-to-end Bell pair can be reconstructed based on the measurement results.}
    \label{fig:ent-swap}
\end{figure*}
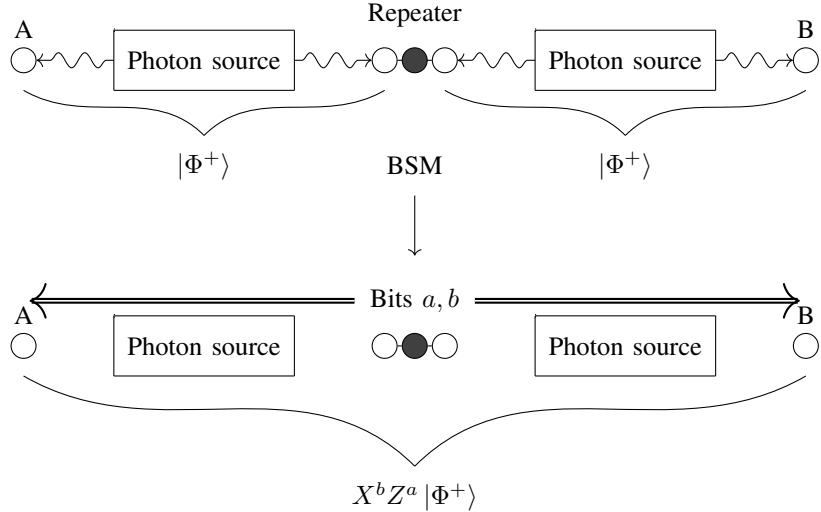

Because protocols like teleportation and QKD consume entanglement, quantum communication is fundamentally a \textbf{hybrid classical-quantum process}. The quantum network's primary job is not to transmit the message itself, but to distribute the shared resource of entanglement, which is then coordinated by classical networks to perform tasks that are not possible for the classical networks alone. 

Due to the no-cloning theorem, a quantum state cannot be copied or amplified, meaning a ``lost" photon is lost forever. This fragility dictates that the strategy for establishing entanglement depends entirely on the target distance. The most basic method is \textbf{direct distribution}, where a source generates an entangled photon pair and sends one photon through an optical fibre to a remote node. While simple and effective for short distances, this method's success rate drops exponentially with distance. The probability of a single photon surviving a long fibre-optic journey is vanishingly small.

To overcome this distance limitation, quantum networks employ \textbf{entanglement swapping} \cite{zukowskiEventreadydetectorsBellExperiment1993}, a technique that forms the basis of a ``quantum repeater'' \cite{briegelQuantumRepeatersCommunication1998}. In the simplest case, two end nodes (A and B) independently establish entanglement with a central midpoint station (M). The midpoint station then performs a joint Bell-state measurement (BSM) on the two photons it received. This BSM ``swaps" the entanglement: it breaks the A-M and B-M links and, in their place, establishes a new entangled link directly between the distant end nodes, A and B, between which no direct quantum interactions occurred. 

One might notice that the entanglement swap and the teleportation procedure are very similar. In fact, they are identical---the entanglement swap is just a teleportation where the qubit being teleported is already entangled with another qubit. The circuit-level equivalence of this procedure is shown in Fig. \ref{fig:ent-swap-circuit}, while its macro-level application between network nodes is illustrated in Fig. \ref{fig:ent-swap}.

\textbf{Classical Analogy:} This is the quantum equivalent of taking two separate, short ethernet cables and physically splicing them together inside a router to create one long, contiguous cable reaching from end to end.

This single-hop swapping technique is the building block for true \textit{multi-hop networks} capable of spanning continents. A long-distance link is broken into a series of shorter, manageable segments connected by quantum repeater stations. Entanglement is first established over each individual hop. Then, at each repeater, a BSM is performed to ``stitch" the adjacent entangled links together. This process is repeated down the line until a single, end-to-end entangled link is created. This is the only known method to create long-distance quantum links without using ``trusted relays," which would compromise the network's security.

In classical networks, routing is primarily a matter of finding the shortest path and avoiding congested queues. However, because entanglement swapping requires the active destruction of intermediate resources via Bell-state measurements, quantum routing is fundamentally a problem of consumable resource allocation. As will be formalized in the Q-NUM framework (Section \ref{sec:q-num}), the SDQN controller cannot simply calculate a path; it must simultaneously schedule the probabilistic generation and consumption of these fragile links across the entire multi-hop chain before the states succumb to decoherence.

\subsection{Gate Teleportation: Performing Remote Unitary Operations}

The principles of teleportation can also be used to enact the effect of multi-qubit gates non-locally~\cite{eisertOptimalLocalImplementation2000}. The concept of gate teleportation is widely used in quantum computing and generally refers to procedures which use a resource state, local operations, measurements and classical feedforward to indirectly perform a quantum gate. In this context, Alice and Bob each hold a qubit, and they wish to perform a controlled-unitary gate (a CNOT, for example) from Alice's qubit to Bob's. They start by sharing a Bell pair $\ket{\Phi^{+}}$. After Alice receives her half of the Bell pair, shes performs a local CNOT gate to entangle with her half of the Bell pair. She then measures the half in the computational basis and the result and sends the result to Bob, who performs a classically controlled $X$ correction on his half of the Bell pair. This projects the joint state of Alice's qubit and Bob's half of the Bell pair into the entangled state $\alpha \ket{00} + \beta \ket{11}$. The state information of Alice's qubit now spans both qubits. Importantly, this means that either Alice's qubit or Bob's half of the Bell pair can be used as the control now in a multi-qubit gate with the same effect on the target.

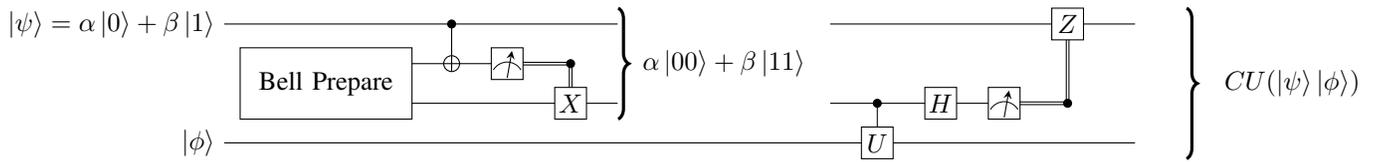
\begin{figure*}
\usetikzlibrary{decorations.pathmorphing,decorations.markings}
\providecommand{\ket}[1]{\left|#1\right\rangle}
\begin{tikzpicture}[scale=1.00000,x=1pt,y=1pt]
\filldraw[color=white] (0.000000, -7.500000) rectangle (450.000000, 52.500000);
\draw[color=black] (0.000000,45.000000) -- (344.500000,45.000000);
\draw[color=black] (0.000000,45.000000) node[left] {${\ket{\psi} = \alpha\ket{0}+\beta\ket{1}}$};
\draw[color=white] (0.000000,30.000000) -- (38.500000,30.000000);
\draw[color=black] (38.500000,30.000000) -- (107.000000,30.000000);
\draw[color=black] (107.000000,29.500000) -- (131.000000,29.500000);
\draw[color=black] (107.000000,30.500000) -- (131.000000,30.500000);
\draw[color=white] (0.000000,30.000000) node[left] {$\ket{0}$};
\draw[color=white] (0.000000,15.000000) -- (38.500000,15.000000);
\draw[color=black] (38.500000,15.000000) -- (295.000000,15.000000);
\draw[color=black] (295.000000,14.500000) -- (319.000000,14.500000);
\draw[color=black] (295.000000,15.500000) -- (319.000000,15.500000);
\draw[color=white] (0.000000,15.000000) node[left] {$\ket{0}$};
\draw[color=black] (0.000000,0.000000) -- (344.500000,0.000000);
\draw[color=black] (0.000000,0.000000) node[left] {$\ket{\phi}$};
\draw (38.500000,30.000000) -- (38.500000,15.000000);
\begin{scope}
\draw[fill=white] (38.500000, 22.500000) +(-45.000000:45.961941pt and 19.091883pt) -- +(45.000000:45.961941pt and 19.091883pt) -- +(135.000000:45.961941pt and 19.091883pt) -- +(225.000000:45.961941pt and 19.091883pt) -- cycle;
\clip (38.500000, 22.500000) +(-45.000000:45.961941pt and 19.091883pt) -- +(45.000000:45.961941pt and 19.091883pt) -- +(135.000000:45.961941pt and 19.091883pt) -- +(225.000000:45.961941pt and 19.091883pt) -- cycle;
\draw (38.500000, 22.500000) node {{Bell Prepare}};
\end{scope}
\draw (86.000000,45.000000) -- (86.000000,30.000000);
\begin{scope}
\draw[fill=white] (86.000000, 30.000000) circle(3.000000pt);
\clip (86.000000, 30.000000) circle(3.000000pt);
\draw (83.000000, 30.000000) -- (89.000000, 30.000000);
\draw (86.000000, 27.000000) -- (86.000000, 33.000000);
\end{scope}
\filldraw (86.000000, 45.000000) circle(1.500000pt);
\draw[fill=white] (101.000000, 24.000000) rectangle (113.000000, 36.000000);
\draw[very thin] (107.000000, 30.600000) arc (90:150:6.000000pt);
\draw[very thin] (107.000000, 30.600000) arc (90:30:6.000000pt);
\draw[->,>=stealth] (107.000000, 24.600000) -- +(80:10.392305pt);
\draw (130.500000,30.000000) -- (130.500000,15.000000);
\draw (131.500000,30.000000) -- (131.500000,15.000000);
\begin{scope}
\draw[fill=white] (131.000000, 15.000000) +(-45.000000:8.485281pt and 8.485281pt) -- +(45.000000:8.485281pt and 8.485281pt) -- +(135.000000:8.485281pt and 8.485281pt) -- +(225.000000:8.485281pt and 8.485281pt) -- cycle;
\clip (131.000000, 15.000000) +(-45.000000:8.485281pt and 8.485281pt) -- +(45.000000:8.485281pt and 8.485281pt) -- +(135.000000:8.485281pt and 8.485281pt) -- +(225.000000:8.485281pt and 8.485281pt) -- cycle;
\draw (131.000000, 15.000000) node {$X$};
\end{scope}
\filldraw (131.000000, 30.000000) circle(1.500000pt);
\draw[fill=white,color=white] (149.000000, 9.000000) rectangle (229.000000, 51.000000);
\draw (189.000000, 30.000000) node {$\alpha\ket{00}+\beta\ket{11}$};
\draw[decorate,decoration={brace,mirror,amplitude = 4.000000pt},very thick] (149.000000,9.000000) -- (149.000000,51.000000);
\draw (247.000000,15.000000) -- (247.000000,0.000000);
\begin{scope}
\draw[fill=white] (247.000000, -0.000000) +(-45.000000:8.485281pt and 8.485281pt) -- +(45.000000:8.485281pt and 8.485281pt) -- +(135.000000:8.485281pt and 8.485281pt) -- +(225.000000:8.485281pt and 8.485281pt) -- cycle;
\clip (247.000000, -0.000000) +(-45.000000:8.485281pt and 8.485281pt) -- +(45.000000:8.485281pt and 8.485281pt) -- +(135.000000:8.485281pt and 8.485281pt) -- +(225.000000:8.485281pt and 8.485281pt) -- cycle;
\draw (247.000000, -0.000000) node {$U$};
\end{scope}
\filldraw (247.000000, 15.000000) circle(1.500000pt);
\begin{scope}
\draw[fill=white] (271.000000, 15.000000) +(-45.000000:8.485281pt and 8.485281pt) -- +(45.000000:8.485281pt and 8.485281pt) -- +(135.000000:8.485281pt and 8.485281pt) -- +(225.000000:8.485281pt and 8.485281pt) -- cycle;
\clip (271.000000, 15.000000) +(-45.000000:8.485281pt and 8.485281pt) -- +(45.000000:8.485281pt and 8.485281pt) -- +(135.000000:8.485281pt and 8.485281pt) -- +(225.000000:8.485281pt and 8.485281pt) -- cycle;
\draw (271.000000, 15.000000) node {$H$};
\end{scope}
\draw[fill=white] (289.000000, 9.000000) rectangle (301.000000, 21.000000);
\draw[very thin] (295.000000, 15.600000) arc (90:150:6.000000pt);
\draw[very thin] (295.000000, 15.600000) arc (90:30:6.000000pt);
\draw[->,>=stealth] (295.000000, 9.600000) -- +(80:10.392305pt);
\draw (318.500000,45.000000) -- (318.500000,15.000000);
\draw (319.500000,45.000000) -- (319.500000,15.000000);
\begin{scope}
\draw[fill=white] (319.000000, 45.000000) +(-45.000000:8.485281pt and 8.485281pt) -- +(45.000000:8.485281pt and 8.485281pt) -- +(135.000000:8.485281pt and 8.485281pt) -- +(225.000000:8.485281pt and 8.485281pt) -- cycle;
\clip (319.000000, 45.000000) +(-45.000000:8.485281pt and 8.485281pt) -- +(45.000000:8.485281pt and 8.485281pt) -- +(135.000000:8.485281pt and 8.485281pt) -- +(225.000000:8.485281pt and 8.485281pt) -- cycle;
\draw (319.000000, 45.000000) node {$Z$};
\end{scope}
\filldraw (319.000000, 15.000000) circle(1.500000pt);
\draw[fill=white,color=white] (364.000000, -6.000000) rectangle (444.000000, 51.000000);
\draw (404.000000, 22.500000) node {$CU(\ket{\psi}\ket{\phi})$};
\draw[decorate,decoration={brace,mirror,amplitude = 4.000000pt},very thick] (364.000000,-6.000000) -- (364.000000,51.000000);
\end{tikzpicture}
    \caption{\textbf{Gate teleportation}. Gate teleportation can be used to perform remote controlled-unitary gates. The initial part of the circuit spreads the control information to the remote location using a shared Bell pair, while the intermediate part uses the remote half of the Bell pair to control the desired unitary operations $U$. After the remote half of the Bell pair is measured out and classical corrections applied, the net effect is the controlled unitary between the two initial qubits.}
    \label{}
\end{figure*}

One may question at this point, does this violate the no-cloning theorem? The answer is no, if we had copied the qubit the joint state would be the unentangled state $(\alpha\ket{0}+\beta\ket{1}) \otimes (\alpha\ket{0}+\beta\ket{1})$. Instead, we have an entangled state in which both qubits contain the state information. However, the information is not independently contained, since measuring either qubit will collapse the joint state, so it is not a real copy. 

To return the qubit to its original state, the remote Bell pair half is then measured in the $X$-basis, followed by a classically controlled $Z$ gate on the original control qubit. Note that if we instead measured the original qubit in the $X$-basis, and performed the $Z$ correction on Bob's half of the Bell pair, this would be equivalent to the state teleportation procedure. Instead what we wish to do is use Bob's half of the Bell pair to perform a controlled-unitary gate onto his target qubit, then measure out his half of the Bell pair to remove it from the system. The net effect is that of the controlled unitary being performed from Alice's qubit to Bob's. 

A number of variations and extensions of gate teleportation exist~\cite{yimsiriwattanaGeneralizedGHZStates2004, wuEntanglementefficientBipartitedistributedQuantum2023}, which together form the backbone of DQC.

\subsection{Multipartite entanglement}\label{sec:multi-entanglement}

\subsubsection{GHZ states}

While the bipartite Bell states are well understood, many interesting entangled states are defined over more than just two qubits. For example, the GHZ (Greenberger-Horne-Zeilinger) states generalize the maximally entangled Bell state $\ket{\Phi^+}$ to the multi-qubit case. An $n$-qubit GHZ state is defined as

\begin{equation}
    \ket{GHZ} = \frac{1}{\sqrt{2}}(\ket{0}^{\otimes n} + \ket{1}^{\otimes n}).
\end{equation}

GHZ states are harder to generate than Bell states, but allow many network protocols to be extended to cases with multiple communicating parties.

\subsubsection{Graph states}

Another interesting class of multipartite entangled states is the \textbf{graph states}~\cite{heinMultipartyEntanglementGraph2004}. The entanglement structure of graph states are defined by a graph $G(V, E)$, where each vertex $v$ in the set of nodes $V$ corresponds to a qubit and each edge $e=(u,v)$ in the set of edges corresponds to entanglement between qubits. A graph state is prepared by initializing all qubits in the $\ket{+}$ state, then applying controlled-$Z$ ($CZ$) gates between each pair of qubits in the edge set $E$. Graph states have structural properties that allow them to be studied, classified and transformed entirely from a graph-theoretic perspective~\cite{heinMultipartyEntanglementGraph2004}. In particular, many graph states are equivalent up to local Clifford operations, and this local Clifford equivalence is captured by local complementation moves on the underlying graph~\cite{heinEntanglementGraphStates2006}. This provides a powerful notion of reconfigurable multipartite entanglement: one can reason about classes of locally equivalent quantum states by reasoning about equivalence classes of graphs.

\subsubsection{Measurement-based quantum computing}

Cluster states, which are graph states defined on regular lattices, form a universal resource for  \textbf{measurement-based quantum computation} (MBQC)~\cite{raussendorfOneWayQuantumComputer2001, raussendorfMeasurementbasedQuantumComputation2003}. MBQC is an alternative to the circuit-based model of quantum computation which we have introduced above. Instead of sequentially applying gates to an array of qubits, MBQC uses large highly-entangled ``cluster states'' as a computational resource. If such a state is possessed, computation can be performed using single-qubit measurements and feed-forward corrections. Such a framework may be advantageous for all-photonic systems, in which entangled photons can be directly generated and measured, but two-qubit interactions between photons is probabilistic~\cite{nielsenOpticalQuantumComputation2004, knillSchemeEfficientQuantum2001}. Measurement-based quantum computation, and a tightly-coupled variant \textbf{fusion-based quantum computation}~\cite{bartolucciFusionbasedQuantumComputation2023}, are leading paradigms for all-photonic quantum computing systems. 

The all-photonic nature of MBQC-based architectures is particularly appealing for quantum communication and DQC, since it removes the need for repeated conversion between matter-based encodings for computation and photonic encodings for communication. Graph states, and cluster states in particular, therefore play a dual role as resources for both fault-tolerant computation~\cite{raussendorfFaulttolerantOnewayQuantum2006} and long-distance entanglement distribution~\cite{raussendorfLongrangeQuantumEntanglement2005}.

In the context of quantum communication, graph states are the foundation of the ``all-photonic'' quantum repeater introduced by Azuma et al.~\cite{azumaAllphotonicQuantumRepeaters2015}. The all-photonic quantum repeater uses large encoded graph states to do away with the need for matter-based quantum memories. This comes at the cost of generating large, high-quality cluster states at the required rates and implementing ultra-fast active feed-forward in the photonic domain.

From a macroscopic networking perspective, the distinction between memory-based repeaters and MBQC-driven all-photonic repeaters is profound. While memory-based architectures act as quantum ``store-and-forward'' routers that buffer states to combat probabilistic losses, MBQC architectures act more akin to ``cut-through'' switches. The quantum data never stops moving; instead, routing and error correction are performed on the fly via continuous measurements on the cluster state. While this elegantly bypasses the decoherence bottleneck of stationary quantum memories, it shifts the engineering burden heavily onto the classical control plane, which must now compute and transmit feed-forward corrections with virtually zero margin for delay.

\subsection{The Role of the Classical Channel and Heralding}
Finally, it is critical to understand that every entanglement distribution scheme requires a \textbf{classical communication channel} running in parallel. Its role differs significantly depending on the method:

\begin{itemize}
    \item \textbf{Confirmation in Direct Distribution:} A source creates an entangled state, and the classical channel simply signals that the photon has successfully arrived at the detector, heralding that the entanglement is now established.
    \item \textbf{Real-Time Feed-Forward in Swapping:} For entanglement swapping, the classical channel is an essential, real-time part of the protocol. The BSM at the midpoint is probabilistic and has a random outcome, projecting the remote qubits into one of four possible Bell states. The classical channel must communicate this random outcome to the end nodes. The end nodes use this classical feedback to perform a specific, local unitary operation (a Pauli gate correction) on their qubit to transform their random state into the single, usable, and agreed-upon final entangled state (e.g., transforming a random $|\Psi^+\rangle$ outcome into the desired target $|\Phi^+\rangle$). The dependence on measurement outcomes is illustrated in Fig. \ref{fig:ent-swap-circuit}.
\end{itemize}

While the entanglement swapping mechanism is functionally consistent, its practical implementation varies based on how a successful swap is confirmed. Inefficient, \textbf{unheralded swapping} approaches rely on post-selection, where measurement results are examined after the fact to find a successful event. The standard for practical networks, however, is \textbf{heralded swapping}. Here, the BSM is designed to produce a classical ``heralding signal" that immediately confirms a successful swap in real-time. This signal is essential for network resource management, telling the distant nodes their entanglement is successfully established and allowing the SDQN control plane to proceed to the next step.

The physical implementation also changes significantly depending on the qubit type. A \textbf{photonic-photonic swap} performs the BSM on two ``flying" photonic qubits using linear optics. This process is inherently probabilistic but is ideal for transmitting entanglement over long-distance optical fibers. Conversely, a \textbf{matter-matter swap}, used for local operations within a repeater node, can be performed by bringing stationary matter qubits (quantum memories) into interaction using a quantum gate. This can be a deterministic process, which is a major advantage for reliable computation. All-photonic systems aim to handle the probabilistic nature of photonic measurements using quantum error correction though rely on the generation of more complex resource states.

The most practical, \textbf{hybrid architecture}, which mirrors classical networks, uses both: photonic qubits serve as the long-haul ``flying" carriers, while stationary matter qubits at the nodes act as memories to store the entanglement and perform the swap, necessitating a robust interface to transfer the quantum state between them, though certain links in the network may benefit from all-photonic capabilities.

\section{QUANTUM NETWORKS}
\label{sec:quantum_networks}

Building upon the foundational mechanisms of entanglement distribution and quantum teleportation established in Section~\ref{sec:entanglement}, we now shift our focus to the macro-level architecture of quantum networks. Quantum networks represent a paradigm shift in how information is transmitted. Unlike classical networks that encode data in binary bits, quantum networks utilize qubits, allowing for entirely new capabilities beyond simply transmitting data. A quantum network is not a wholesale replacement for the classical Internet; rather, it is designed to work alongside it as a specialized, parallel infrastructure. Classical networks will continue to handle bulk data transmission like streaming and web browsing, while quantum networks will be reserved for specific tasks that leverage quantum properties for superior security or computational efficiency. This section explores the three primary applications of quantum networking—Quantum Key Distribution, DQC, and Distributed Quantum Sensing—and examines the architectural demands they place on the network.

\subsection{The Hybrid Architecture: Control and Data Planes}

Before detailing these specific applications, it is crucial from a network engineering perspective to understand the high-level architecture that enables them. Because quantum information cannot be copied or easily buffered, quantum networks operate on a strictly \textbf{hybrid quantum-classical architecture}. This architecture is logically and physically divided into distinct operational planes—a concept that will be formalized rigorously in the later sections on the Quantum Protocol Stack (Section~\ref{sec:protocol-stack}) and Software-Defined Quantum Networking (Section~\ref{sec:sdqn}). Broadly, the architecture consists of four interacting layers:

\begin{itemize}
    \item \textbf{Quantum Data Plane:} The physical channels (e.g., optical fibers or free-space links) dedicated exclusively to the transmission of fragile "flying" qubits (photons) and the distribution of entanglement.
    \item \textbf{Quantum Control Plane:} The local classical hardware at the network nodes (e.g., lasers, microwave sources, optical switches) that directly manipulates the quantum states to perform measurements, gate operations, or entanglement swapping.
    \item \textbf{Classical Data Plane:} Traditional networking infrastructure (e.g., Ethernet, TCP/IP) used to transmit robust classical payloads, such as encrypted ciphertexts or final sensor readout data.
    \item \textbf{Classical Control Plane:} The high-level network "brain." It utilizes classical channels to exchange the critical timing signals, basis choices, heralding confirmations, and measurement outcomes required to orchestrate the quantum nodes.
\end{itemize}

The primary engineering challenge across all quantum network applications is managing the extreme synchronization overhead between these planes, as classical control commands must execute with nanosecond precision to match the fleeting coherence times of the quantum data plane.

\subsection{Quantum Key Distribution (QKD)}\label{sec:qkd}

QKD is a method for securely creating and sharing a secret encryption key between two parties, like two people who want to send a private message. It's not a way to transmit the message itself, but rather to establish a key which is then used to encrypt the message. Its purpose is to solve the problem of a potential eavesdropper, often called ``Eve", by guaranteeing that any attempt to intercept the key is detected \cite{Gisin2002QCrypto}. The security of QKD is based on the laws of physics, making the key ``unbreakable" from a cryptographic standpoint, unlike traditional methods that rely on complex math problems that could one day be solved by powerful quantum computers. The main protocols used in QKD are BB84 \cite{BennettBrassard1984BB84} and E91 \cite{Ekert1991E91}. Both use fundamental principles of quantum mechanics to ensure a secure key exchange, but they do so in different ways.

\subsubsection{BB84 Protocol}
BB84 stands for Bennett-Brassard 1984, named after its creators \cite{BennettBrassard1984BB84}. It is the most widely implemented QKD protocol and is based on the principles of the Heisenberg Uncertainty Principle and the No-Cloning Theorem.
\begin{itemize}
\item \textbf{Superposition Preparation:} The process begins with the sender, traditionally called ``Alice," creating qubits, typically individual photons of light. The ``key" is encoded in the photon's polarization, such as vertical, horizontal, or diagonal. Alice randomly chooses the polarization for each photon from one of two bases. In the \textbf{rectilinear basis}, she can choose either vertical (90°) or horizontal (0°) polarization. In the \textbf{diagonal basis}, she can choose either 45° or 135° polarization. For each qubit, she randomly selects a basis and a corresponding polarization, creating a random sequence of bits.
\item \textbf{Transmission and Measurement:} Alice sends this stream of photons to the receiver, ``Bob," over a secure quantum channel. Bob doesn't know which basis Alice used to prepare each photon, so for every photon he receives, he \textbf{randomly chooses} one of his two measurement bases (rectilinear or diagonal) and measures the photon. He'll get the correct state (i.e., polarization) only if his chosen basis matches the one Alice used.
\item \textbf{Sifting:} After the transmission, Alice and Bob communicate over a public, \textbf{classical network}. They don't reveal the measurement results; instead, they \textbf{publicly announce which basis they used for each photon}. They then discard all the photons where their bases didn't match. The remaining sequence of bits is their shared, preliminary key.
\item \textbf{The Spy Detector (Verification):} This is the most crucial step, where quantum mechanics guarantees security. Alice and Bob publicly sacrifice a small, random portion of their key and compare the bits. If an eavesdropper, ``Eve," tried to intercept the photons, her measurements would have disturbed their quantum states. Since Eve does not know which basis Alice used, she is forced to guess. If she guesses incorrectly, her measurement will irreversibly collapse the qubit to a random state. These changes would appear as a higher-than-expected error rate in the compared bits. If errors are detected above a certain threshold, they know the key has been compromised and they discard it, starting over.
\item \textbf{Privacy Amplification (Hashing)}: Even if Eve's eavesdropping was subtle enough to remain undetected, she may have gained a tiny amount of information. This step is performed after Alice and Bob have used classical error correction to account for errors caused by noise, ensuring their unhashed keys are identical. To neutralize any remaining partial information Eve might have gained, they perform a final step called privacy amplification. They first publicly agree on the universal hash function they will use. Then, they use this function to ``compress" their long, identical, but slightly-insecure key into a much shorter, but perfectly secret, final key. This process mathematically eliminates any partial information Eve might have gained, making the final key cryptographically ``unhackable".
\end{itemize}

\subsubsection{E91 Quantum Key Distribution Protocol}
The E91 protocol, proposed by Artur Ekert \cite{Ekert1991E91}, is an entanglement-based QKD protocol that uses quantum correlations and Bell's theorem to establish a secure key between two parties, traditionally named Alice and Bob.
\begin{itemize}
\item\textbf{Entangled Pair Preparation and Distribution: } Unlike BB84, this protocol doesn't have a sender and a receiver. The process begins with a trusted, centralized source creating pairs of entangled qubits (typically photons). Each photon in the pair is in a shared quantum state, meaning their properties are perfectly correlated, no matter how far apart they are. The source then sends one photon from each entangled pair to Alice and the other to Bob.
\item\textbf{Random Measurement: } After receiving their stream of entangled photons, Alice and Bob perform their measurements independently and simultaneously. For each photon they receive, they each randomly and independently choose a measurement basis from a set of three possible bases. They record both their chosen basis and the outcome of the measurement.
\item\textbf{Verification and Sifting: } Once all the measurements are complete, Alice and Bob communicate over a public, classical network. They publicly announce which bases they used for each measurement, but they do not reveal their measurement outcomes. They then discard all the measurements where their bases did not match. The remaining bits form their shared, preliminary key.
\item\textbf{The Spy Detector (Bell Test): } This is the most crucial step and the primary difference from BB84. Instead of detecting an increase in errors from a disturbed state, security is guaranteed by a \textbf{Bell test}. Alice and Bob publicly sacrifice a small, random portion of their measurements to perform a Bell test (e.g., the CHSH inequality test). This test checks for a specific type of correlation between their results that can only exist if the particles were truly entangled. If an eavesdropper, ``Eve," has interfered, her measurements would break the entanglement, and the Bell test will fail. This failure reveals the presence of a spy, and the key is discarded.
\item\textbf{Privacy Amplification: } Even with the powerful security check, a subtle eavesdropper may have gained a tiny amount of information. To neutralize this, Alice and Bob perform a final step called privacy amplification, which works the same way as in the BB84 protocol. They use classical error correction to remove noise from their key, then apply a universal hash function to ``compress" their long, slightly-insecure key into a much shorter, but perfectly secret, final key. This process mathematically eliminates any partial information Eve might have gained, making the final key cryptographically ``unhackable".
\end{itemize}

Applying the hybrid architecture to QKD, the \textbf{Quantum Data Plane} serves strictly as the dedicated physical channel transporting the photons from sender to receiver. The \textbf{Classical Data Plane} handles the subsequent robust communication, such as basis comparisons, key verification, and carrying the final AES-encrypted message. Meanwhile, the \textbf{Classical Control Plane} acts as the network's command center, orchestrating the entire QKD process from qubit preparation to the final hashing algorithms.

\subsection{Distributed Quantum Computing (DQC)}
A DQC network is a way to link multiple smaller quantum computers together to form a much more powerful, distributed system \cite{barralReviewDistributedQuantum2025, caleffiDistributedQuantumComputing2024}. Leveraging the principles of teleportation, early works explored the concept of using entanglement as a resource to perform distributed quantum operations \cite{groverQuantumTelecomputation1997, eisertOptimalLocalImplementation2000, huelgaQuantumRemoteControl2001, cleveSubstitutingQuantumEntanglement1997}, while Cirac et al.~\cite{ciracDistributedQuantumComputation1999} highlighted the potential of the approach.

Its purpose is to solve problems that are too complex for a single quantum computer to handle alone. This addresses a major challenge in the field: building a single, large-scale quantum computer is incredibly difficult due to the fragility and complexity of the hardware~\cite{vanmeterLocalDistributedQuantum2016}. By networking smaller machines, we can create a collective system with far greater computational power and a larger number of qubits~\cite{caleffiDistributedQuantumComputing2024}. Beyond this, a DQC system has the added benefit of being more resistant to catastrophic errors that might corrupt large numbers of qubits in a single quantum device \cite{xuDistributedQuantumError2022}. The security and power of a DQC network are based on its ability to synchronize and manage entangled qubits across a distributed network. The entire DQC process is not a one-time event, but rather a dynamic and iterative feedback loop between the classical and quantum layers.

\begin{itemize}
\item \textbf{Preparation and Instruction: } A centralized controller (i.e, the ``brain" that orchestrates the entire operation) first breaks down a complex computational problem into smaller, manageable subproblems. It then determines the specific gate instructions and entanglement requirements needed to solve them. Acting as a compiler, this controller translates the high-level quantum algorithm into a set of precise instructions for each distributed quantum computer (see Sec.~\ref{sec:partitioning}). It then sends these instructions to the quantum computers and, simultaneously, directs a centralized entanglement source to generate and forward the corresponding entangled qubits to the quantum computers, which act as the foundational resource for the computation. 

\item \textbf{Quantum Execution:} After being set up with entangled links and gate instructions, the quantum computers execute the computation. This process involves applying a sequence of quantum gates to the local qubits -- some local (intra-QPU) and others non-local, or ``remote'' (inter-QPU). Through the application of local and remote gates, qubits through the whole system may become entangled. As a results, certain local operations can affect the global state of the system. In this light, it is more appropriate to view the network of machines as a single, unified computational system, since the state space of the system grows exponentially, just as it would for a single machine containing the qubits of all the machines. If, on the other hand, the machines are only classically connected, the state space grows only exponentially within each machine~\cite{cuomoDistributedQuantumComputing2020}.

\item \textbf{Classical Feedback:} After the quantum computation is complete, the qubits are measured, collapsing their quantum states into classical bits. These measurement results are sent back over a classical network to the centralized controller. Since the program has been partitioned and compiled for the distributed system, the compiler will need to post-process the classical results to recover the output relative to the original circuit.

\item \textbf{Analysis and Adaptation:} The centralized controller uses a classical processor to analyze the measurement data received from the quantum computers during and after the computation. Since teleportation operations require classical feedback, the controller needs to send and receive classical measurement results to orchestrate feed-forward corrections during the computation. In applications using parametrized quantum circuits (PQCs) such as quantum machine learning (QML)~\cite{biamonteQuantumMachineLearning2017}, the measurement results at the end of a circuit may have to be processed and used to optimize parameters for the next round of gates. The overall network's performance is heavily dependent on the speed of this classical feedback loop, as any latency directly impacts the efficiency of the hybrid algorithm.

\item \textbf{Iteration:} Many quantum algorithms require repeated iteration. This is partly to combat noise, though is also statistically necessary for many algorithms to guarantee a correct solution even in the ideal case. Moreover, applications using PQCs require iterative update of the parameters, which are typically optimized by a classical optimizer at the central controller. This iterative nature allows the network to overcome the fragility of quantum states and synchronize its operations with extreme precision. The bottleneck in this system is not the ability to make dynamic changes, but rather the latency of the classical communication.
\end{itemize}

In the context of DQC, the architectural demands on the hybrid planes are significantly more intense than in QKD \cite{cacciapuotiQuantumInternetNetworking2020}. The \textbf{Classical Control Plane} acts as the master compiler, formulating the problem and dispatching sub-circuit instructions. The \textbf{Quantum Data Plane} must act as an on-demand "quantum bus," distributing high-fidelity entanglement to perfectly correlate the disparate processors. The \textbf{Quantum Control Plane} is the local classical hardware executing the precise gate operations. Because DQC involves actively maintaining fragile superpositions during computation, the synchronization overhead is extreme, demanding that classical control commands be generated, transmitted, and executed with microsecond precision before the quantum states lose their coherence.

\subsection{Distributed Quantum Sensing (DQS)}
A DQS network is a system that connects multiple quantum sensors to perform highly precise measurements over a large area~\cite{zhangDistributedQuantumSensing2020}. A single quantum sensor, no matter how sensitive, is inherently limited in what it can detect due to fundamental quantum noise. By working together, a network of entangled quantum sensors can achieve a level of precision that is impossible for one sensor alone, which is particularly useful for detecting extremely weak signals like gravitational waves or faint magnetic fields. The main challenge is creating and distributing a shared quantum state among the sensors without it being destroyed by environmental noise. 

To quantify this advantage, we must compare the mathematical scaling behavior of classical versus quantum sensor networks~\cite{giovannettiQuantumMetrology2006}. In a classical distributed array with $N$ independent sensors, precision is improved by taking independent measurements and averaging the results. According to classical statistics and the central limit theorem, the measurement error (the standard deviation of the estimate) scales proportionally to $1/\sqrt{N}$. This boundary is known as the \textbf{Standard Quantum Limit (SQL)} or shot-noise limit. Under the SQL, improving a network's precision by a factor of 10 requires deploying 100 times as many sensors.

Quantum sensor networks bypass this classical bottleneck by sharing multipartite entangled states (such as the GHZ states introduced in Sec. \ref{sec:multi-entanglement}) across the $N$ sensing nodes. Because the sensors are entangled, they do not act as $N$ separate devices; rather, they behave as a single, unified, macroscopic quantum probe. When exposed to the external phenomenon, the quantum phase shift accumulates coherently and additively across all nodes simultaneously. This collective quantum interaction allows the measurement precision to scale proportionally to $1/N$, reaching the ultimate physical boundary known as the \textbf{Heisenberg Limit (HL)}~\cite{giovannettiQuantumMetrology2006}. This quadratic enhancement means that a distributed network of just 100 entangled quantum sensors could theoretically achieve the same measurement precision as 10,000 independent classical sensors.

A DQS network operates by harnessing these quantum properties to link its sensors and perform a collective measurement:

\begin{itemize}
\item \textbf{Preparation and Entanglement:} The process begins with the initialization of individual quantum sensors, which are typically stable matter-based qubits like nitrogen-vacancy (NV) centres in diamond or trapped ions. In parallel, a centralized entanglement source generates entangled photons. These photons are then distributed to the sensor nodes, often via optical fibre. At each node, the incoming photon interacts with the local matter-based sensor, and through a process known as entanglement swapping, the quantum link is transferred from the photons to the sensors. This procedure establishes a shared entangled state between the matter-based sensors themselves. The advanced goal is to create multipartite states, linking many sensors $(N > 2)$ to achieve even greater precision, though this is exponentially more difficult to create and maintain.

\item \textbf{Collective Measurement:} Once the sensors are entangled, they are collectively prepared in a \textbf{quantum state of superposition}, making the network acutely sensitive to an external signal. The entangled network is then exposed to the phenomenon being measured, such as a magnetic field. Because of the shared entanglement, the sensors perform a collective measurement. The correlated noise that would normally obscure a weak signal in each sensor can be effectively ``squeezed out" or filtered, allowing the network to detect a signal that would have been lost if the sensors had acted independently~\cite{eldredgeOptimalSecureMeasurement2018}. 

\item \textbf{Classical Processing:} After the collective measurement, the quantum states collapse, and the information is now in a classical format. The raw measurement data is collected and sent over a classical network to a centralized classical processor. Here, sophisticated algorithms are used to combine the sensor readings and extract the final, highly accurate result.
\end{itemize}

It is important to note an alternative, more common DQS model where the "flying" photons are the sensors themselves. In this \textbf{``all-photonic" model}, there are no local matter qubits:

\begin{itemize}
\item \textbf{Preparation and Entanglement:} The process begins at a centralized entanglement source, which generates entangled photons. These photons act as the sensing probes themselves and are immediately distributed into separate ``sensing arms" of the network (e.g., different optical fibres). They are routed through the environment to be measured and then converge at a separate, central collection point.
\item \textbf{Collective Measurement:} As the entangled photons travel along their separate paths, they are directly exposed to the phenomenon being measured. This interaction induces a subtle, collective change in their quantum state, most commonly a phase shift. Upon converging at the collection point, they are recombined (e.g., on a beam splitter) and immediately strike a set of single-photon detectors. This joint detection filters out quantum noise and reveals signals far weaker than a single photon could resolve.
\item \textbf{Classical Processing:} This joint measurement collapses the photons' quantum state into classical data (e.g., coincidence counts). This raw data, containing the crucial phase shift information, is read out by classical electronics and sent to a classical processor to extract the final sensing value.
\end{itemize}

An continuous variable alternative all-photonic model has also been experimentally demonstrated by Guo et al.~\cite{guoDistributedQuantumSensing2020}.

When mapping DQS to the hybrid network architecture, the utilization of the planes depends on the specific sensing model. In a hybrid matter-based DQS, the \textbf{Quantum Data Plane} distributes the initial flying entanglement, while the local \textbf{Quantum Control Plane} performs the swaps to transfer that entanglement into the stationary sensors. In an all-photonic DQS, the \textbf{Quantum Data Plane} \textit{is} the sensing region—the physical paths the photons traverse to probe the environment—while the \textbf{Quantum Control Plane} handles the optical recombination and detection at the terminus. In both models, the \textbf{Classical Control Plane} must precisely synchronize the photon generation with the final detection windows, and the \textbf{Classical Data Plane} gathers the final readouts for the centralized processor.

\subsection{Conclusion: Transitioning to Physical Realities}

While the applications described above—QKD, DQC, and DQS—promise revolutionary capabilities across security, computation, and metrology, their real-world deployment is severely restricted by the fragility of the underlying quantum hardware. The theoretical operation of these hybrid control and data planes assumes that entanglement can be reliably generated, transmitted, and maintained over long periods. In physical reality, the quantum data plane is relentlessly degraded by environmental interactions. The following section explores these physical limitations, detailing the core challenges of quantum noise, photon loss, and the inherently probabilistic nature of quantum operations that the classical network architecture must actively combat to make the Quantum Internet a reality.


\section{Core Challenges in Quantum Networking: Noise, Loss, and Probability}
\label{sec:core_challenges}

In both classical and quantum networking, a core set of challenges related to distance, interference, and hardware limitations must be addressed to ensure reliable communication. The similarities between the two realms, while not identical in their underlying physics, provide a useful initial framework for understanding the obstacles faced by a quantum network.

\textbf{Signal Attenuation due to Distance:} In classical networks, signals lose strength over long distances, a process known as attenuation. This weakening necessitates the use of amplifiers and repeaters to boost the signal. A direct parallel exists in quantum networking, where the primary form of signal attenuation is photon loss. As photons carrying quantum information travel through a medium like an optical fiber, they have a certain probability of being absorbed or scattered. This probability increases exponentially with distance, causing a rapid decline in the success rate of entanglement distribution. Just as classical networks rely on repeaters, quantum networks need quantum repeaters to overcome this exponential loss by segmenting the long-distance link into shorter, more manageable hops.

\textbf{Environmental Interference:} Both classical and quantum networks are susceptible to environmental interference that corrupts the transmitted information. In classical systems, this might be electromagnetic interference, crosstalk, or other sources of noise that degrade the signal. The quantum equivalent is decoherence, where a qubit's fragile superposition or entangled state is disturbed by unintended interactions with its environment,
such as thermal vibrations or stray light.
Both classical and quantum interference fundamentally reduce the integrity of the information being transmitted and require robust countermeasures to be effective.

\textbf{Component Imperfections:} Perfect components are a rarity in any engineering field. In both classical and quantum systems, imperfections in hardware introduce errors. In classical networks, this includes things like noisy amplifiers, jitter in clocks, or manufacturing defects in integrated circuits. These imperfections lead to bit errors and degraded performance. The same holds true for quantum networks. Noisy quantum sources may produce entangled pairs of lower quality, and imperfect detectors can introduce false signals (dark counts) or fail to register a photon, leading to errors that must be accounted for by the network's protocols.

While these high-level similarities provide a useful conceptual bridge for the network engineer, they mask a profound paradigm shift. In classical systems, attenuation and interference degrade a robust, replicable signal; in the quantum domain, they irreversibly destroy a fragile, uncopyable state. To truly engineer a quantum network, we must abandon classical approximations and confront how the underlying physics fundamentally alters the nature of noise, error propagation, and network capacity.

\subsection{Quantum Noise}

Additive White Gaussian Noise (AWGN) is a cornerstone model in classical communication theory. It describes noise that is simply added to a signal, is statistically random (white), and follows a normal (Gaussian) distribution. In quantum mechanics, a direct, one-to-one analogy to AWGN does not exist due to the fundamental principles of quantum information. One cannot simply ``add" a noise vector to a qubit's state. 

This ``additive" model is physically impossible because a qubit's state is not arbitrary; it is rigidly constrained by the laws of quantum mechanics. A qubit's coefficients, as in $|\psi\rangle = c_0|0\rangle + c_1|1\rangle$, must always satisfy the normalization condition ($|c_0|^2 + |c_1|^2 = 1$), which ensures the total probability of all outcomes is 100\%. If noise were simply ``added" to these coefficients (e.g., $c_0 \to c_0 + n$), this rule would be violated, resulting in an ``unphysical" state with a total probability not equal to 1. The Bloch sphere provides an apt analogy: a valid qubit state is a point on the surface of the sphere. Classical additive noise would ``nudge" this point off the surface, which is physically forbidden.

Therefore, any interaction with the environment (the source of noise) must be a physical transformation—like an unwanted rotation—that inherently preserves this probability rule, transforming the qubit's state itself.
In other words, instead of being an added signal, quantum noise is modelled as a \textbf{quantum channel},
i.e., a mathematical operation that transforms an input quantum state into an output state. There are many types of quantum channels that model different physical noise processes.

\subsubsection{Quantum Channel Models}

A primary model for this is the \textbf{depolarizing channel}. This models a symmetric, randomizing interaction with the environment where the qubit's state is, with some probability $p$, ``forgotten" and replaced by a state of complete randomness. With probability $1-p$, the state passes through unaffected. With probability $p$, it is transformed into a \textbf{maximally mixed state}, which represents a state of complete ignorance (i.e., an equal 50/50 probability of being measured as 0 or 1, regardless of the measurement basis). This channel is a common, general-purpose model for a qubit interacting with a very noisy, chaotic environment.

A second model is the \textbf{amplitude damping channel}, which is the quantum mechanical description of \textbf{energy loss} (dissipation). It models the spontaneous decay of a qubit's ``excited" state, $|1\rangle$, to its ``ground" state, $|0\rangle$. This is an asymmetric process: a qubit in the $|0\rangle$ state is stable and unaffected, but a qubit in the $|1\rangle$ state has a defined probability of decaying to $|0\rangle$.

From a classical communications perspective, these channels have familiar effects, even if the underlying physics is different. The depolarizing channel is analogous to a classical symmetric channel where a bit is randomly flipped, destroying its information content. The amplitude damping channel is analogous to an asymmetric channel where a ``1" can be flipped to a ``0" (representing energy loss), but a ``0" remains a ``0". The critical distinction, however, is that these are not simple bit-flips on a classical signal but irreversible transformations of the quantum state itself, which is why classical, cloning-based techniques for signal amplification and data retransmission are physically impossible. Instead, as explored in Section~\ref{sec:Error_management}, the network control plane must rely on specialized quantum techniques, such as entanglement purification and fault-tolerant Quantum Error Correction (QEC), to manage these errors.

While the depolarizing and amplitude damping channels model the two most common noise sources (information randomization and energy loss, respectively), they are not exhaustive. Other significant models include the \textbf{phase-damping channel}, which is critical for understanding decoherence. This channel models the loss of \textbf{quantum phase information}, destroying the superposition without any energy loss (i.e., the probabilities of measuring 0 or 1 remain unchanged, but the relative phase between them is randomized
so that measurement in the ${\lvert\pm\rangle}$ basis approaches a ${50/50}$ mixture). Furthermore, the \textbf{generalized amplitude damping channel} extends the basic energy-loss model to account for a non-zero temperature, modeling the effects of a thermal environment~\cite{nielsen2000quantum}.

Quantum channel models are often treated as independent because they model distinct physical processes. In real-world hardware, however, a stationary memory qubit rarely experiences just one type of noise. It concurrently suffers from both energy relaxation (quantified by the $T_1$ time, which measures how long a qubit can hold its excited energy state) and phase randomization (quantified by the $T_2$ time, which measures how long the delicate quantum superposition is preserved). The severe network-level constraints imposed by these $T_1$ and $T_2$ coherence limits are formally detailed later in Section~\ref{sec:performance_metrics}.

While advanced quantum network simulators utilize composite noise models to capture this simultaneous degradation (e.g., NetSquid~\cite{coopmans2021netsquid} and SquidASM~\cite{squidasm_docs_T1T2}), high-level classical routing algorithms frequently simplify reality by assuming a single, symmetric noise channel (e.g., a depolarizing channel)~\cite{liu_quantum_bgp_2024, abane2025survey, glisic2024}. As will be explored, this mathematical simplification creates a severe ``AWGN Fallacy'' blind spot for the network control plane. Optimizing for a generic, symmetric error will fail to protect the state against the highly asymmetric physical realities of $T_1$ and $T_2$ decay.

\subsubsection{Quantum Vacuum and the Unavoidable Noise Floor}
A final difference from classical systems is that quantum devices are subject to fundamental noise constraints arising from zero-point fluctuations and measurement back-action. In classical networks, noise from thermal energy or electromagnetic interference can be minimized through engineering, and an idealized noiseless channel is theoretically possible as temperature tends to $0$. In quantum mechanics, however, observables exhibit irreducible fluctuations even at absolute zero due to zero-point motion. These fluctuations are typically benign for an isolated system but become operationally relevant—and limit performance—whenever the system is measured, amplified, or controlled~\cite{clerkIntroductionQuantumNoise2010}. This fundamental constraint implies that quantum channels and devices have irreducible noise floors tied to the physics of measurement and control, rather than merely external imperfections. The engineering goal thus shifts from eliminating noise to operating protocols within these unavoidable quantum limits.

\subsubsection{Noise and Physical Qubit Mediums}
The specific noise channels affecting a quantum network are not universal; they are fundamentally dependent on both the \textbf{physical medium} and the choice of encoding used for the qubit. This leads to a core concept in quantum networking: the distinction between ``flying" and ``stationary" qubits.

\textbf{Photonic qubits}, or ``flying" qubits, are the de facto standard for transmitting quantum information. They are ideal for this task as they travel at the speed of light and interact very weakly with the environment. However, their primary weakness is that they are extremely difficult to store and are highly susceptible to photon loss. Depending on the type of encoding, this loss can have drastically different effects. If the quantum information is encoded in the presence or absence of the photon, this corresponds to the \textbf{amplitude damping channel}. Conversely, if a polarization encoding is used, then neither of the computational states are occupied after loss. This is known as \textbf{leakage}, since the qubit has physically ``leaked'' out of the computational subspace. In some protocols, the network may receive a classical herald that the loss has happened; in this case, the leakage becomes an \textbf{erasure}. This can happen, for example, when a BSM gives a forbidden result, allowing the network to infer a prior qubit loss. Finally, if the loss goes unheralded, its randomizing effect on future measurements is often approximated in network performance models using a \textbf{depolarizing channel}.

\textbf{Matter-based qubits}, on the other hand, are ``stationary" qubits used for storage and computation at the network nodes (e.g., trapped ions, neutral atoms, or superconducting circuits). Their strength is the ability to store quantum information, but their primary vulnerability is decoherence due to unwanted interactions with their local environment. This noise is best modeled by the \textbf{depolarizing and phase-damping channels}.  

\subsubsection{Effects on Network Architecture and Performance}
The specific noise channels dominating a system have a direct impact on the design and performance of a quantum network. A network dominated by \textbf{photon loss} must employ entanglement distribution methods like quantum repeaters. Conversely, a network dominated by \textbf{phase damping} at its nodes would prioritize the fault-tolerant Quantum Error Correction (QEC) protocols detailed in Section~\ref{sec:Error_management}. These noise channels directly determine the Key Performance Indicators (KPIs) of the network, dictating the trade-offs between entanglement distribution rates and final state fidelity.

This hybrid noise challenge is perfectly illustrated in the architecture of a quantum repeater. During the transmission phase, a ``flying" photonic qubit is sent through an optical fiber (combating amplitude damping). In the storage and swapping phase, the photon arrives at a repeater and its state is mapped into a ``stationary" matter-based qubit (combating depolarization and phase-damping). This hybrid approach uses the right qubit for the right job to overcome the distinct limitations of each medium.

\subsection{Foundational Differences Between Quantum and Classical Networks}

Beyond the unique nature of quantum noise, the very operational principles of quantum networks diverge sharply from classical paradigms. These differences render classical protocol solutions unworkable and demand entirely new, counter-intuitive strategies for managing network operations.

\subsubsection{The ``No-Copy" and ``No-Peek" Constraints} 
The first and most significant operational divergence from classical networks stems from the \textbf{No-Cloning Theorem} \cite{Wootters1982ASQ}. This fundamental law of quantum mechanics forbids the creation of an identical, independent copy of an unknown quantum state. In classical networks, a decaying signal is easily managed: a repeater simply reads the bitstream, creates a fresh copy, and re-transmits it. This ability to copy enables all forms of data redundancy and retransmission protocols, such as Automatic Repeat reQuest (ARQ). The No-Cloning Theorem (often referred to informally in ICT literature as the ``no-copy'' rule) makes this entire classical paradigm physically impossible. 

This constraint is strictly compounded by the \textbf{Information-Disturbance Theorem}, a direct consequence of the quantum measurement postulate \cite{nielsen2000quantum} . In a classical network, a router can passively measure a data packet's integrity (e.g., via a checksum) or read its routing header without altering the payload. In a quantum network, however, any attempt to extract information from an unknown quantum state fundamentally disturbs it; a direct measurement invariably collapses its delicate superposition. 

Informally termed the ``no-peek'' rule, this theorem dictates that classical techniques for non-destructive error checking, mid-span signal monitoring, and passive packet sniffing are physically forbidden. Because the network cannot copy a state to save it, nor peek at a state to check it, quantum networks must abandon classical ARQ entirely and instead rely on entirely different paradigms, such as classical heralding, entanglement purification, and fault-tolerant Quantum Error Correction (QEC).

\subsubsection{Error Propagation} 
The second profound difference lies in the nature and propagation of errors. Classical systems face correlated noise that is external to the signal. Quantum noise, however, can be intrinsic to the state itself, and more critically, it can be \textbf{non-local}. For example, if two qubits are in a shared Bell state $\ket{\Phi^{+}}$, a local phase error results in a phase error on the joint state. By the same principle from which quantum sensors derive their power, many local phase errors on entangled qubits can accumulate into larger phase errors on the shared state. Analogously, single-qubit errors occurring \textit{before} entanglement is generated can propagate through the entangling operations onto all the qubits. This phenomenon, known as \textbf{error propagation}, means a single local environmental interaction on one qubit can result in much larger errors on the global state.

Furthermore, the non-local nature of entanglement fundamentally blurs the classical architectural boundary between ``communication'' noise (link degradation) and ``storage'' noise (memory decoherence). In a classical network, a corrupted memory buffer at a destination router does not degrade the integrity of the fiber-optic cable leading to it. In a quantum network, however, local noise mechanisms have highly non-local network effects. 

For instance, imagine Alice and Bob successfully establish a pristine, end-to-end entangled link across a continent. The active communication phase is complete. However, if Bob's stationary qubit subsequently suffers a phase-flip due to a fluctuating magnetic field in his local memory (a purely local ``storage'' error), the shared joint state is corrupted. Alice's qubit, despite being perfectly isolated in a pristine environment, is now part of a degraded network link. For the network controller, this means node reliability and link reliability cannot be modeled independently; a localized physical hardware fault manifests as a global network-layer failure.

\subsubsection{Determinism vs. Probabilism}
Perhaps the most counter-intuitive difference for a networking researcher is that quantum networks are inherently \textbf{probabilistic}. In a classical network, uncertainty comes from noise—a random, external factor that engineers try to eliminate to operate as close to 100\% determinism as possible. In quantum networking, the probabilistic nature is a \textbf{built-in feature of the underlying physics}. This forces a complete shift in engineering philosophy: from minimizing errors to managing probabilities.

For example, the Bell-state measurement (BSM), introduced in Section~\ref{sec:entanglement}, is an essential part of many quantum network protocols. In linear-optical implementations, the BSM can only be implemented probabilistically, since the physics of the system only allows two of the four Bell states to be distinguished~\cite{calsamigliaMaximumEfficiencyLinearoptical2001}. This gives a $50\%$ success probability. Nonlinear interactions between photons are extremely weak, making the linear-optical BSM the standard in quantum networks. Though additional resources can be used to boost the success probabilities~\cite{griceArbitrarilyCompleteBellstate2011}, non-determinism remains an unavoidable feature of the network.

For matter-based systems, this is not typically a problem, since strong interactions permit deterministic two-qubit gates, which is why this doesn't pose an issue in matter-based quantum computers. Though matter-matter BSMs can be performed deterministically, stationary matter cannot be transmitted over macroscopic distances. Consequently, the network's long-haul communication links must rely on photonic ``flying'' qubits, making this probabilism an inescapable reality at the link layer.

This non-determinism permeates the entire protocol stack, drastically reducing end-to-end throughput and demanding a far more complex control plane. Unlike a classical network that assumes a high success rate, a quantum control plane must meticulously manage a pipeline of probabilistic events, waiting for classical ``heralding" signals to confirm success before proceeding. To combat these low, physics-bound success rates, networks must multiplex their resources, performing many probabilistic operations in parallel to make an unreliable process viable at the network layer.

\subsection{The Non-Additivity of Quantum Channel Capacity}

These probabilistic and non-cloning constraints also precipitate a profound breakdown in classical network capacity planning. In classical network design, Claude Shannon’s channel capacity theorem provides a reliable, additive mathematical foundation. If a network engineer bundles two independent classical cables, the total data capacity is simply the sum of their individual capacities ($C_{A+B} = C_A + C_B$). If both cables are severed or too noisy to carry data ($C_A = 0$ and $C_B = 0$), their combined capacity remains zero. 

In quantum networking, this foundational additive logic completely fails. Quantum information theory reveals that the quantum capacity of channels is profoundly non-additive. The most extreme manifestation of this is \textit{superactivation}. It has been theoretically proven that two completely noisy quantum channels, each with a strict quantum capacity of zero, can be used jointly to achieve a strictly positive quantum communication rate~\cite{smithQuantumCommunicationZeroCapacity2008}. Furthermore, the theory of \textit{causal activation} demonstrates that sending a quantum state through two noisy channels in a quantum superposition of different orders can establish a perfect communication link, even if the channels independently destroy all quantum information~\cite{chiribellaQuantumComputationsDefinite2013}.

For the network engineering community, this represents a massive core challenge. It means that the fundamental capacity of a quantum network cannot be calculated simply by analyzing individual links in isolation. Network routing, capacity planning, and resource allocation algorithms must eventually account for these deep, non-linear quantum synergies, rendering classical graph-theory network optimization models fundamentally inadequate.

\subsection{Physical Imperfections and Quantum Side-Channels}

While theoretical capacity limits assume mathematically perfect operations, deploying a physical quantum network introduces severe hardware-specific vulnerabilities. Classical networks cleanly separate the physical layer from the security layer; a fiber-optic cable simply transports bits, while mathematical encryption algorithms (like AES or RSA) executed at the application layer secure the payload. In contrast, quantum communication applications like QKD rely directly on the physical layer—the laws of quantum mechanics—to guarantee security. 

While the physics of QKD are theoretically perfectly secure, the physical hardware is not. This introduces a core challenge unique to quantum networks: hardware-induced side-channel attacks. Real-world Single-Photon Avalanche Diodes (SPADs) suffer from physical imperfections such as dark counts (registering a photon when none exists) and dead times (a brief recovery period after a detection where the sensor is blind)~\cite{scarani2009security}. Eavesdroppers can exploit these classical hardware vulnerabilities. For instance, in a \textit{blinding attack}~\cite{makarovControllingPassivelyQuenched2009}, an eavesdropper shines a bright classical light into the receiver's detectors, forcing them into a classical linear mode, allowing the hacker to intercept the key without triggering the quantum error threshold.

Consequently, a major challenge in deploying functional quantum networks is developing \textit{Device-Independent (DI)} or \textit{Measurement-Device-Independent (MDI)} protocols~\cite{loMeasurementDeviceIndependentQuantumKey2012}. These protocols represent a paradigm shift where the network can mathematically guarantee the security of an entangled link even if the physical measurement hardware itself was manufactured or compromised by a malicious adversary.

\subsection{The Wavelength Mismatch and Quantum Transduction}

Compounding the challenge of imperfect hardware is the fundamental physical incompatibility of the hardware itself across different network segments. The global classical telecommunications infrastructure is heavily optimized for optical transmission in the C-band (around 1550 nm), where fiber-optic loss is minimal. 

However, unlike classical routers that process uniform electronic signals, the nodes of a future quantum internet will be built from wildly heterogeneous quantum hardware. The most advanced Quantum Processing Units (QPUs), such as superconducting circuits, operate in the microwave frequency regime (around 5 GHz)~\cite{laddQuantumComputers2010}. Meanwhile, the most stable stationary quantum memories, such as trapped ions or neutral atoms, operate using visible or near-infrared light~\cite{monroeScalingIonTrap2013}.

To connect a microwave QPU to a 1550 nm fiber-optic network, the system requires a \textit{Quantum Transducer}. This device must convert the quantum state from a microwave photon into an optical photon while perfectly preserving its superposition and entanglement -- a process that currently suffers from devastatingly low efficiencies and high noise. Until highly efficient quantum transduction is mastered, this wavelength mismatch remains a fundamental physical bottleneck preventing the integration of quantum computers with long-distance quantum optical networks.

\subsection{Summary: The Network Engineering Perspective on Quantum Challenges}

In summary, the core challenges of quantum networking extend far beyond simply dealing with higher noise levels. From a network architecture perspective, these physical phenomena dictate the entire architecture of the protocol stack. The non-additivity of capacity breaks standard routing algorithms; the no-cloning theorem breaks ARQ and necessitates complex error correction; the probabilistic nature of Bell-state measurements requires massive parallel multiplexing and real-time heralding; and the heterogeneity of wavelengths demands lossy transduction layers. Consequently, the control plane cannot merely be a passive observer assigning bandwidth. It must act as a highly active, microsecond-precise orchestrator, explicitly designed to manage probabilism, navigate side-channel vulnerabilities, and dynamically route around fundamental physical fragility.


\section{QUANTUM MEMORY: THE HARDWARE BOTTLENECK}
\label{sec:quantum_memory}

As established in Section~\ref{sec:core_challenges}, the exponential loss of flying qubits (photons) over long distances makes direct quantum communication practically impossible beyond modest distances, necessitating the use of quantum repeaters. The fundamental hardware component that enables a quantum repeater to function—and indeed, the cornerstone of any scalable quantum network—is the \textit{quantum memory}. 

A quantum memory acts as a ``quantum buffer." It catches a fragile flying qubit, safely stores its quantum state as a stationary qubit (e.g., within a trapped ion or neutral atom), and retrieves it on demand. This ``store-and-stitch" capability allows a repeater to hold a successfully generated entanglement link on one side while probabilistically attempting to generate a link on the other side. It should be noted that alternative architectural paradigms, such as the all-photonic repeaters based on cluster states discussed in Section~\ref{sec:entanglement}, attempt to bypass stationary memories entirely. However, these architectures trade the memory decoherence bottleneck for extreme spatial complexity (requiring massive arrays of synchronized photon generators) and highly restrictive classical feed-forward bandwidth requirements. Consequently, for the vast majority of generalized, multi-hop quantum network designs, the stationary quantum memory remains the foundational, albeit highly volatile, building block. However, unlike classical RAM, which stores robust and static bits indefinitely, quantum memory is a highly volatile, perishable, and error-prone environment. To understand the limits of a quantum network, the control plane and higher-layer protocols must be designed around the physical constraints of the memories that power it, making the mapping from device-level metrics to network-level behaviour explicit throughout this section.

\subsection{Fidelity and the Classical Limit}
Before evaluating the performance of an entire network, we must define the baseline quality of the memory itself. While classical communications measure link quality using the Bit Error Rate (BER)---and while a specific Quantum Error Rate (QER) is actively monitored at the network layer (as detailed in Section~\ref{sec:performance_metrics})---the foundational, device-level metric of quantum accuracy is \textit{fidelity}. This is a continuous value from 0 to 1 that quantifies how well a quantum state is preserved after being written to and read from the memory.

For a quantum memory device, it is critical to distinguish between its static and dynamic quality:
\begin{itemize}
    \item \textit{Initial Fidelity ($F_0$):} This is the memory’s best-case performance, often cited as its ``spec-sheet'' number. It represents the fidelity of a write-and-retrieve operation assuming the shortest possible storage time. Crucially for network engineers, this metric is typically measured under pristine, highly isolated laboratory conditions. When deployed in noisy, real-world infrastructure, the baseline operational $F_0$ is often significantly lower.
    \item \textit{Real-Time Fidelity ($F(t)$):} In a functional network, fidelity is a perishable resource. The real-time fidelity is the actual quality of the stored qubit, which continually decays from its initial value $F_0$ due to physical decoherence over time $t$.
\end{itemize}

This decay has a critical ``pass/fail'' threshold for network applications: the classical fidelity limit of $2/3$ ($\approx 66.7\%$)~\cite{massar1995optimal}. This threshold is not derived from the memory's device physics but from quantum information theory: for uniformly distributed single-qubit states, it represents the maximum average fidelity achievable by any ``classical impostor'' that measures the input state and then prepares an output state based only on the measurement result. For many common single-qubit noise models (such as depolarizing noise), the parameter regime in which the memory channel becomes \textit{entanglement-breaking (EB)} coincides with fidelities at or below this classical limit, meaning that any entanglement initially shared across the memory is necessarily destroyed \cite{horodecki2003general}. From a network perspective, memory modes that operate in this EB regime must therefore be treated as unusable links in routing, scheduling, and resource-allocation decisions.

\subsection{Coherence and Storage Time}
While fidelity measures the quality of the state, the speed at which that quality degrades is dictated by the memory's coherence time. This fidelity decay is caused by two primary physical noise processes: \textit{energy relaxation} ($T_1$), the decay of an excited state, and \textit{dephasing} ($T_2$), the loss of fragile phase information. Because quantum information relies heavily on phase, and because $T_2 \le 2T_1$ in many platforms of interest, the $T_2$ limit is typically the most restrictive physical bottleneck.

For network protocols, the physical decay timescale characterized by $T_2$ is less important than the functional lifetime it induces. The most precise functional metric at the network level is the \textit{entanglement-preserving lifetime} ($T_{\text{non-EB}}$). This is not a fundamental physical constant, but a derived quantity defined as the total operational time window during which the real-time fidelity $F(t)$ of a stored qubit remains above the critical $2/3$ classical fidelity limit.

In practice, $T_{\text{non-EB}}$ is obtained by solving the fidelity decay model for time, i.e., finding $T_{\text{non-EB}}$ such that $F(T_{\text{non-EB}}) = 2/3$. For example~\cite{coopmans2021netsquid, sangouard2011quantum}, consider a simple exponential dephasing model where one half of an entangled pair is held in memory. The \textit{Bell-state fidelity} (a measure of entanglement fidelity) of the stored pair decays as
\[
F(t) = \tfrac{1}{2}\bigl(1 + (2F_0 - 1)\,e^{-t/T_2}\bigr),
\]
where $F_0$ is the initial fidelity at $t=0$ and $T_2$ is the coherence time. Solving $F(T_{\text{non-EB}}) = 2/3$ yields
\[
T_{\text{non-EB}} = T_2 \,\ln\!\left(\frac{2F_0 - 1}{1/3}\right),
\]
which explicitly shows that even for a fixed $T_2$, a modest initial fidelity (e.g., $F_0 = 0.8$) can reduce $T_{\text{non-EB}}$ to a small fraction of $T_2$. Thus, simulators and performance models should treat $T_{\text{non-EB}}$ as a derived, hardware- and configuration-specific parameter rather than substituting $T_2$ directly.

From the perspective of the control plane, $T_{\text{non-EB}}$ provides the hard time budget within which all operations that rely on a stored entangled state—from waiting for classical heralding signals, to running entanglement swapping and purification, to making routing decisions—must complete if that state is to remain usable for the network.

\subsection{Efficiency and Loss Scaling}
If fidelity dictates the \textit{quality} of the stored information, \textit{efficiency} dictates the \textit{probability} of successfully accessing it. Storage-and-retrieval efficiency quantifies the probability that a photon sent into the memory can be successfully absorbed and later retrieved on demand. It is the direct quantum equivalent of packet loss in a classical network buffer.

While memory efficiency is a critical individual bottleneck, it is just one component in a larger, system-wide loss budget. The total end-to-end efficiency of any operation is the product of the efficiencies of every probabilistic step: the source efficiency (photon generation), coupling efficiency (fiber-to-chip connection), propagation loss (photon survival in the optical fiber), the memory efficiency itself, and detector efficiency (measurement). From a protocol perspective, this compounded efficiency directly controls the rate at which entangled links can be generated and refreshed, and thus bounds the throughput available to higher-layer applications.

Because these losses compound multiplicatively, the ``store-and-stitch'' operation of a repeater is highly lossy. The most challenging of these losses is the propagation loss in the optical fiber, which scales exponentially with distance, quickly making direct, long-distance quantum communication impractical. However, the purpose of the repeater is not to eliminate loss, but to fundamentally alter how it scales with distance. By using memories to connect multiple shorter segments, suitably designed repeater architectures can, in principle, transform the single, catastrophic \textit{exponential} loss of direct fiber transmission into more favorable, sub-exponential or effectively \textit{polynomial} scaling under appropriate assumptions about resources and protocols. This foundational principle is what makes long-distance quantum networks theoretically achievable.

\subsection{Operating Wavelength}
The operating wavelength is the specific ``color'' of light that a quantum component is built to use. This metric is not primarily about performance but about fundamental compatibility with the surrounding infrastructure. The global fiber-optic network is engineered for minimum loss in the telecom C-band (around 1550 nm~\cite{miya1979ultimate}). However, many of the best-performing quantum memories demonstrated to date operate at visible or near-infrared wavelengths (e.g., 780 nm for certain atomic systems~\cite{radnaev2010quantum}), creating a wavelength mismatch between memories and deployed fiber.

Bridging this mismatch generally requires inserting a \textit{Quantum Frequency Converter (QFC)} to act as a translator between the memory and the telecom-band fiber. A QFC stage is an additional, non-ideal network component that introduces its own penalties to both overall efficiency (through several dB of extra loss) and fidelity (through added noise at the single-photon level). System and architecture designs must therefore choose between integrating QFC stages at memory interfaces—accepting these penalties—or adopting native telecom-band platforms (such as Erbium-doped solid-state memories) that directly match the C-band but may currently lag in other metrics like efficiency or multimode capacity.

\subsection{Bandwidth and the ``Clock Speed''}
Bandwidth defines the operational frequency range of a quantum device. While analogous to a classical channel width (e.g., a 20 MHz Wi-Fi channel), its role in a quantum network is fundamentally different. In classical communications, information is encoded into the bandwidth using modulation schemes on a strong carrier wave. In a quantum network, the information is carried by single, discrete quantum states, and the effective ``throughput'' is governed by the repetition rate at which these single-qubit wavepackets can be generated, transmitted, and processed.

Bandwidth dictates this maximum ``clock speed.'' In the time–frequency domain, sending more wavepackets per second (a higher repetition rate) requires shorter temporal pulses with duration $\Delta t$, and such pulses necessarily occupy a broader frequency bandwidth, with a characteristic relation $\Delta t\,\Delta \nu \gtrsim 1$. Therefore, a high-throughput network must employ components that support sufficiently large bandwidths, so that pulses can be short enough to sustain the desired attempt rate without excessive temporal overlap. For link-layer protocols, this bandwidth-limited clock speed directly bounds how fast entanglement generation attempts, refresh operations, and link maintenance procedures can be scheduled.

The quantum memory is typically the primary bottleneck in this respect. In almost all real-world physical platforms demonstrated to date, there are strong trade-offs between bandwidth and fidelity: increasing bandwidth generally requires faster, stronger control fields, which tend to be noisier and less precise, thereby degrading fidelity. Conversely, highly resonant, finely tuned memories that achieve excellent efficiencies and coherence times are often intrinsically narrow-band. This creates a persistent physical tension between speed, quality, and success probability that network designs must account for when targeting specific rates and service levels.

\begin{table*}[htbp]
\caption{Quantum Memory Metrics, Classical Equivalents, and Network Impact}
\begin{center}
\renewcommand{\arraystretch}{1.25}
\begin{tabularx}{\textwidth}{%
  >{\raggedright\arraybackslash}p{2.5cm}%
  >{\raggedright\arraybackslash}p{2.5cm}%
  >{\raggedright\arraybackslash}p{4.0cm}%
  >{\raggedright\arraybackslash}X}
\hline
\textbf{Quantum Memory Metric} & \textbf{Classical Concept} & \textbf{Physical Definition} & \textbf{Network \& Control Plane Impact} \\ \hline

\textbf{Fidelity ($F$)} &
Signal integrity / BER &
Accuracy/purity of the stored quantum state. &
Dictates the usability of the state for entanglement-based tasks. For many common single-qubit noise models, if $F(t) \lesssim 2/3$, the channel effectively behaves as entanglement-breaking and cannot be used to distribute entanglement. \\ \hline

\textbf{Coherence time ($T_2$)} &
Buffer / TTL lifetime &
Physical timescale over which phase information is randomized. &
Determines, together with $F_0$ and the noise model, the derived entanglement-preserving lifetime $T_{\text{non-EB}}$ that bounds all classical control signalling and round-trip heralding delays. \\ \hline

\textbf{Efficiency} &
Packet loss rate &
Probability of successfully absorbing and retrieving a photon. &
Multiplies across hops and components. Low efficiency forces many repeated generation attempts, severely limiting the usable entanglement rate and end-to-end throughput. \\ \hline

\textbf{Multimode capacity} &
Queue depth / WDM &
Number of isolated qubits (modes) that can be stored simultaneously. &
Turns a memory into a network queue; enables spatial/temporal multiplexing, advanced scheduling, and q-datagram–style separation of quantum data and classical metadata. \\ \hline

\textbf{Operating wavelength} &
Carrier frequency &
Optical frequency of the interfacing photons. &
Mismatch with the telecom C-band (1550 nm) typically forces the use of lossy Quantum Frequency Converters (QFC), adding extra loss and noise at memory–fiber interfaces. \\ \hline

\textbf{Bandwidth} &
Clock speed &
Usable frequency range, related to the minimum pulse duration via the time–bandwidth relation. &
Determines the maximum repetition rate for entanglement generation attempts and refresh operations (the network’s raw link-layer ``clock speed''). \\ \hline

\end{tabularx}
\label{tab:memory_metrics}
\end{center}
\end{table*}

\subsection{Multimode Capacity}
\textit{Multimode capacity} is the ``parallelism'' metric, answering the question: \textit{How many qubits can the memory store and process simultaneously?} This value varies dramatically across platforms, from tens of modes in high-fidelity register-based systems (such as trapped ions or small spin registers) to hundreds or thousands of modes in temporally multiplexed ensemble and solid-state memories.

This capacity acts as a direct multiplier on the network's overall throughput, analogous to increasing the number of channels in a classical Wavelength-Division Multiplexing (WDM) system. More importantly, sufficient multimode capacity physically transforms the memory from a simple component into a true \textit{network queue}. It allows a repeater to multiplex its resources in space, time, or frequency, enabling more sophisticated scheduling, blocking avoidance, and congestion-aware routing strategies in the control plane.

In classical packet switching, routers perform logic by reading identifiers directly from packet headers. This is impossible for quantum information, as a qubit cannot be read or duplicated without being destroyed. The equivalent solution, as proposed in ``q-datagram'' architectures, is to separate the quantum data from its classical metadata. The qubit itself is held in a specific memory mode, while a parallel classical packet is transmitted and processed alongside it, containing all the necessary network identifiers and, crucially, an explicit link-level label pointing to the physical memory mode where the corresponding qubit resides. This separation allows all network logic (queuing, scheduling, and routing) to be performed on the classical, readable metadata, while the fragile quantum state itself is never disturbed.

This highlights that multimode capacity is a crucial design parameter that must be carefully balanced, often forcing a choice between the high-fidelity, low-mode operation of register-based systems and the high-capacity, noisier operation of temporally multiplexed platforms. To synthesize these multidimensional constraints for network design, Table~\ref{tab:memory_metrics} provides a conceptual mapping between the physical quantum memory metrics, their classical networking equivalents, and their direct impact on the network's control plane.

\subsection{Critical Assumptions and Modeling Pitfalls}
Because quantum memory physics are incredibly complex, network models and protocol designs frequently rely on abstracted representations of memory behavior. However, these abstractions often make highly optimistic assumptions about memory capabilities, creating a significant gap between simulated network performance and physical reality. Four recurring pitfalls are particularly relevant:

\begin{enumerate}
    \item \textbf{Using $T_2$ as the Memory Lifetime.} Many works use the physical coherence time ($T_2$) as the functional lifespan of the memory. Because $T_{\text{non-EB}}$ is a derived quantity that depends on both the initial fidelity $F_0$ and the detailed noise model, treating $T_2$ as a direct proxy is only a good approximation when $F_0$ is very close to unity (e.g., $F_0 \gtrsim 0.99$). In realistic regimes with more modest initial fidelities (e.g., $F_0 \approx 0.75$–$0.9$), the true $T_{\text{non-EB}}$ can be a small fraction of $T_2$, sometimes shorter by orders of magnitude. Network simulators and performance models should therefore compute $T_{\text{non-EB}}$ explicitly under the assumed noise model, rather than optimistically assuming qubits remain useful for the entire $T_2$ duration.

    \item \textbf{Cherry-Picking a ``Frankenstein'' Memory.} Because of the physical tension between speed, quality, and efficiency, theoretical models sometimes cherry-pick the best reported specifications from entirely different and incompatible platforms—combining, for example, the record-high fidelity of a slow trapped-ion system, the record-high bandwidth of a noisy solid-state system, and the massive multimode capacity of an atomic ensemble. As illustrated in Table~\ref{tab:hardware_platforms_transposed}, these metrics are strongly coupled in practice: platforms with very long coherence and high fidelities typically operate with limited bandwidth and mode count, while highly multimode or broadband systems often pay a price in efficiency or fidelity. Resource-allocation and routing algorithms should parameterize concrete hardware families with their internal trade-offs, rather than assuming an idealized ``Frankenstein'' component whose core specifications are physically contradictory.

    \item \textbf{Assuming Perfect Resource Fungibility.} Networking algorithms often treat the modes within a multimode memory as perfectly fungible, isolated queue slots, implicitly assuming that retrieving a qubit from any mode or swapping qubits between modes is instantaneous and error-free. In practice, addressing a specific mode out of hundreds or thousands typically requires complex optical or microwave switchyards, which introduce additional loss and mode-dependent latency. Moreover, as more modes are packed into a single device, control fields addressing one mode can ``nudge'' neighboring modes, creating crosstalk that lowers fidelity across the array. To remain realistic, control-plane designs and mapping algorithms should become explicitly hardware-aware, modeling mode-dependent loss, crosstalk, and switching penalties when assigning logical flows to physical memory layouts.

        \item \textbf{Treating Entanglement as a Classical Buffer.} Because classical networks can store data during off-peak hours to absorb peak traffic, it is tempting to apply the same load-balancing logic to quantum networks. For instance, overlay proposals have suggested generating and storing entangled pairs in intermediate nodes during low-demand periods so they can be consumed later during peak demand~\cite{pouryousef2023}. This paradigm effectively treats a quantum memory like a classical hard drive, ignoring the continuous, exponential decay of real-time fidelity. For most near-term memory platforms operated in realistic network conditions, the entanglement-preserving lifetime $T_{\text{non-EB}}$ is on the order of milliseconds to seconds, making macroscopic ``caching'' across traffic epochs physically infeasible even if longer coherence has been demonstrated in carefully controlled lab settings. Scheduling and routing algorithms should therefore treat entanglement as an on-demand, perishable resource tied to current link conditions, not as a storable asset over long timescales.
\end{enumerate}

\begin{table*}[htbp]
\caption{Indicative Performance Trade-Offs Across Quantum Memory Platforms}
\begin{center}
\renewcommand{\arraystretch}{1.25}
\begin{tabularx}{\textwidth}{%
  >{\raggedright\arraybackslash}l
  >{\raggedright\arraybackslash}X
  >{\raggedright\arraybackslash}X
  >{\raggedright\arraybackslash}X
  >{\raggedright\arraybackslash}X}
\hline
\textbf{Metric} &
\textbf{Trapped Ions} &
\textbf{Atomic Ensembles} &
\textbf{NV Centers (Diamond)} &
\textbf{Superconducting Circuits} \\ \hline

\textbf{Native wavelength} &
Visible / UV &
Near-IR (NIR) &
Visible / NIR &
Microwave ($\sim$5 GHz) \\ \hline

\textbf{Coherence time ($T_2$)} &
Excellent (sec to hours) &
Moderate (msec to sec) &
Good (msec to sec) &
Low ($\mu$sec to msec) \\ \hline

\textbf{Multimode capacity} &
Low (tens of modes) &
Very high (hundreds to thousands of temporal modes via multiplexing) &
Low to moderate (local nuclear-spin registers) &
Moderate (limited by footprint and wiring) \\ \hline

\textbf{Fidelity / efficiency} &
Very high / very high &
Moderate / moderate &
High / low to moderate &
Very high / high \\ \hline

\textbf{Primary network challenge} &
Slow clock speeds and narrow bandwidth; complex vacuum setup and QFC needed for telecom fiber. &
Moderate fidelity and efficiency; probabilistic interactions complicate large-scale deterministic entanglement distribution. &
Low photon collection and non-telecom wavelengths; needs efficient interfaces and QFC for long-distance links. &
Microwave-only operation; needs efficient microwave–optical transduction for room-temperature, long-distance networks. \\ \hline

\end{tabularx}
\label{tab:hardware_platforms_transposed}
\end{center}
\end{table*}

\subsection{The Path to the Quantum Internet: Required Memory Advances}

The quantum memory is the physical anchor of the quantum network. Its efficiency dictates the probability that entanglement-generation attempts succeed, its multimode capacity dictates the degree of multiplexing available to the network, and its coherence time (through $T_{\text{non-EB}}$) dictates the strict temporal budget for all classical control operations. Within these boundaries, current memory technologies—while sufficient for localized, proof-of-concept demonstrations—remain one of the dominant hardware bottlenecks preventing the deployment of a large-scale, wide-area Quantum Internet.

For the Quantum Internet to transition from experimental physics to a scalable, networked infrastructure managed by a protocol stack and control plane, quantum memories will need to evolve along several key dimensions. At least four particularly impactful directions can be identified:

\begin{enumerate}
    \item \textbf{More Deterministic Light–Matter Interfaces.} Current memories suffer from severe storage-and-retrieval efficiency losses because the interaction between a flying photon and a stationary emitter is inherently weak and probabilistic. Realizing a practical Quantum Internet will require interfaces (for example, based on advanced nanophotonic or cavity-QED structures) that can catch and emit single photons with efficiencies approaching unity. Such near-deterministic interfaces would shift the network from a regime of ``constant retries'' to one in which entanglement generation behaves much closer to a reliable, on-demand primitive.

    \item \textbf{Native Telecom-Band Operation.} As detailed in Table~\ref{tab:memory_metrics}, the mismatch between memory wavelengths and the 1550 nm telecom fiber grid currently forces the use of lossy Quantum Frequency Converters (QFC) at memory–fiber interfaces. A major step forward will be the maturation of solid-state memories (such as Erbium-doped crystals and fibers) that natively operate in the telecom C-band while maintaining long coherence times and useful multimode capacity, allowing them to plug directly into existing dark-fiber infrastructure without incurring transduction penalties.

    \item \textbf{On-Chip, Scalable Multiplexing.} To function as true quantum routers, memories must scale from handling tens of qubits to thousands of well-isolated modes. This requires a shift away from bulky, room-sized optical tables toward photonic integrated circuits (PICs) and chip-scale architectures. The key advance will be high-density, on-chip spatial, temporal, and spectral multiplexing with low loss and minimal crosstalk, enabling the large ``queue depths'' needed for sophisticated SDQN-style routing, path diversity, and Quality-of-Service (QoS) management.

    \item \textbf{Relaxed Bandwidth–Fidelity Trade-Offs.} Ultimately, the most transformative progress would come from physical platforms that substantially relax the traditional trade-offs between bandwidth, coherence, and multimode capacity. From a network perspective, the ideal memory would combine the rapid, high-bandwidth absorption rates characteristic of solid-state systems (supporting gigahertz-scale clock speeds) with the ultra-long, high-fidelity $T_{\text{non-EB}}$ lifetimes characteristic of isolated atomic systems, thereby enabling both fast link refresh and deep classical control-plane orchestration on the same hardware platform.
\end{enumerate}

Until such advances are realized, the network control plane and protocol stack must carry much of the burden of hardware fragility. The following section explores how these localized, imperfect memory constraints compound across multiple hops, defining the end-to-end performance metrics and design space of the overall quantum network.


\section{PERFORMANCE METRICS: EVALUATING THE END-TO-END NETWORK}
\label{sec:performance_metrics}

As established in Section~\ref{sec:quantum_memory}, the physical constraints of a quantum network are anchored by the local limitations of its quantum memories and optical sources. However, a functional Quantum Internet is a complex, multi-hop chain of these components, connected by optical fibers and orchestrated by classical control planes. Classical networking metrics like Bit Error Rate (BER) and packet latency are insufficient for evaluating this chain because classical errors are discrete events that can be easily detected and corrected via data retransmission. Quantum information, however, is encoded in fragile superpositions that suffer from continuous, irreversible degradation, and the no-cloning theorem renders classical retransmission obsolete. 

Evaluating a quantum network requires a specialized, multi-dimensional metric space. Furthermore, for a network engineer, it is crucial to distinguish between two distinct classes of metrics, as summarized in Table \ref{tab:network_metrics}. \textit{Real-Time Operational Metrics} (such as End-to-End Fidelity, Quantum Delay, and Secret Key Rates) are actively monitored by the control plane to make dynamic routing, scheduling, and resource allocation decisions. In contrast, \textit{Offline Characterization Metrics} (such as Quantum Channel Capacity) are used primarily to benchmark the theoretical limits and baseline health of the physical hardware.

\begin{table*}[htbp]\caption{Taxonomy of Quantum Network Performance Metrics}
\centering
\renewcommand{\arraystretch}{1.4}
\begin{tabularx}{\textwidth}{llXX}
\hline
\textbf{Metric Class} & \textbf{Performance Metric} & \textbf{Classical Equivalent} & \textbf{Network Impact \& Operational Role} \\ \hline
\multirow{6}{*}{\textbf{Real-Time Operational}} & \textbf{End-to-End Fidelity ($F_{e2e}$)} & Signal Integrity / SNR & The primary routing constraint; dictates if a path is viable or if costly purification is required. \\ \cline{2-4}
 & \textbf{Quantum Error Rate (QER)} & Bit Error Rate (BER) & Multi-dimensional (X/Z errors); dictates which specific Quantum Error Correction code to deploy. \\ \cline{2-4}
 & \textbf{Quantum Delay} & Network Latency / RTT & Must remain strictly below the global coherence budget, otherwise the state decays entirely. \\ \cline{2-4}
 & \textbf{Timing Jitter} & Packet Delay Variation & A signal-level constraint; sub-nanosecond jitter physically destroys quantum interference and swaps. \\ \cline{2-4}
 & \textbf{Entanglement Distribution Rate} & Effective Goodput & The final network throughput (usable pairs/sec), heavily penalized by intermediate swapping failures. \\ \cline{2-4}
 & \textbf{Secret Key Rate (SKR)} & IPSec/TLS Payload Rate & The application-layer throughput for QKD; collapses to zero if the Quantum Bit Error Rate is too high. \\ \hline
\textbf{Offline Characterization} & \textbf{Quantum Channel Capacity} & Shannon-Hartley Limit & Defines the absolute physical ceiling for entanglement distribution over a given lossy channel. \\ \hline
\end{tabularx}
\label{tab:network_metrics}
\end{table*}

\subsection{End-to-End Fidelity ($F_{e2e}$)}

While local fidelity defines the quality of a single quantum memory, the primary metric of concern at the network layer is the End-to-End Fidelity ($F_{e2e}$). This continuous value, ranging from 0 to 1, quantifies the mathematical ``closeness'' of the final, distributed entangled state shared between two distant end-nodes to the ideal, error-free target state. It serves as the quantum analogue to classical Signal Integrity or Signal-to-Noise Ratio (SNR). However, while a degraded classical signal can be passively regenerated or amplified by a standard repeater, a loss in quantum fidelity represents an irreversible destruction of the state's phase and superposition.

End-to-end fidelity is a cumulative, non-additive metric. Every time a repeater node performs an entanglement swap to extend a link, the algorithmic noise of the Bell-state measurement (BSM) and the physical noise of the local gates degrade the state. In a practical network, $F_{e2e}$ acts as the primary real-time routing constraint. If $F_{e2e}$ drops below the classical threshold of $2/3$, the end-to-end channel becomes entanglement-breaking and the link is functionally useless. To illustrate, if a DQC application requires an $F_{e2e}$ of 0.85 to function, and the SDQN controller routes the connection over five hops where each hop introduces a 4\% fidelity penalty due to imperfect gates, the final state will arrive below the required threshold, rendering the generated entanglement unusable.

\subsection{Quantum Error Rate (QER)}

While fidelity offers a high-level, single-number summary of a state's overall quality, the Quantum Error Rate (QER) provides the granular diagnostic data required to understand exactly \textit{why} that quality is degrading. QER breaks down the total error into specific physical types, serving as the multi-dimensional counterpart to the classical Bit Error Rate (BER). Unlike a classical bit, which only has one axis of error (flipping from 0 to 1), a qubit can flip its state ($X$-error), flip its quantum phase ($Z$-error), or suffer a simultaneous combination of both ($Y$-error).

The evaluation of these specific error rates is critical for the engineering of robust control planes. Standardized methods like \textit{Randomized Benchmarking (RB)} provide a scalable way to measure the average error rate of network nodes in real-time. For instance, if a network operator notices $F_{e2e}$ dropping on a specific path, they will run RB. If RB reveals a high average gate error rate of 5\%, engineers will deploy \textit{Gate Set Tomography (GST)} to drill down and discover that 4.5\% of those errors are specifically phase-flips caused by fluctuating magnetic fields at a repeater node. This interaction directly dictates the network's operational strategy: a routing path dominated by phase-flip errors requires the SDQN controller to allocate entirely different Quantum Error Correction (QEC) codes than a path dominated by amplitude damping.

\subsection{Diagnostic Instrumentation: The Scalability Bottleneck}
\label{subsec:Tomography}

To accurately measure $F_{e2e}$ and diagnose specific QERs, network operators rely on a hierarchy of tomographic techniques. The fundamental process of quantum tomography is a systematic procedure: a known state is prepared, evolved through the system, measured across multiple bases, and classically reconstructed into a mathematical density matrix. 

Historically, this is applied at the component level in two distinct ways:
\begin{itemize}
    \item \textbf{Quantum State Tomography (QST)} \cite{Altepeter2004QST, Lvovsky2009CVQST} characterizes the final state of a qubit, answering: \textit{``What exact state did our network deliver?''}
    \item \textbf{Quantum Process Tomography (QPT)} \cite{Childs2001QPT, OBrien2004QPTCNOT} characterizes how a specific channel or gate affects a state, answering: \textit{``How does this specific link behave?''} The highly detailed noise models derived from QPT are an absolute prerequisite for designing the error mitigation and error correction protocols discussed in the following section.
\end{itemize}

While highly detailed, QST and QPT suffer from an exponential sample complexity in relation to the state dimension. Consequently, attempting to perform brute-force QPT on every individual link and node to maintain a global, real-time view of the network is practically infeasible. Using these tools to continuously monitor a multi-hop network would consume massive amounts of entanglement generation time, effectively paralyzing the network's data throughput. This creates a severe operational bottleneck: how can the real-time link metrics required for network routing and resource allocation be gathered without destroying the network's payload capacity?

A scalable solution to this bottleneck is \textbf{Quantum Network Tomography (QNT)}~\cite{deandrade2024_QNT}. Unlike QPT, which isolates single components, QNT draws inspiration from classical network inference. It relies on end-to-end state distribution (such as routing multipartite entangled states across the network) and uses the correlated measurement statistics at the edge nodes to mathematically infer the error parameters of internal links. Crucially, QNT achieves this with polynomial sample complexity, bypassing the exponential overhead of QPT. The potential operational integration of this technique into a network's control loop will be discussed further in Section~\ref{sec:q-num}.

\subsection{Network Decoherence Rate and the Global Coherence Budget}

In Section~\ref{sec:quantum_memory}, the \textit{Coherence Times} ($T_1$ and $T_2$) were introduced as localized constraints of a physical memory. In the context of the entire network, these local decay times combine to create the \textit{Network Decoherence Rate}, which imposes a strict, global time limit on all network operations. The real-time fidelity of a stored state roughly decays exponentially as a function of the dephasing time, often modeled heuristically as $F(t) \approx F_0 e^{-t/T_2}$. 

In long-distance communication, a qubit is continuously exposed to environmental noise during transmission and while buffered in intermediate repeaters. Because quantum states cannot be passively amplified, this time-dependent exponential decay fundamentally limits the maximum physical distance between nodes and the maximum number of intermediate hops the network can support. If the accumulated time for all multi-hop operations exceeds the minimum $T_2$ coherence limit of the path, the quantum states will completely decohere before the end-to-end link is established. The global coherence budget therefore acts as the ultimate operational deadline that the classical control plane must beat.

\subsection{Quantum Delay}

The time consumed against this global coherence budget is measured by the \textit{Quantum Delay}. In classical networks, latency is primarily a QoS issue causing buffering or lag. In quantum networks, latency actively destroys the payload. Quantum Delay measures the total time required to successfully distribute entanglement between end nodes. 

This metric is not a simple measure of travel time; it is the sum of three key components: transmission time (the speed-of-light propagation in the fiber), generation time (the time required by the source to emit a pair), and protocol time. Protocol time is often the largest and most volatile component, accounting for all the probabilistic retries of entanglement generation, the execution of local swapping gates, and the round-trip classical heralding delays required to confirm success. For practical context, if two nodes are 100 km apart, the speed-of-light transmission delay is roughly 0.5 milliseconds. However, if the generation fails 9 times out of 10, and the nodes must wait for classical heralding signals to confirm each attempt, the actual Quantum Delay to secure one successful pair might balloon to 5 milliseconds. Minimizing Quantum Delay is therefore not just an optimization for speed, but a fundamental prerequisite for network connectivity, as exceeding the coherence budget drops $F_{e2e}$ below the classical limit.

\subsection{Timing Jitter: Packet-Level vs. Signal-Level}

\textit{Timing Jitter} functions as the quantum analogue to classical jitter, but the physical consequences are far more severe. In classical networks, jitter is generally a \textit{packet-level} issue caused by variable queuing delays at routers, resulting in uneven data delivery times that can be easily smoothed out by an application-layer buffer. 

In quantum networks, jitter is exclusively a \textit{signal-level} issue that actively destroys quantum information. It represents the random, unwanted temporal deviation in the arrival of a photon or the execution of a control pulse, primarily caused by electronic noise in classical synchronization clocks and detectors. Because quantum information is encoded directly into a single photon wavepacket, any jitter in the underlying classical electronics directly corrupts the state. For example, entanglement swapping relies on the Hong-Ou-Mandel effect, which requires two independent photons from different links to arrive at a beam splitter within the exact same femtosecond window to undergo quantum interference. If classical clock jitter causes one photon to arrive just a picosecond late, they will not interfere, destroying the interference and collapsing the operation entirely.

\subsection{Entanglement Swapping Success Probability ($P_{swap}$)}

The destructive impact of timing jitter directly degrades the \textit{Entanglement Swapping Success Probability} ($P_{swap}$). This is the critical link-layer operational metric for building long-distance networks, measuring the likelihood (a unitless probability between 0 and 1) that an intermediate quantum repeater will successfully connect two adjacent entangled segments. 

Because standard linear-optical BSMs are fundamentally limited by quantum interference rules, the maximum theoretical $P_{swap}$ for basic photonic setups is capped at 50\%. At the network level, $P_{swap}$ acts as a massive bottleneck because the success of an entire $N$-hop chain is proportional to $(P_{swap})^{N-1}$. For example, in a 4-hop network utilizing linear optics, the entanglement must be swapped 3 times. If $P_{swap}$ is 50\%, the probability of successfully establishing the end-to-end link on any given attempt is only $0.5^3 = 12.5\%$. A low $P_{swap}$ forces the network into a continuous loop of probabilistic retries, which drastically drives up the Quantum Delay and consumes the strict coherence budget.

\subsection{Entanglement Distribution Rate (Network Throughput)}

The cumulative effect of these physical and temporal constraints ultimately dictates the network's throughput. It is crucial to distinguish between the local \textit{Entanglement Generation Rate} (EGR) and the end-to-end \textit{Entanglement Distribution Rate}. EGR is the raw, localized speed at which a single physical source can emit entangled pairs (analogous to the clock rate of a network interface card). 

However, the ultimate capacity metric for the network is the Entanglement Distribution Rate, measured in \textbf{E}ntangled \textbf{P}airs per \textbf{S}econd (EPS). To account for all physical and protocol losses, we define this final operational throughput as the \textit{Effective EPS}, which serves as the direct equivalent of application-layer ``Goodput.'' It quantifies the final rate at which high-fidelity, usable entangled pairs are reliably delivered between distant end-nodes. The Distribution Rate is calculated by taking the raw generation rate and filtering it through all the network's inefficiencies: fiber attenuation, the compounding penalty of $P_{swap}$, and the time overhead of classical heralding. This mathematical bottleneck can be formalized as:
\begin{equation}
\text{Effective EPS} \propto \frac{\text{EGR} \cdot (P_{swap})^{N-1} \cdot \eta_{\text{link}}^N}{O_{\text{protocol}}}
\end{equation}
where $\eta_{\text{link}}$ represents the transmissivity (survival probability) of the photons over the intermediate fiber hops, and $O_{\text{protocol}} \ge 1$ is a dimensionless overhead factor accounting for the operational inefficiencies of the network, primarily the time penalties of classical heralding and probabilistic retries. 
In practice, a network path might start with an EGR of 1,000,000 pairs per second, but after accounting for exponential fiber attenuation, the 50\% $P_{swap}$ limit, and the sacrifice of multiple pairs to satisfy purification thresholds, the Effective EPS might be a mere 100 usable pairs per second. Consequently, the effective network throughput is mathematically constrained by a strict, inverse relationship with the desired end-to-end fidelity.

\subsection{Secret Key Rate (SKR)}

While the Distribution Rate (EPS) measures raw quantum capacity, the \textit{Secret Key Rate (SKR)} is the ultimate application-layer operational metric for QKD systems. It quantifies the final rate, in bits per second, at which two parties can extract a truly secure, shared cryptographic key mathematically guaranteed to be safe from eavesdropping.

The SKR ($R$) is derived directly from the raw throughput by applying strict cryptographic security conditions. For a typical protocol like BB84, the lower bound for the secret key rate is calculated as:
\begin{equation}
R \ge q \left[ 1 - h(\text{QBER}) - f_{\text{EC}} \cdot h(\text{QBER}) \right]
\end{equation}
In this formulation, $q$ is the raw transmission rate of photons, while the \textit{Quantum Bit Error Rate} (QBER) represents the fraction of differing bits between the sender and receiver. Crucially, the QBER is inversely proportional to the end-to-end fidelity of the channel. The function $h(\text{QBER})$ is the binary Shannon entropy, which quantifies the maximum amount of information an eavesdropper could have theoretically extracted, and $f_{\text{EC}} \ge 1$ represents the classical error correction overhead. If network fidelity drops too low due to natural decoherence, the protocol cannot mathematically distinguish between environmental noise and an active eavesdropper. The resulting $h(\text{QBER})$ penalty becomes so large that the privacy amplification step shrinks the key to zero, safely aborting the compromised connection.

\subsection{Quantum Channel Capacity}

Finally, \textit{Quantum Channel Capacity} serves as the ultimate offline characterization metric, defining the theoretical maximum rate at which quantum information can be reliably sent through a noisy communication channel. Measured in \textit{qubits per second (QPS)}, it represents the absolute physical ceiling of the network. 

This metric is not measured experimentally; it is calculated using formulas derived from quantum information theory, such as the \textit{P-LOB (Pirandola-Laurenza-Ottaviani-Banchi) bound}~\cite{pirandola}. For a pure loss channel (like an optical fiber) with transmissivity $\eta$, the ultimate quantum capacity is bounded by:
\begin{equation}
C_{\text{PLOB}} = -\log_2(1-\eta) \approx 1.44 \eta \quad \text{(bits/use)}
\end{equation}
The calculation accounts for the fundamental physical limits of the channel, particularly exponential photon loss and intrinsic vacuum noise. By providing the theoretical ceiling, the Channel Capacity offers an essential benchmark, revealing to network engineers exactly how close their operational Entanglement Distribution Rates and SKR are to achieving the absolute physical limits of the universe over a given distance.

\subsection{Critical Assumptions and Misunderstandings}

Because classical network engineers are accustomed to highly deterministic metrics, the transition to modeling quantum performance often results in severe architectural misunderstandings. When the classical networking community designs quantum routing and scheduling algorithms, they frequently make idealized assumptions about these metrics that fail in physical deployment:

\begin{enumerate}
    \item \textbf{The AWGN Fallacy (Uniform Error Rates).} Because classical communications engineers are used to AWGN, they frequently model the Quantum Error Rate as a uniform, symmetric probability of a qubit simply ``flipping." As established in Section~\ref{sec:core_challenges}, quantum errors are highly asymmetric. Energy relaxation ($T_1$) only flips a state from $|1\rangle$ to $|0\rangle$, never the reverse. Furthermore, phase-flips ($Z$-errors) occur much more frequently than bit-flips ($X$-errors) in most memory platforms. Network models must abandon symmetric error assumptions and utilize realistic, asymmetric noise channels derived from actual Quantum Process Tomography.
    
    \item \textbf{Jitter as a Manageable Queuing Delay.} Classical network models often treat quantum jitter as just another variable in the overall Quantum Delay, assuming it merely slows down the protocol's execution time. However, because quantum interference requires perfect indistinguishability, quantum jitter is a destructive physical force. Simulators must model jitter as a direct, severe penalty to the Entanglement Swapping Success Probability ($P_{swap}$). If jitter exceeds the photon's temporal wavepacket duration, $P_{swap}$ becomes 0\%, and the link fails entirely regardless of the allowed delay.
    
    \item \textbf{The Independence of Rate and Fidelity.} Network simulators frequently treat the Entanglement Distribution Rate and End-to-End Fidelity as independent variables that can be optimized separately in a multi-objective routing algorithm. In reality, these metrics are tightly coupled in a strict, zero-sum trade-off. Increasing effective throughput by skipping entanglement purification explicitly destroys fidelity. Routing models must adopt coupled cost-functions where the achievable operational rate is mathematically penalized by the required fidelity threshold.
\end{enumerate}

\subsection{Summary: The Need for Active Error Management}

The performance metrics detailed in this section reveal a harsh reality for network design: because quantum errors compound multiplicatively across every hop and continuously over time, end-to-end fidelity will inevitably collapse over long distances. Relying solely on the raw capabilities of the hardware is insufficient to build a scalable Quantum Internet. To support complex applications, the network control plane must actively fight back against these degrading metrics. The following section explores the protocols designed to achieve this: Quantum Error Mitigation (QEM) and Quantum Error Correction (QEC).


\section{MANAGING ERRORS IN QUANTUM NETWORKS}
\label{sec:Error_management}

The central challenge in building functional quantum networks is ensuring the integrity of fragile quantum information as it is generated, stored, and transmitted. In classical networks, this challenge is met by a robust, layered reliability stack. At the link layer, Forward Error Correction (FEC) adds redundant bits to detect and correct errors \cite{lin1983error}, while at the transport layer, protocols like TCP \cite{rfc793} use a combination of acknowledgments, timeouts, and retransmissions (ARQ) to guarantee bit-perfect, in-order delivery over unreliable channels \cite{tanenbaum2011computer}. 

While the end-goal is similar, the physical principles of quantum mechanics render this entire classical playbook obsolete. The ``no-copy'' rule of the No-Cloning Theorem \cite{Wootters1982ASQ} makes classical retransmission protocols like ARQ physically impossible, as an unknown quantum state cannot be buffered and re-sent \cite{briegel1998quantum}. Similarly, the ``no-peek'' rule of the Information-Disturbance Theorem \cite{nielsen2000quantum} means classical error detection, such as verifying a checksum, is forbidden, as any direct measurement would collapse the qubit's superposition and destroy the information. This fundamental departure forces the development of entirely new strategies for achieving reliability, which in turn dictate the hardware requirements, the real-time demands on the classical control plane, and the ultimate performance trade-offs between fidelity, rate, and latency \cite{gottesman1997stabilizer}. 

Table \ref{tab:error_management_summary} provides a high-level taxonomy of these quantum reliability paradigms before we analyze their specific network-level mechanics.

\begin{table*}[htbp]
\caption{Taxonomy of Quantum Error Management Strategies}
\begin{center}
\renewcommand{\arraystretch}{1.4}
\begin{tabularx}{\textwidth}{llXXl}
\hline
\textbf{Paradigm} & \textbf{Network Layer} & \textbf{Primary Mechanism} & \textbf{Classical Analogy} & \textbf{Primary Application} \\ \hline
\textbf{Heralding} & Link Layer & Real-time classical signaling to confirm probabilistic success and trigger post-selection. & TCP ACK (but drops failed attempts instead of retransmitting). & All Quantum Networks \\ \hline
\textbf{Purification} & Link / Network & Distilling one high-fidelity state from an ensemble of low-fidelity pairs. & None (Destructive statistical filtering). & DQC, Repeater-QKD \\ \hline
\textbf{QEM (ZNE/PEC)} & Application Layer & Statistical post-processing and inverse gate injections to extrapolate noise-free results. & Equalizer optimization (but has exponential sampling overhead). & DQC, DQS \\ \hline
\textbf{QEC} & Transport / App & Active, real-time syndrome measurement and localized gate correction on logical qubits. & Hybrid ARQ (HARQ) (but triggers local gate fixes, not retransmissions). & Fault-Tolerant DQC \\ \hline
\end{tabularx}
\label{tab:error_management_summary}
\end{center}
\end{table*}

\subsection{Link-Layer Reliability: Heralding and Purification}

Before computational errors can be addressed at the application layer, the network must first establish a reliable communication channel over noisy physical media. Because quantum signals cannot be passively amplified, the network relies on a sequential mechanism of operational confirmation and statistical distillation. 

\subsubsection{Heralding}
Because entanglement generation over optical fiber is probabilistic and highly lossy, the network's first line of defense is heralding. When a Bell-state measurement (BSM) is performed to swap entanglement at a repeater, a classical signal—the herald—is immediately sent to the distant nodes to confirm success. Conceptually, this is the closest quantum equivalent to a TCP handshake's acknowledgment (ACK) packet \cite{rfc793}. When a classical client sends a SYN packet, the server sends a SYN-ACK back to confirm receipt; if the packet is lost, it is simply retransmitted. In quantum networking, however, a failed entanglement attempt cannot be retransmitted at the qubit level. Instead, heralding acts as a strict post-selection filter. If no herald is received within the expected time window, the distant nodes know the operation failed, immediately discard their noisy memory states, and wait for a completely new generation attempt \cite{sangouard2011quantum, abane2025survey}.
q
\subsubsection{Entanglement Purification}
While heralding confirms that an entangled pair successfully exists, that pair may still suffer from low fidelity due to environmental noise. To combat this, networks rely on a complementary technique known as entanglement purification (or distillation) \cite{bennettPurificationNoisyEntanglement1996, deutsch1996quantum}. Unlike classical FEC, which repairs a single data stream by reading redundant bits, purification works probabilistically on \textit{ensembles} of quantum states. It takes two or more low-fidelity entangled pairs shared between two nodes and uses a set of local quantum operations and classical communication (LOCC) to distill them into a single, higher-fidelity entangled pair \cite{dur1999RepeatersPurification}. The noisy pairs are explicitly sacrificed and discarded to produce a single, more robust resource. From a telecommunications perspective, this makes purification fundamentally different from classical FEC: instead of embedding redundancy within a single codeword and correcting errors \emph{in place}, the network consumes multiple end-to-end entangled pairs as expendable ``fuel'' to probabilistically upgrade one surviving pair to a higher fidelity. This is conceptually reminiscent of diversity combining in classical wireless systems \cite{tse2005fundamentals}, but here the additional pairs are irreversibly consumed by the LOCC protocol rather than coherently combined at the receiver.

\subsubsection{Key Trade-offs and Application to Quantum Networks}
For network architects, purification represents the ultimate manifestation of the rate-fidelity trade-off. Distilling one high-fidelity pair from two low-fidelity pairs cuts the effective Entanglement Distribution Rate (EPS) by more than half. The decision to invoke this costly process depends entirely on the application's utility threshold. Short-distance, Prepare-and-Measure QKD networks \cite{Gisin2002QCrypto} often avoid purification entirely to maximize raw key generation rates, relying instead on classical post-processing to handle noise. Conversely, multi-hop networks—like those linking DQC or enabling continental QKD via long-distance repeaters—absolutely require deep purification \cite{durQuantumRepeatersBased1999}. Without it, the compounding noise of multiple swaps would cause the end-to-end fidelity to collapse below the critical $2/3$ entanglement-breaking threshold \cite{horodecki2003general}, rendering the link mathematically useless.

\subsection{Application-Layer Reliability: Quantum Error Mitigation (QEM)}

Once the underlying transmission links are secured, the network must address the algorithmic errors that occur when DQC actually process those entangled states. This is the domain of Quantum Error Mitigation (QEM) \cite{caiQuantumErrorMitigation2023a}. Unlike classical error mitigation (such as passively optimizing an equalizer to improve a signal's SNR), QEM does not correct errors in real-time. Instead, it is a set of statistical strategies used to obtain a more accurate final result in the presence of noise. Crucially, QEM does not involve any form of qubit retransmission. Instead, it relies heavily on the retransmission of \textit{classical instructions} to the quantum hardware, requiring a tight, two-way control loop between the classical orchestrator and the stationary memory qubits. 

\subsubsection{Zero Noise Extrapolation (ZNE)}
One primary QEM strategy is Zero Noise Extrapolation (ZNE), which works by intentionally applying a controlled amount of extra noise to a quantum circuit \cite{temmeErrorMitigationShortDepth2017, liEfficientVariationalQuantum2017}. The classical network controller commands the quantum hardware to run the exact same computational circuit multiple times, each with an artificially increased noise scaling factor. The measurement results from these varying noise levels are collected and sent to a classical processor. The classical computer then uses statistical curve-fitting to perform a numerical extrapolation backward, mathematically estimating what the computational outcome would have been at a theoretical zero-noise level. 

\subsubsection{Probabilistic Error Cancellation (PEC)}
Alternatively, Probabilistic Error Cancellation (PEC) takes a proactive approach by modifying the circuit before the final measurement \cite{temmeErrorMitigationShortDepth2017}. PEC attempts to invert the effects of quantum noise by applying sequences of operations which, on average, cancel the effect of the noise. The success of PEC is dependent on the accuracy of the noise model assumed, which highlights the importance of good noise characterization. Additionally, since the effects of noise are typically non-unitary, the correction operations must be applied probabilistically over many runs of the circuit.

\begin{figure}
    \centering
    \usetikzlibrary{decorations.pathmorphing,decorations.markings}
\providecommand{\ket}[1]{\left|#1\right\rangle}
\begin{tikzpicture}[scale=1.000000,x=1pt,y=1pt]
\begin{scope}[shift={(60,0)}]
\filldraw[color=white] (0.000000, -7.500000) rectangle (123.000000, 52.500000);
\draw[color=black] (0.000000,45.000000) -- (123.000000,45.000000);
\draw[color=black] (0.000000,30.000000) -- (123.000000,30.000000);
\draw[color=black] (0.000000,15.000000) -- (123.000000,15.000000);
\draw[color=black] (0.000000,0.000000) -- (123.000000,0.000000);
\draw (16.000000,45.000000) -- (16.000000,0.000000);
\begin{scope}
\draw[fill=white] (16.000000, 22.500000) +(-45.000000:14.142136pt and 40.305087pt) -- +(45.000000:14.142136pt and 40.305087pt) -- +(135.000000:14.142136pt and 40.305087pt) -- +(225.000000:14.142136pt and 40.305087pt) -- cycle;
\clip (16.000000, 22.500000) +(-45.000000:14.142136pt and 40.305087pt) -- +(45.000000:14.142136pt and 40.305087pt) -- +(135.000000:14.142136pt and 40.305087pt) -- +(225.000000:14.142136pt and 40.305087pt) -- cycle;
\draw (16.000000, 22.500000) node {$\tilde{\mathcal{U}}$};
\end{scope}
\draw[fill=white,color=white] (38.000000, -6.000000) rectangle (53.000000, 51.000000);
\draw (45.500000, 22.500000) node {$=$};
\draw (75.000000,45.000000) -- (75.000000,0.000000);
\begin{scope}
\draw[fill=white] (75.000000, 22.500000) +(-45.000000:14.142136pt and 40.305087pt) -- +(45.000000:14.142136pt and 40.305087pt) -- +(135.000000:14.142136pt and 40.305087pt) -- +(225.000000:14.142136pt and 40.305087pt) -- cycle;
\clip (75.000000, 22.500000) +(-45.000000:14.142136pt and 40.305087pt) -- +(45.000000:14.142136pt and 40.305087pt) -- +(135.000000:14.142136pt and 40.305087pt) -- +(225.000000:14.142136pt and 40.305087pt) -- cycle;
\draw (75.000000, 22.500000) node {$\mathcal{U}$};
\end{scope}
\draw (107.000000,45.000000) -- (107.000000,0.000000);
\begin{scope}
\draw[fill=white] (107.000000, 22.500000) +(-45.000000:14.142136pt and 40.305087pt) -- +(45.000000:14.142136pt and 40.305087pt) -- +(135.000000:14.142136pt and 40.305087pt) -- +(225.000000:14.142136pt and 40.305087pt) -- cycle;
\clip (107.000000, 22.500000) +(-45.000000:14.142136pt and 40.305087pt) -- +(45.000000:14.142136pt and 40.305087pt) -- +(135.000000:14.142136pt and 40.305087pt) -- +(225.000000:14.142136pt and 40.305087pt) -- cycle;
\draw (107.000000, 22.500000) node {$\mathcal{E}$};
\end{scope}
\end{scope}
\begin{scope}[shift={(0,-70)}]
\filldraw[color=white] (0.000000, -7.500000) rectangle (246.000000, 52.500000);
\draw[color=black] (0.000000,45.000000) -- (246.000000,45.000000);
\draw[color=black] (0.000000,30.000000) -- (246.000000,30.000000);
\draw[color=black] (0.000000,15.000000) -- (246.000000,15.000000);
\draw[color=black] (0.000000,0.000000) -- (246.000000,0.000000);
\draw (16.000000,45.000000) -- (16.000000,0.000000);
\begin{scope}
\draw[fill=white] (16.000000, 22.500000) +(-45.000000:14.142136pt and 40.305087pt) -- +(45.000000:14.142136pt and 40.305087pt) -- +(135.000000:14.142136pt and 40.305087pt) -- +(225.000000:14.142136pt and 40.305087pt) -- cycle;
\clip (16.000000, 22.500000) +(-45.000000:14.142136pt and 40.305087pt) -- +(45.000000:14.142136pt and 40.305087pt) -- +(135.000000:14.142136pt and 40.305087pt) -- +(225.000000:14.142136pt and 40.305087pt) -- cycle;
\draw (16.000000, 22.500000) node {$\tilde{\mathcal{U}}$};
\end{scope}
\draw (48.000000,45.000000) -- (48.000000,0.000000);
\begin{scope}
\draw[fill=white] (48.000000, 22.500000) +(-45.000000:14.142136pt and 40.305087pt) -- +(45.000000:14.142136pt and 40.305087pt) -- +(135.000000:14.142136pt and 40.305087pt) -- +(225.000000:14.142136pt and 40.305087pt) -- cycle;
\clip (48.000000, 22.500000) +(-45.000000:14.142136pt and 40.305087pt) -- +(45.000000:14.142136pt and 40.305087pt) -- +(135.000000:14.142136pt and 40.305087pt) -- +(225.000000:14.142136pt and 40.305087pt) -- cycle;
\draw (48.000000, 22.500000) node {$\mathcal{E}^{-1}$};
\end{scope}
\draw[fill=white,color=white] (70.000000, -6.000000) rectangle (85.000000, 51.000000);
\draw (77.500000, 22.500000) node {$=$};
\draw (107.000000,45.000000) -- (107.000000,0.000000);
\begin{scope}
\draw[fill=white] (107.000000, 22.500000) +(-45.000000:14.142136pt and 40.305087pt) -- +(45.000000:14.142136pt and 40.305087pt) -- +(135.000000:14.142136pt and 40.305087pt) -- +(225.000000:14.142136pt and 40.305087pt) -- cycle;
\clip (107.000000, 22.500000) +(-45.000000:14.142136pt and 40.305087pt) -- +(45.000000:14.142136pt and 40.305087pt) -- +(135.000000:14.142136pt and 40.305087pt) -- +(225.000000:14.142136pt and 40.305087pt) -- cycle;
\draw (107.000000, 22.500000) node {$\mathcal{U}$};
\end{scope}
\draw (139.000000,45.000000) -- (139.000000,0.000000);
\begin{scope}
\draw[fill=white] (139.000000, 22.500000) +(-45.000000:14.142136pt and 40.305087pt) -- +(45.000000:14.142136pt and 40.305087pt) -- +(135.000000:14.142136pt and 40.305087pt) -- +(225.000000:14.142136pt and 40.305087pt) -- cycle;
\clip (139.000000, 22.500000) +(-45.000000:14.142136pt and 40.305087pt) -- +(45.000000:14.142136pt and 40.305087pt) -- +(135.000000:14.142136pt and 40.305087pt) -- +(225.000000:14.142136pt and 40.305087pt) -- cycle;
\draw (139.000000, 22.500000) node {$\mathcal{E}$};
\end{scope}
\draw (171.000000,45.000000) -- (171.000000,0.000000);
\begin{scope}
\draw[fill=white] (171.000000, 22.500000) +(-45.000000:14.142136pt and 40.305087pt) -- +(45.000000:14.142136pt and 40.305087pt) -- +(135.000000:14.142136pt and 40.305087pt) -- +(225.000000:14.142136pt and 40.305087pt) -- cycle;
\clip (171.000000, 22.500000) +(-45.000000:14.142136pt and 40.305087pt) -- +(45.000000:14.142136pt and 40.305087pt) -- +(135.000000:14.142136pt and 40.305087pt) -- +(225.000000:14.142136pt and 40.305087pt) -- cycle;
\draw (171.000000, 22.500000) node {$\mathcal{E}^{-1}$};
\end{scope}
\draw[fill=white,color=white] (193.000000, -6.000000) rectangle (208.000000, 51.000000);
\draw (200.500000, 22.500000) node {$=$};
\draw (230.000000,45.000000) -- (230.000000,0.000000);
\begin{scope}
\draw[fill=white] (230.000000, 22.500000) +(-45.000000:14.142136pt and 40.305087pt) -- +(45.000000:14.142136pt and 40.305087pt) -- +(135.000000:14.142136pt and 40.305087pt) -- +(225.000000:14.142136pt and 40.305087pt) -- cycle;
\clip (230.000000, 22.500000) +(-45.000000:14.142136pt and 40.305087pt) -- +(45.000000:14.142136pt and 40.305087pt) -- +(135.000000:14.142136pt and 40.305087pt) -- +(225.000000:14.142136pt and 40.305087pt) -- cycle;
\draw (230.000000, 22.500000) node {$\mathcal{U}$};
\end{scope}
\end{scope}
\end{tikzpicture}
    \caption{\textbf{Probabilistic error cancellation.} A quantum computation implements a desired unitary $\mathcal{U}$, but errors in the hardware cause it to implement a noisy version $\tilde{\mathcal{U}} = \mathcal{E} \circ \mathcal{U} $, disturbed by the error channel $\mathcal{E}$. The ideal unitary $\mathcal{U}$ can be recovered by applying the inverse error channel $\mathcal{E}^{-1}$ after the noisy $\tilde{\mathcal{U}}$.}
    \label{fig:placeholder}
\end{figure}
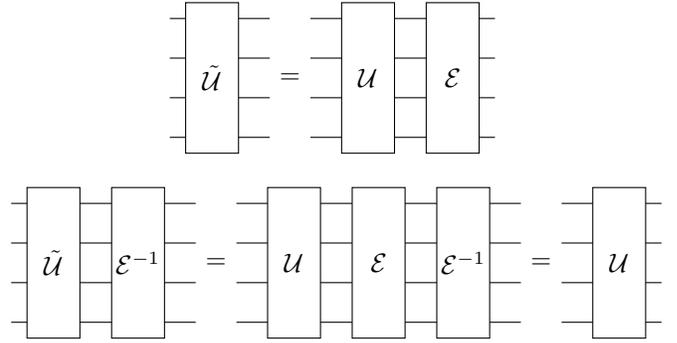

\subsubsection{Key Trade-offs}
The use of these QEM techniques is not a ``free'' solution to noise but rather a complex series of engineering trade-offs, fundamentally balancing the desired quality of the final quantum state against the consumption of finite network resources. The most critical network-level trade-off is between fidelity and effective throughput. QEM schemes directly purchase higher fidelity at the cost of throughput. In both ZNE and PEC, a circuit must be run many times to infer an accurate result for the circuit. Moreover, any error mitigation protocol which suppresses the error below given threshold requires has a sampling overhead that scales exponentially with the circuit depth \cite{takagiFundamentalLimitsQuantum2022}. For PEC, this comes on top of additional circuit depth increase due to the cancellation operations.

This trade-off is further constrained by the consumption of physical network resources, with qubit coherence time being the most valuable and finite. All QEM schemes are in a race against decoherence: the multiple runs of ZNE or the deeper, gate-heavy circuits of PEC must all be completed before the stationary qubits decohere. Moreover, there is a significant trade-off between the QEM schemes themselves concerning upfront versus operational costs. PEC requires a precise, pre-calculated noise model obtained through extensive Quantum Process Tomography (QPT), presenting an extremely high upfront cost to the network operator. ZNE, in contrast, is ``model-free'' and has a low upfront cost but incurs a much higher operational cost in terms of latency and repeated resource usage.

\subsubsection{Application to Quantum Networks}
Because of this heavy computational overhead, the application of ZNE and PEC is mainly reserved for DQC and DQS. These applications rely on performing complex quantum circuits across multiple nodes, which are highly sensitive to noise; the final computational or sensing result would be rendered useless without significant mitigation. In contrast, QKD is far less reliant on computationally intensive QEM techniques. QKD protocols are simpler and more linear, focusing on the distribution of a shared secret key rather than performing a complex quantum computation. Because QKD protocols are designed to tolerate a certain level of physical noise mathematically, the benefits of running repeated mitigation circuits do not outweigh the significant cost in terms of network throughput and key generation rates.

\subsection{Fault-Tolerant Reliability: Quantum Error Correction (QEC)}

While QEM mitigates the effects of noise statistically, Quantum Error Correction (QEC) is the long-term route to true fault tolerance. Rather than estimating and cancelling noise in post-processing, QEC aims to continuously detect errors during computation and keep the encoded quantum information intact.

\subsubsection{Logical Qubits and the Syndrome Cycle}

The basic idea behind QEC is redundancy. In a classical repetition code, a single bit can be protected by encoding it across several physical bits and using majority vote to infer the original value if one bit is corrupted. Quantum error correction follows the same high-level principle, but the implementation is fundamentally different: a single \textbf{logical qubit} is encoded across many noisy \textbf{physical qubits}, and the encoded state cannot be directly read out without disturbing the quantum information.

This challenge was first overcome in the 1990s through foundational work by Shor, Steane, Calderbank, and others, and later unified by Gottesman's \textbf{stabilizer formalism}~\cite{shorSchemeReducingDecoherence1995,steaneErrorCorrectingCodes1996,gottesman1997stabilizer}. In this framework, the encoded state is characterized by a set of multi-qubit Pauli operators, known as \textbf{stabilizers}, for which the ideal encoded state is a $+1$ eigenstate. Errors are detected by measuring these operators and checking whether their expected values have changed.

\begin{figure}[h]
    \centering
    \usetikzlibrary{decorations.pathmorphing,decorations.markings}
\providecommand{\ket}[1]{\left|#1\right\rangle}
\usetikzlibrary{decorations.pathmorphing}
\providecommand{\ket}[1]{\left|#1\right\rangle}
\begin{tikzpicture}[scale=1.000000,x=1pt,y=1pt]
\filldraw[color=white] (0.000000, -7.500000) rectangle (120.000000, 37.500000);
\draw[color=black] (0.000000,30.000000) -- (108.000000,30.000000);
\draw[color=black] (108.000000,29.500000) -- (120.000000,29.500000);
\draw[color=black] (108.000000,30.500000) -- (120.000000,30.500000);
\draw[color=black] (0.000000,30.000000) node[left] {$\ket{0}$};
\draw[color=black] (0.000000,15.000000) -- (120.000000,15.000000);
\draw[color=black] (0.000000,15.000000) node[left] {$q_1$};
\draw[color=black] (0.000000,0.000000) -- (120.000000,0.000000);
\draw[color=black] (0.000000,0.000000) node[left] {$q_2$};
\begin{scope}
\draw[fill=white] (12.000000, 30.000000) +(-45.000000:8.485281pt and 8.485281pt) -- +(45.000000:8.485281pt and 8.485281pt) -- +(135.000000:8.485281pt and 8.485281pt) -- +(225.000000:8.485281pt and 8.485281pt) -- cycle;
\clip (12.000000, 30.000000) +(-45.000000:8.485281pt and 8.485281pt) -- +(45.000000:8.485281pt and 8.485281pt) -- +(135.000000:8.485281pt and 8.485281pt) -- +(225.000000:8.485281pt and 8.485281pt) -- cycle;
\draw (12.000000, 30.000000) node {$H$};
\end{scope}
\draw (36.000000,30.000000) -- (36.000000,15.000000);
\begin{scope}
\draw[fill=white] (36.000000, 15.000000) +(-45.000000:8.485281pt and 8.485281pt) -- +(45.000000:8.485281pt and 8.485281pt) -- +(135.000000:8.485281pt and 8.485281pt) -- +(225.000000:8.485281pt and 8.485281pt) -- cycle;
\clip (36.000000, 15.000000) +(-45.000000:8.485281pt and 8.485281pt) -- +(45.000000:8.485281pt and 8.485281pt) -- +(135.000000:8.485281pt and 8.485281pt) -- +(225.000000:8.485281pt and 8.485281pt) -- cycle;
\draw (36.000000, 15.000000) node {$P_1$};
\end{scope}
\filldraw (36.000000, 30.000000) circle(1.500000pt);
\draw (60.000000,30.000000) -- (60.000000,0.000000);
\begin{scope}
\draw[fill=white] (60.000000, -0.000000) +(-45.000000:8.485281pt and 8.485281pt) -- +(45.000000:8.485281pt and 8.485281pt) -- +(135.000000:8.485281pt and 8.485281pt) -- +(225.000000:8.485281pt and 8.485281pt) -- cycle;
\clip (60.000000, -0.000000) +(-45.000000:8.485281pt and 8.485281pt) -- +(45.000000:8.485281pt and 8.485281pt) -- +(135.000000:8.485281pt and 8.485281pt) -- +(225.000000:8.485281pt and 8.485281pt) -- cycle;
\draw (60.000000, -0.000000) node {$P_2$};
\end{scope}
\filldraw (60.000000, 30.000000) circle(1.500000pt);
\begin{scope}
\draw[fill=white] (84.000000, 30.000000) +(-45.000000:8.485281pt and 8.485281pt) -- +(45.000000:8.485281pt and 8.485281pt) -- +(135.000000:8.485281pt and 8.485281pt) -- +(225.000000:8.485281pt and 8.485281pt) -- cycle;
\clip (84.000000, 30.000000) +(-45.000000:8.485281pt and 8.485281pt) -- +(45.000000:8.485281pt and 8.485281pt) -- +(135.000000:8.485281pt and 8.485281pt) -- +(225.000000:8.485281pt and 8.485281pt) -- cycle;
\draw (84.000000, 30.000000) node {$H$};
\end{scope}
\draw[fill=white] (102.000000, 24.000000) rectangle (114.000000, 36.000000);
\draw[very thin] (108.000000, 30.600000) arc (90:150:6.000000pt);
\draw[very thin] (108.000000, 30.600000) arc (90:30:6.000000pt);
\draw[->,>=stealth] (108.000000, 24.600000) -- +(80:10.392305pt);
\end{tikzpicture}
    \caption{\textbf{Pauli product measurement.} An ancilla qubit is used to extract the outcome of the Pauli product $P_1 \otimes P_2$. The same circuit structure generalizes to arbitrary Pauli products $\bigotimes_i P_i$.}
    \label{fig:stabilizer}
\end{figure}
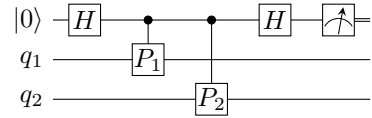

These stabilizer measurements are akin to classical parity checks, though they must be performed without destroying the logical state. This is typically done using auxiliary \textbf{ancilla qubits}: the ancilla interacts with the data qubits through a sequence of controlled operations, and is then measured. The resulting classical outcomes form the \textbf{syndrome}, which indicates where errors may have occurred without revealing the encoded quantum information itself.

\begin{figure}
    \centering
    \begin{tikzpicture}
	\begin{pgfonlayer}{nodelayer}
	\end{pgfonlayer}
	\begin{scope}[tikzit layer=data]
		\begin{pgfonlayer}{nodelayer}
			\node [style=small black] (0) at (0, 4) {};
			\node [style=small black] (7) at (2, 4) {};
			\node [style=small black] (8) at (4, 4) {};
			\node [style=small black] (9) at (3, 3) {};
			\node [style=small black] (10) at (1, 3) {};
			\node [style=small black] (11) at (0, 2) {};
			\node [style=small black] (12) at (2, 2) {};
			\node [style=small black] (13) at (4, 2) {};
			\node [style=small black] (15) at (1, 1) {};
			\node [style=small black] (21) at (3, 1) {};
			\node [style=small black] (22) at (4, 0) {};
			\node [style=small black] (23) at (2, 0) {};
			\node [style=small black] (24) at (0, 0) {};
		\end{pgfonlayer}
		\begin{pgfonlayer}{edgelayer}
		\end{pgfonlayer}
	\end{scope}
	\begin{scope}[tikzit layer=ancilla]
		\begin{pgfonlayer}{nodelayer}
			\node [style=small white] (1) at (0, 3) {};
			\node [style=small white] (2) at (2, 3) {};
			\node [style=small white] (3) at (1, 4) {};
			\node [style=small white] (4) at (1, 2) {};
			\node [style=small white] (5) at (3, 4) {};
			\node [style=small white] (6) at (3, 2) {};
			\node [style=small white] (14) at (4, 3) {};
			\node [style=small white] (16) at (0, 1) {};
			\node [style=small white] (17) at (2, 1) {};
			\node [style=small white] (18) at (4, 1) {};
			\node [style=small white] (19) at (1, 0) {};
			\node [style=small white] (20) at (3, 0) {};
		\end{pgfonlayer}
		\begin{pgfonlayer}{edgelayer}
		\end{pgfonlayer}
	\end{scope}
	\begin{scope}[tikzit layer=X]
		\begin{pgfonlayer}{nodelayer}
			\node [style=none] (25) at (0, 4) {};
			\node [style=none] (26) at (1, 3) {};
			\node [style=none] (27) at (0, 2) {};
			\node [style=none] (28) at (0, 2) {};
			\node [style=none] (29) at (1, 1) {};
			\node [style=none] (30) at (0, 0) {};
			\node [style=none] (31) at (4, 4) {};
			\node [style=none] (32) at (3, 3) {};
			\node [style=none] (33) at (4, 2) {};
			\node [style=none] (34) at (4, 2) {};
			\node [style=none] (35) at (3, 1) {};
			\node [style=none] (36) at (4, 0) {};
			\node [style=none] (37) at (0, 2) {};
			\node [style=none] (38) at (1, 3) {};
			\node [style=none] (39) at (2, 2) {};
			\node [style=none] (40) at (1, 1) {};
			\node [style=none] (41) at (1, 1) {};
			\node [style=none] (42) at (2, 2) {};
			\node [style=none] (43) at (3, 1) {};
			\node [style=none] (44) at (2, 0) {};
			\node [style=none] (45) at (1, 3) {};
			\node [style=none] (46) at (2, 4) {};
			\node [style=none] (47) at (3, 3) {};
			\node [style=none] (48) at (2, 2) {};
			\node [style=none] (49) at (2, 2) {};
			\node [style=none] (50) at (3, 3) {};
			\node [style=none] (51) at (4, 2) {};
			\node [style=none] (52) at (3, 1) {};
			\node [style=none] (53) at (0, 0) {};
			\node [style=none] (54) at (1, 1) {};
			\node [style=none] (56) at (2, 0) {};
			\node [style=none] (57) at (2, 0) {};
			\node [style=none] (58) at (3, 1) {};
			\node [style=none] (60) at (4, 0) {};
			\node [style=none] (61) at (2, 4) {};
			\node [style=none] (62) at (3, 3) {};
			\node [style=none] (63) at (4, 4) {};
			\node [style=none] (64) at (0, 4) {};
			\node [style=none] (65) at (1, 3) {};
			\node [style=none] (66) at (2, 4) {};
		\end{pgfonlayer}
		\begin{pgfonlayer}{edgelayer}
			\draw [style=grey fill] (25.center) to (27.center) to (26.center) to cycle;
			\draw [style=grey fill] (28.center) to (30.center) to (29.center) to cycle;
			\draw [style=grey fill] (31.center) to (33.center) to (32.center) to cycle;
			\draw [style=grey fill] (34.center) to (36.center) to (35.center) to cycle;
			\draw (38.center) to (37.center) to (40.center) to (39.center) to cycle;
			\draw [style=grey fill] (42.center) to (41.center) to (44.center) to (43.center) to cycle;
			\draw [style=grey fill] (46.center) to (45.center) to (48.center) to (47.center) to cycle;
			\draw [style=grey fill] (49) to (52.center);
			\draw [style=grey fill] (54.center) to (53.center);
			\draw [style=grey fill] (56) to (54);
			\draw [style=grey fill] (58.center) to (57.center);
			\draw [style=grey fill] (60) to (58);
			\draw [style=grey fill] (62.center) to (61.center);
			\draw [style=grey fill] (63) to (62);
			\draw [style=grey fill] (65.center) to (64.center);
			\draw [style=grey fill] (66) to (65);
		\end{pgfonlayer}
	\end{scope}
	\begin{scope}[tikzit layer=Z]
		\begin{pgfonlayer}{nodelayer}
			\node [style=none] (67) at (0, 0) {};
			\node [style=none] (68) at (1, 1) {};
			\node [style=none] (69) at (2, 0) {};
		\end{pgfonlayer}
		\begin{pgfonlayer}{edgelayer}
			\draw (24.center) to (68) to (23) to[out=180, in=360] (67);
			\draw (0) to (7);
			\draw (7) to (8);
			\draw (23) to (22);
			\draw (23) to (24);
		\end{pgfonlayer}
	\end{scope}
	\begin{pgfonlayer}{edgelayer}
	\end{pgfonlayer}
\end{tikzpicture}
    \caption{\textbf{Planar surface code.} Dark nodes represent data qubits and light nodes ancillas. Dark and light faces indicate where data qubits share stabilizers.}
    \label{fig:placeholder}
\end{figure}
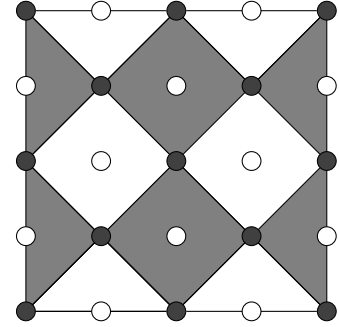

The syndrome does not directly specify the error; it must be interpreted by a classical \textbf{decoder}, which infers the most likely fault pattern and determines the appropriate response. In many architectures, the correction is not applied immediately as a physical gate, but instead tracked in software as a so-called Pauli frame. Either way, QEC is fundamentally a repeated quantum--classical feedback loop: quantum hardware produces syndrome data, classical hardware decodes it, and the result is fed back into the ongoing computation.

\subsubsection{QEC vs. Classical HARQ}

At a high level, this feedback structure makes QEC loosely analogous to Hybrid ARQ (HARQ) in classical communication systems, since both combine redundancy with active classical processing. The crucial difference is that HARQ requests retransmission of classical data, whereas QEC preserves a quantum state already inside the hardware by repeatedly diagnosing and compensating for local faults. The quantum information is never copied or retransmitted; instead, it is actively stabilized throughout the computation.

Among the many QEC architectures proposed, \textbf{surface codes} are one of the leading candidates for large-scale systems because they require only local interactions on a two-dimensional lattice and have relatively high fault-tolerant thresholds~\cite{kitaevFaulttolerantQuantumComputation2003a,bravyiQuantumCodesLattice1998}. More recently, there has also been growing interest in quantum low-density parity-check (qLDPC) codes~\cite{breuckmannQuantumLowDensityParityCheck2021}, which offer better asymptotic encoding rates, although their practical implementation remains less mature.

\subsubsection{The Threshold Theorem and Spatial Overhead}

The viability of QEC depends on a central result known as the \textbf{threshold theorem}: fault-tolerant quantum computation is possible only if the physical error rate $p$ is below a code- and architecture-dependent threshold $p_{th}$~\cite{aharonovFaultTolerantQuantum1996,knillResilientQuantumComputation1998,preskillFaulttolerantQuantumComputation1997}. If this condition is met, increasing the size of the code suppresses the logical error rate. For surface-code-style families, this suppression is often summarized by the heuristic scaling~\cite{fowlerSurfaceCodesPractical2012}.
\begin{equation}
P_L \approx C \left( \frac{p}{p_{th}} \right)^{\frac{d+1}{2}},
\end{equation}
where $d$ is the code distance and $C$ is a code-dependent constant.

The key implication is simple. If the underlying hardware is good enough that $p < p_{th}$, then increasing the code distance makes logical failure exponentially less likely. If instead $p > p_{th}$, adding more qubits does not help: the extra redundancy introduces more faults than the code can remove. QEC therefore creates a large spatial overhead, but only pays off once the underlying hardware is sufficiently reliable.

\subsubsection{Error Correction in DQC}

Both quantum error correction and DQC are widely expected to play important roles in the path toward large-scale quantum advantage~\cite{ caleffiDistributedQuantumComputing2024,barralReviewDistributedQuantum2025}. In a distributed setting, however, fault tolerance becomes more demanding: processors must suppress their own local errors while also coping with noise, loss, and latency on the interconnects used to distribute entanglement and operations across the network.

We organize existing approaches into three broad paradigms, distinguished by how widely a single code block is spread across the network.

\begin{itemize}
    \item \textbf{Local codes.} One or more logical qubits are encoded entirely within individual QPUs, and quantum networking is used only to perform operations between them. Communication may be mediated by physical Bell pairs shared between modules, or by encoded logical entanglement created between code patches, for example using lattice surgery~\cite{horsmanSurfaceCodeQuantum2012,jacintoNetworkRequirementsDistributed2025,rametteConnectionErrorCorrectedQubits2024,yoderTourGrossModular2025}. In this model, error correction remains primarily local to each processor, while the network is responsible for enabling distributed logical gates. From a networking perspective, this is the least demanding regime: the network is used primarily to support occasional inter-module logical operations, so communication overhead is comparatively sparse and does not usually scale with every local syndrome-extraction cycle. However, in cases where the number of logical qubits per processor is small, a high proportion of logical operations will require use of the network.

    \item \textbf{Partially distributed codes.} Here, a single logical qubit or code block is spread across multiple processors, but each processor still hosts a non-trivial local portion of the code. This may arise because individual QPUs are too small to host a complete code patch, or because the code itself naturally spans module boundaries. Examples include distributed surface-code patches with non-local stabilizer checks, as well as modular realizations of LDPC or hyperbolic-code ideas~\cite{xuDistributedQuantumError2022,strikisQuantumLDPCCodes2023,sutcliffeDistributedQECFloquet2025}. In these schemes, decoding must account for both local faults and errors introduced by the inter-module links. The network load is correspondingly higher, because inter-module communication is no longer reserved for logical gates alone: it is also required during repeated syndrome-extraction cycles, so link fidelity, entanglement-generation rate, synchronization, and classical decoding latency directly affect fault-tolerant performance.

    \item \textbf{Fully distributed codes.} In the most extreme case, each module hosts only one data qubit---or a very small number of data qubits---from the overall code block~\cite{nickersonFreelyScalableQuantum2014}. Quantum communication is then required for essentially every non-trivial stabilizer-extraction cycle and, in the limit, for almost every entangling operation needed to maintain the code~\cite{deBoneThresholdsDistributedSurface2024,singhModularArchitecturesEntanglement2024}. This is the most communication-intensive regime: the network effectively becomes part of the error-correcting code, and its capacity, latency, synchronization, and control overhead scale with the ongoing maintenance of the logical state rather than only with higher-level algorithmic communication.
\end{itemize}

This final regime is particularly relevant from a networking perspective, because it makes the communication substrate part of the error-correction machinery rather than merely a support layer for distributed logical gates. It therefore provides a natural bridge to the broader question of what kinds of quantum-network architecture are needed to support large-scale DQC.

A related class of architectures lies slightly outside this taxonomy, namely measurement-based and fusion-based photonic schemes in which fault tolerance is built around the generation, distribution, storage, and measurement of photonic resource states rather than around matter-qubit code patches connected by explicit inter-module entangling links~\cite{raussendorfOneWayQuantumComputer2001,bartolucciFusionbasedQuantumComputation2023}. These approaches do not map cleanly onto a picture in which a fixed code block is partitioned across QPUs, but from a networking perspective they place demands most similar to the partially distributed and fully distributed regimes. In particular, they can reduce or bypass the need for direct matter--photonic interfaces between remote processors, replacing this with a requirement for high-rate photonic resource-state generation, low-loss optical delay or transport, precise synchronization, and repeated fusion or measurement operations across the distributed architecture.

\subsection{Critical Assumptions and Misunderstandings}

Because classical network engineers are accustomed to deterministic, protocol-driven error correction, the transition to modeling quantum reliability often results in severe architectural misunderstandings. When the networking community designs routing, scheduling, and simulation algorithms for the Quantum Internet, they frequently make idealized assumptions about error management that fail in physical deployment:

\begin{itemize}
    \item \textbf{Unconditional Error Correction Gain:} Network simulators frequently model QEC as a black-box function that universally improves link fidelity, akin to how classical Forward Error Correction (FEC) always provides some degree of coding gain. In reality, as dictated by the Quantum Threshold Theorem, if a routing algorithm selects a path where the raw physical error rate exceeds a specific threshold ($p_{th}$), deploying QEC will actively accelerate the destruction of the quantum state. Consequently, network control planes must be designed to mathematically verify that a link operates safely below the fault-tolerant threshold before allocating the massive physical qubit overhead required for QEC.
    
    \item \textbf{Negligible Classical Control Overhead:} Several models treat QEM and QEC as purely quantum operations, assuming the classical network running parallel to the quantum data plane possesses infinite bandwidth and zero latency. However, QEC generates massive, continuous streams of classical telemetry (syndrome bit strings), and QEM requires constant two-way instruction loops. This places an immense bandwidth and ultra-low-latency burden on the classical control plane. If the classical network experiences congestion, the syndrome decoding delays will exceed the logical qubit's coherence time, causing the fault-tolerant quantum link to collapse. The SDQN controller must therefore actively co-schedule classical bandwidth alongside quantum entanglement generation.
    
    \item \textbf{Decoupled Routing and Error Management:} Theoretical networking papers frequently separate routing algorithms from error management, assuming the controller can first compute the shortest path and then arbitrarily apply purification to fix the fidelity. In practice, purification fundamentally alters the effective capacity of a link. Because distilling higher fidelity consumes raw pairs, a path's Entanglement Distribution Rate (EPS) is highly dynamic and inversely coupled to the purification depth. While recent works, such as Li et al.~\cite{Li2022FidelityRouting}, have successfully demonstrated that coupling path selection with purification decisions is mandatory to minimize raw entanglement consumption, they often restrict the solution to strict cost-minimization under hard fidelity thresholds. Network architects must generalize this tightly coupled approach, treating joint routing and purification scheduling as a dynamic, application-aware optimization problem governed by the Q-NUM framework.
    
    \item \textbf{Separation of Computation and Error Correction:} In classical systems, protected data can often be decoded, processed in unencoded form, and then re-encoded when needed. This encourages the intuition that error correction is simply a wrapper around computation rather than part of it. In fault-tolerant quantum computing, this separation usually fails. Decoding a logical qubit into its underlying physical qubits removes the protection of the code and exposes the state directly to environmental noise. As a result, computation must typically be carried out on encoded logical qubits using fault-tolerant logical operations, many of which are substantially more expensive than their unencoded counterparts. Reliability, communication, and computation are therefore tightly coupled: in distributed architectures, the network may be required not only to help maintain encoded states but also to actively enable fault-tolerant logical operations between them.
\end{itemize}


\section{The Quantum Protocol Stack and Network Architecture}
\label{sec:protocol-stack}

The preceding sections have established the physical realities of quantum networking: qubits decohere, photons are lost, entanglement swapping succeeds only probabilistically, and the performance of any end-to-end quantum link is characterised by a multidimensional set of fragile metrics. The natural question for a network engineer is: how is this physical substrate actually managed and controlled at the network level? 

In classical networks, this question is answered by decades of accumulated protocol design, structured into well-known layers from the physical up to the application. In quantum networks, the same question must be asked, but the answers are fundamentally different in several critical respects. To provide a rigorous architectural framing, this section develops a network-centric view of how a quantum network is organised. We adopt a three-dimensional protocol reference model---an approach that distinguishes between vertical layers (services) and horizontal planes (control functions)---mapping the classical layered model onto its quantum counterpart to highlight where standard networking paradigms apply, and more importantly, where they physically break down.

\subsection{The Classical Protocol Stack: A Reference Model}

To appreciate the structure of a quantum network's hierarchy, it is worth briefly recalling how classical networks are organised. The classical Internet is governed by a layered protocol stack (e.g., the OSI/TCP-IP models \cite{ISO7498-1, KuroseRossBook}), where each layer provides well-defined services to the layer above it and relies on well-defined services from the layer below it:

\begin{itemize}
    \item \textbf{The Physical Layer} is responsible for transmitting raw bits over a physical medium (optical fibre, radio waves, or copper wire). Crucially, the physical layer has no concept of addressing or routing; its sole responsibility is faithful bit transmission via signal regeneration.
    \item \textbf{The Link Layer} operates between adjacent nodes, framing data into packets, detecting transmission errors, and providing reliable communication service over a single physical hop (e.g., via ARQ~\cite{tanenbaum2011computer} and FEC~\cite{lin1983error} retransmission).
    \item \textbf{The Network Layer} is responsible for end-to-end delivery across multiple hops. It performs routing (the selection of a path from source to destination), manages addressing, and handles resource allocation among competing flows.
\end{itemize}

Underlying all layers is the \textbf{Control Plane}~\cite{KuroseRossBook} ---the set of protocols and processes that establish and maintain the forwarding tables and topology information that the data plane uses to forward traffic. In classical networks, the control plane operates on classical, copyable, inspectable information, and its decisions can be updated at any time without affecting the data already in transit.

\subsection{The Quantum Protocol Stack: A Synthesized Reference Model}

Currently, there is no formally standardized protocol stack for the Quantum Internet equivalent to the IETF TCP/IP model~\cite{li2024,zhang202x}. Over the past decade, the research community has proposed various conceptual mappings that align quantum physical operations with the familiar OSI model to tame system complexity.

However, existing surveys on quantum networking frequently attempt to force all operations into a single, flat vertical stack mimicking the OSI model. This purely vertical approach often fails to capture the physical reality of quantum systems. For instance, in their comprehensive layered survey of quantum Internet protocols, Li et al.~\cite{li2024} explicitly concede that ``a complete and universal layered model may not exist for quantum networks.'' They note that quantum communication inevitably involves cross-layer interactions, which conflict with the conventional protocol stack concept in which only adjacent layers interact.

This conflict is evident in how the community categorizes core operations. Entanglement purification, for example, is debated as either a link-layer protocol (to ensure robust single-hop fidelity) or a network-layer protocol (to manage end-to-end routing fidelity). Similarly, Abane et al. \cite{abane2025survey} emphasize that a ``one-size-fits-all'' layered architecture has yet to emerge for intermediate-scale quantum networks. Their taxonomy separates a ``routing phase'' (path computation and route installation) from a ``forwarding phase'' (entanglement swapping, purification, and reliability mechanisms), but they also highlight that swapping strategies and fidelity support unavoidably couple node-level physical capabilities with end-to-end path selection.

Furthermore, the concept of ``scheduling'' appears at multiple locations in the literature: Li et al.~\cite{li2024} discuss purification scheduling and distributed-queue request scheduling at the link and network layers, while other works treat the timing of swaps and link usage as part of the routing or transport logic. As a result, terms such as routing, scheduling, and resource allocation are sometimes used for distinct mechanisms and sometimes conflated, depending on the chosen vertical layering.

By attempting to isolate functions into strict vertical layers, these surveys risk masking the critical cross-layer trade-offs between rate, fidelity, and time that must be managed jointly. To bridge this conceptual gap for networking researchers, we present a synthesized reference model. Rather than proposing a competing standard, this stack abstracts complex physical-layer quantum operations into familiar networking paradigms while acknowledging their inherent ``layer-leaky'' physics.

A useful starting point is to recognise that a quantum network is inherently a hybrid system: it possesses a \textit{quantum data plane}, carrying fragile entangled states, and a \textit{classical control plane}, carrying heralding outcomes, measurement results, and synchronization instructions. By explicitly separating the vertical service stack from the cross-cutting control functions that manage it, this tutorial adopts an architectural abstraction that is better aligned with the realities of quantum hardware.

The core quantum layers in our synthesized stack are detailed below (summarised alongside their classical counterparts in Table \ref{tab:stack_comparison}):

\begin{table*}[htbp]
\centering
\caption{Layer-by-Layer Comparison of Classical and Synthesized Quantum Protocol Stacks}
\label{tab:stack_comparison}
\renewcommand{\arraystretch}{1.4}
\begin{tabularx}{\textwidth}{lXXX}
\hline
\textbf{Layer} & \textbf{Classical Network Function} & \textbf{Quantum Network Function} & \textbf{Key Differences \& Constraints} \\ \hline
\textbf{Application} & Generates data payloads and consumes network services (e.g., HTTP, FTP). & Consumes entangled states for quantum tasks (e.g., QKD, DQC, Sensing). & Classical requests data delivery; Quantum requests specific entangled states with strict fidelity/timing requirements. \\ \hline
\textbf{Transport} & Ensures reliable end-to-end data delivery, flow control, and congestion management (e.g., TCP). & Manages end-to-end state fidelity, entanglement distillation, and logical qubit scheduling. & Quantum errors cannot be solved by simply re-transmitting data; requires complex, resource-heavy purification. \\ \hline
\textbf{Network} & Calculates paths and forwards discrete packets across multi-hop topologies (e.g., IP, OSPF). & Executes multi-hop entanglement routing, swapping, and multipartite state generation. & Qubits cannot carry readable routing headers without measurement; all path setup must be coordinated via classical signaling. \\ \hline
\textbf{Link} & Manages point-to-point framing, MAC, and localized error detection across a single physical medium. & Establishes robust, heralded bipartite entanglement (Bell states) across a single physical link. & Quantum link creation is probabilistic. Requires real-time classical heralding signals to confirm success before proceeding. \\ \hline
\textbf{Physical} & Transmits classical bits (0s and 1s) via voltage, RF, or optical pulses. & Transmits flying qubits (e.g., photons) and interfaces with stationary quantum memories. & Governed by no-cloning and decoherence. Lost qubits are permanently destroyed; operations bounded by coherence times. \\ \hline
\end{tabularx}
\end{table*}

\begin{itemize}
    \item \textbf{The Quantum Physical Layer} corresponds directly to the classical physical layer; it governs how qubits are prepared, transmitted over a quantum channel (typically optical fibre), and received at the far end of a single hop. However, in contrast to classical physical layers, the quantum physical layer has no way to amplify or regenerate a signal. A lost photon is simply lost. In fiber-based photonic links, amplitude damping (photon loss) is typically dominant, and the primary figure of merit is the probability of successfully delivering a photon across a single link~\cite{sangouard2011quantum,briegel1998quantum}.
    \item \textbf{The Quantum Link Layer} is responsible for converting an unreliable physical connection into a reliable entangled link between two adjacent nodes. Its primary function is heralded entanglement generation---running a physical-layer protocol repeatedly until a Bell-state measurement (BSM) confirms that two adjacent quantum memories hold an entangled pair, and then signalling this success upward. It also manages entanglement purification, spending rate to recover fidelity when link quality degrades, and manages the local quantum memory by deciding which physical memory qubits to use and for how long~\cite{briegel1998quantum,sangouard2011quantum,meterDesigningQuantumRepeater2013}.
    \item \textbf{The Quantum Network Layer} is where routing and long-distance entanglement distribution reside. Its fundamental operation is entanglement swapping: a repeater node holding entanglement with Node A and Node B performs a BSM on its two local qubits, stitching the two short links into a single end-to-end link. This is the quantum analogue of classical routing and packet forwarding, but with a crucial difference: it is a destructive, irreversible operation that consumes the short-range pairs. The network layer therefore has to manage not just paths, but the entire lifecycle of entanglement~\cite{meterDesigningQuantumRepeater2013,Pant2019RoutingEntanglement}.
\end{itemize}

\subsection{Fundamental Architectural Differences}
While both classical and quantum stacks are hierarchical and deal with lossy channels, the constraints of quantum mechanics reshape the underlying operational paradigm in five profound ways:

\begin{itemize}

    \item \textbf{The resource is consumed, not forwarded:} In classical networks, packet forwarding is a non-destructive process; a router reads a header and passes the data along. In quantum networks, multi-hop routing requires entanglement swapping, a process that inherently consumes the quantum resource at every hop to extend the link~\cite{briegel1998quantum,dur1999RepeatersPurification}.
    
    \item \textbf{No amplification or retransmission:} Classical link layers ensure reliability by simply duplicating and retransmitting lost packets. In quantum networks, the no-cloning theorem strictly prohibits the duplication of unknown quantum states. If a qubit is lost or decoheres, the network layer cannot request a retransmission of that specific state; it must command the physical layer to regenerate the foundational entanglement from scratch.
    
    \item \textbf{Probabilism is intrinsic, not incidental:} Classical networks treat packet loss as a failure condition to be engineered away. Conversely, quantum networks are built on physical processes—such as photon detection and linear-optical BSMs—that are fundamentally probabilistic. The protocol stack must treat probabilistic success as a normal, continuous operating condition.
    
    \item \textbf{Quality is a network-layer concern:} In classical networks, signal quality (e.g., SNR) is abstracted away by the physical and link layers, allowing higher layers to deal with bits that are essentially perfect. In quantum networks, the fidelity of entangled states continuously degrades with every hop due to swapping imperfections and memory decoherence~\cite{briegel1998quantum,sangouard2011quantum}. Consequently, the network layer must actively manage fidelity, making complex routing decisions that trade off generation rate against state quality.
    
    \item \textbf{The control plane is time-critical at the quantum level:} In classical networks, control-plane latency (e.g., a slow routing table update) is inconvenient but rarely catastrophic to the data payload. In quantum networks, quantum memories have strict coherence times. A classical control-plane response—such as a heralding signal—that arrives after the quantum memory has decohered results in a total protocol failure~\cite{meterDesigningQuantumRepeater2013,sangouard2011quantum}. 
\end{itemize}
\subsection{The Separation of Planes: A Physical Necessity}
The differences outlined above reveal why strict vertical encapsulation fails in quantum systems. In a classical router, the data plane and the control plane intersect natively: the router physically inspects the headers of the packets passing through it to execute routing logic. In a distributed quantum network, this is physically impossible. The quantum data plane is constrained by the observer effect; it cannot inspect headers, queue packets based on readable IDs, or execute embedded routing logic without destroying the fragile superposition of the qubits. Furthermore, operations such as entanglement purification and swapping are inherently cross-layer: they manipulate physical-layer photons but are dictated by network-layer routing thresholds. To maintain mathematical tractability under rigid vertical models, current routing literature often adopts idealized hardware assumptions, for example by omitting purification
~\cite{caleffi2017}, or ignores the massive classical signaling delays required to herald these operations across multiple nodes~\cite{abane2025survey}.

Therefore, the separation of planes is not merely an optional, software-defined convenience as it is in classical networks---it is a physical necessity. The mechanisms that decide how quantum services are used must sit in a logically separate classical control plane. This classical plane operates over conventional optical or electronic channels to orchestrate entanglement generation, heralding, swapping, and path selection at timescales commensurate with the coherence of the underlying memories. By explicitly separating the vertical service stack from the cross-cutting control functions that manage it, we adopt an architectural abstraction that aligns precisely with the realities of quantum hardware.

Having established the layered architecture and its fundamental departures from classical networking, we must now address the critical question: how is this unreadable, probabilistic, and highly perishable data plane orchestrated and controlled in practice? The answer lies in the adaptation of centralized network intelligence, known as Software-Defined Quantum Networking (SDQN).


\section{Software-Defined Quantum Networking (SDQN)}
\label{sec:sdqn}

As established in Section~\ref{sec:protocol-stack}, the fundamental constraints of quantum mechanics---specifically the observer effect and the no-cloning theorem---make it physically impossible for quantum data to carry its own readable routing headers~\cite{nielsen2000quantum}. Therefore, the separation of the data plane and the control plane is a physical necessity. The architectural paradigm that addresses this management challenge is often described as Software-Defined Quantum Networking (SDQN), extending software-defined control to quantum data planes~\cite{ornl2018sdqn_switch, aguado2018sdqkd,tessinari2023sdqn_qkd}. 

Prior work has already explored software-defined control for quantum and QKD networks,
including software-defined quantum switches~\cite{ornl2018sdqn_switch},
software-defined QKD and quantum-aware SDN architectures that abstract QKD devices into
operator networks~\cite{aguado2020enablingQKDsdn,zhao2019sdqkdn}, SDN-controlled QKD relay, survivability, and key-management mechanisms~\cite{shirko2023sdqkrf, mehic2022sdqkdDoS, tessinari2023sdqn_qkd}, QoS- and slice-oriented QKD networks~\cite{wang2023sliceQKDN}, software-defined QKD network architectures that reduce key consumption and enable more flexible routing~\cite{yu2017sqn}, and studies of routing-protocol choice and key usage in SDN-based QKD networks~\cite{monita2022routingSDQKD}. 

In this tutorial, we use the term SDQN to denote a more general, stack-wide abstraction that explicitly separates a classical control plane from a dual (quantum + classical) data plane. Throughout this tutorial, SDQN should therefore be understood as an architectural template informed by these SDN-enabled QKD examples, rather than as a fully specified protocol or standardized implementation. In contrast to existing SDN-enabled QKD architectures, which primarily focus on key distribution and trusted-relay management~\cite{aguado2020enablingQKDsdn, zhao2019sdqkdn,yu2017sqn, wang2023sliceQKDN, shirko2023sdqkrf,monita2022routingSDQKD}, our SDQN abstraction is service-agnostic and explicitly targets entanglement generation, routing, and rate–fidelity trade-offs for a broader class of quantum applications.

SDQN is not a complete departure from classical networking paradigms but rather a sophisticated adaptation of Software-Defined Networking (SDN)~\cite{mckeown2008openflow,nunes2014sdn}, a framework that has already revolutionized the management of classical data centers and telecommunication networks. By applying the core principles of SDN to the quantum domain, SDQN aims to provide the programmability, flexibility, and centralized intelligence required to orchestrate delicate quantum resources effectively.

\subsection{From Classical SDN to the Dual Data Plane}

Classical SDN logically centralizes a network's intelligence by decoupling the control plane (which makes routing decisions) from the data plane (the switches that forward the packets). This separation allows network administrators to program and manage the entire network from a single, centralized point via open APIs (such as the OpenFlow protocol\cite{mckeown2008openflow}). 

SDQN adopts this exact principle of plane separation but adapts it to a profoundly more complex environment. The most critical architectural modification in SDQN is the mandatory implementation of a dual data plane architecture:

\begin{itemize}
    \item \textbf{The Classical Data Plane:} Operates exactly like a conventional network, carrying robust classical bits used for control signaling, heralding, transmission of measurement results, and the exchange of routing metadata.
    \item \textbf{The Quantum Data Plane:} A physically separate channel (e.g., dedicated dark fiber or distinct WDM optical channels) strictly reserved for the transmission of fragile quantum states, or qubits. 
\end{itemize}

This dual-plane architecture protects fragile quantum information from the decoherence
and state collapse that would be induced by the measurement inherent in classical control
channels~\cite{nielsen2000quantum}. In existing software-defined quantum and QKD architectures, controllers already use classical SDN control traffic carrying quantum-related metadata (e.g., QKD service identifiers and keying requirements) to steer specific quantum resources~\cite{tessinari2023sdqn_qkd,ornl2018sdqn_switch}.

In this tutorial, we extend this idea under the term SDQN to describe a stack-wide
abstraction with a dual data plane (quantum + classical), a hardware-agnostic QNOS layer,
and a control plane that treats entanglement rate and fidelity as first-class routing
objectives. In this abstraction, quantum metadata includes not only the service type
(e.g., QKD, distributed sensing, or teleportation) but also target figures of merit such
as minimum end-to-end entanglement fidelity, which the SDQN controller uses when
configuring quantum switches and repeaters. A logically centralized SDQN controller then
intercepts these classical packets and, inspired by OpenFlow-style southbound APIs in
SDN~\cite{mckeown2008openflow}, sends configuration commands to quantum switches,
instructing them how to establish the requested quantum circuits without ever directly
observing the qubits themselves. Table~\ref{tab:sdqn_comparison} summarizes these shifts.

Similar ideas already appear in software-defined QKD networks, where a QKD layer and a
classical data layer are jointly controlled by an SDN controller and quantum keys are
pooled and allocated as a consumable resource via a quantum key pool
abstraction~\cite{zhao2019sdqkdn}. SDQN lifts this dual-plane and pooled-resource view
from QKD-specific key material to a broader class of quantum resources, with entanglement
as the primary currency.

\begin{table}[htbp]
\centering
\caption{SDQN vs. Classical SDN Architecture}
\label{tab:sdqn_comparison}
\renewcommand{\arraystretch}{1.4}
\begin{tabularx}{\columnwidth}{lXX}
\hline
\textbf{Aspect} & \textbf{Classical SDN} & \textbf{Quantum SDQN} \\ \hline
\textbf{Data Plane} & Single (classical bits) & Dual (quantum + classical) \\ \hline
\textbf{Resource Type} & Bandwidth (renewable capacity) & Entanglement (perishable, consumed) \\ \hline
\textbf{Packet Handling} & Copy, buffer, and retransmit & Generate-once, no-cloning \\ \hline
\textbf{Monitoring} & Direct (read packet headers) & Blind (avoid observer effect) \\ \hline
\textbf{Control Timing} & Best-effort latency & Hard real-time ($T_2$ bound) \\ \hline
\textbf{Failure Mode} & Packet drop (triggers retransmission) & State destruction (requires regeneration) \\ \hline
\end{tabularx}
\end{table}

\subsection{Hardware-Agnostic Abstraction: The Role of QNOS}
A fundamental orchestration challenge for the SDQN controller is managing the severe hardware heterogeneity of the physical layer. A future global Quantum Internet will not comprise a single qubit technology; instead, it will consist of a disparate mixture of different physical modalities. This includes photonic qubits for long-distance transmission over optical fiber, alongside various matter-based qubits (e.g., trapped ions, neutral atoms, or superconducting circuits) serving distinct roles in local memory and processing nodes. In practice, even QKD-only networks already face protocol heterogeneity, where software-defined heterogeneous QKD chaining is used to orchestrate multi-protocol QKD chains over optical networks~\cite{cao2022hqkdc}. 

To manage this complexity, SDQN relies on a hardware-agnostic abstraction layer, often conceptualized as a \textit{Quantum Network Operating System (QNOS)}, analogous in spirit to emerging operating systems for quantum networks such as QNodeOS~\cite{delleDonne2025qnodeos}. This conceptual shift draws direct inspiration from the evolution of classical Software-Defined Networking (SDN), where centralized Network Operating Systems (NOS)—such as ONOS or OpenDaylight—abstract away vendor-specific hardware details~\cite{berde2014onos}. In our abstraction, QNOS generalizes the heterogeneity challenge seen in QKD-only networks~\cite{cao2022hqkdc} to a broader space of quantum technologies and services. In the classical domain, a NOS presents a standardized, unified southbound API (e.g., OpenFlow) to control the packet forwarding behavior of heterogeneous classical switches, regardless of the underlying application-specific integrated circuit (ASIC) architecture.

Building upon this inspiration, QNOS serves a similar foundational purpose by decoupling the high-level quantum application from the underlying physics of the endpoint hardware. Through a standardized API, an application can request a high-level resource---such as ``establish a Bell pair between Node A and Node B with $F > 0.90$''---without needing to know the physical medium. The QNOS translates these technology-agnostic intents into the precise sequence of technology-specific Southbound controls---such as the low-level microwave or optical pulse sequences---required to execute the operation on the unique endpoint hardware (e.g., manipulating a trapped ion vs. a superconducting
qubit, where software-defined optimal-control pulses are already used to implement
high-fidelity gates~\cite{wu2020softwareDefinedQudit}). This abstraction becomes unequivocally mandatory when considering the macroscopic architectural differences between repeater types. For instance, if a network path traverses both a memory-based ``store-and-forward'' node and an MBQC-driven ``all-photonic'' node, the SDQN controller cannot manage them with the same physical-layer logic. The QNOS must abstract this massive disparity, translating a single high-level routing intent into either a memory-swapping schedule or an intensive cluster-state generation and feed-forward schedule, depending entirely on the local node's physical architecture. Efforts such as the \textbf{IEEE P1913} Software-Defined Quantum Communication standard already move in this direction by defining YANG-based data models and control protocols for configuring networked quantum devices over classical TCP/IP networks~\cite{ieee1913standard}.

However, while the abstract goal of hardware decoupling remains similar to classical networks, the operational mechanics of QNOS are fundamentally different due to the rules of quantum physics. These similarities and distinctions are summarized below:
\begin{itemize}
    \item \textbf{Similarity: Abstraction and Resource Management.} Both classical and quantum NOS aim to present a unified API to the control plane, managing limited resources (bandwidth and buffers vs. coherence time and memory) without exposing physical hardware specifics to the applications.
    \item \textbf{Difference: Buffering and Determinism.} A classical NOS manages robust, copyable bits and relatively deterministic links. If congestion occurs, a classical NOS can buffer data packets indefinitely. Conversely, QNOS operates under the observer effect and the no-cloning theorem; quantum states cannot be readily buffered or copied. Therefore, while a classical NOS can wait for resources to become available, QNOS must orchestrate resources with sub-microsecond precision within the strict coherence time limits of the hardware.
    \item \textbf{Difference: Probabilism vs. Bandwidth.} A classical NOS allocates resources based on a relatively stable bandwidth metric. QNOS must account for the inherently probabilistic nature of entanglement generation and swapping attempts. A successful quantum operation is not guaranteed; rather, QNOS must continuously manage the \textbf{Rate-Fidelity trade-off}, treating coherence time as a decaying, non-renewable resource rather than a static capacity.
\end{itemize}

Ultimately, as discussed in Section~\ref{subsec:Tomography}, QNOS must continuously infer and track these quantum parameters—such as environmental noise, decoherence rates, and link degradation probabilities—using quantum-compatible monitoring techniques (e.g., network tomography and calibration routines), and expose them to the SDQN controller as time-varying metrics for routing and scheduling.

\subsection{The Conceptual Challenge: Managing Probabilistic Resources}

While SDQN inherits the centralized architecture of SDN, its operational reality is entirely different. In classical SDN, translating a high-level intent into switch commands is deterministic. In SDQN, the controller acts as a dynamic, stateful orchestrator of distributed quantum experiments. It must initiate probabilistic quantum operations, wait for classical success feedback, and trigger corrective actions via continuous Local Operations and Classical Communication (LOCC) loops.

At a high level of abstraction, the challenges of managing quantum networks appear similar to those of dealing with unreliable channels in classical wireless networks. However, these analogies can be misleading, as quantum mechanics introduces constraints that are qualitatively more complex:

\begin{itemize}
    \item \textbf{Entanglement Fidelity vs. Wireless Signal Attenuation:} While Signal-to-Noise Ratio (SNR) degradation in wireless links is analogous to entanglement fidelity loss, the crucial difference is the impossibility of amplification in quantum systems due to the no-cloning theorem. A weak classical signal can be amplified and corrected; conversely, entanglement degradation is a loss of quantum correlation that cannot be restored, but must instead be overcome by complex protocols like entanglement purification that consume existing pairs.
    
    \item \textbf{Probabilistic Generation vs. Random Wireless Capacity:} The fluctuation in a wireless link's capacity due to fading is often compared to the probabilistic success rate of establishing an entangled pair. However, a classical wireless channel is a persistent resource that fluctuates, allowing protocols to adapt. In contrast, entanglement is a generated and consumed resource; a failure means the link simply fails to come into existence, forcing the entire generation process to restart, resulting in a distinct ``success by repetition'' operational model.
    
    \item \textbf{Qubit Coherence Time vs. Queuing/QoS Delays:} Though a classical real-time packet has an application-imposed delay budget similar to a qubit's limited lifespan in memory, coherence time is a hard physical lifetime of the quantum information itself. A qubit physically decays over milliseconds, making this an intrinsic constraint that dictates maximum node distance and synchronization precision.
\end{itemize}

Ultimately, SDQN is forced to operate under three severe constraints. First, \textit{no retransmission}: because qubits cannot be copied, classical robustness by buffering is impossible. Second, the \textit{observer effect}: the controller must operate as a ``black box'' manager, inferring network health solely from classical metadata because inspecting the data plane destroys the payload. Third, and most critically, the \textit{latency-coherence gap}. Network operations require classical heralding signals to be sent between nodes; if the classical control channel is too slow, the qubit decays before the instruction arrives. A protocol is guaranteed to fail if $t_{\text{classical\_feedback}} > T_{\text{coherence}}$.

\subsection{Critical Assumptions and Network Enginnering Challenges}

While the SDQN architecture provides an elegant conceptual framework for managing quantum resources, its practical viability currently rests on a set of highly optimistic assumptions specific to the classical control paradigm. These assumptions are not merely theoretical caveats; they are critical engineering roadblocks that currently render large-scale SDQN impractical. For the networking community, these represent the primary grand challenges that must be solved, as the entire field hinges on the engineering challenge of making a ``classical brain'' effectively and universally manage a ``quantum body''. The three foundational assumptions—and the corresponding challenges they present to network researchers—are as follows:

\begin{itemize}
    \item \textbf{Perfect Classical Abstraction.} SDQN assumes that the highly complex and fragile quantum data plane can be accurately and universally abstracted using simple classical metadata and topology reports. Because the observer effect prevents the classical controller from directly inspecting the qubits, the system assumes that historical, probabilistic reports are sufficient to maintain an accurate global view for real-time routing. Existing SDN-enabled QKD survivability, routing, and key-management schemes, such as SDQKRF, SDN-based routing studies, and KMS DoS mitigation mechanisms in software-defined QKD networks~\cite{shirko2023sdqkrf,monita2022routingSDQKD,mehic2022sdqkdDoS}, illustrate this trend: they extend the classical control plane with additional functions to handle failures and attacks, but still reason in terms of key buffers, session identifiers, and QoS rather than directly observable quantum-state information. The networking community must move beyond simple statistical tracking and develop advanced, quantum-native telemetry and prediction algorithms. The network must be able to accurately infer and predict the degradation of quantum states on the fly, without triggering state collapse.

    \item \textbf{Universal Southbound Interfaces.} 
    SDQN assumes that a standardized, simple set of commands (such as OpenFlow extensions) can successfully translate high-level routing decisions into the precise, low-level, time-sensitive quantum operations required by disparate hardware. This implies that a single abstraction layer can control wildly different physical systems (e.g., cryogenic superconducting circuits vs. room-temperature photonic circuits) without losing vital operational details. Early software-defined QKD architectures such as SQN~\cite{yu2017sqn} already leverage an SDN controller with a global view of trusted-relay QKD links to optimize key consumption and routing, but still operate on classical control abstractions rather than explicit quantum-state dynamics. Existing standardization efforts, such as IEEE P1913, already propose YANG-based models and NETCONF-style control protocols for configuring quantum devices and QKD modules over classical networks~\cite{ieee1913standard}. These initiatives illustrate how a common, implementation-agnostic device model can enable interoperability, but they also highlight how much work remains to translate high-level SDQN routing decisions into tightly timed, platform-specific quantum operations across heterogeneous hardware. Network engineers must develop robust, hardware-agnostic southbound protocols (analogous to a true QNOS) that can handle highly specialized, platform-specific pulse translations while maintaining strict cross-platform timing synchronization.
    
    \item \textbf{The Latency-Coherence Guarantee.} The entire architecture assumes that the classical control plane orchestrating the Local Operations and Classical Communication (LOCC) is fast and reliable enough that the command-to-operation latency is minor compared to the qubit's short coherence time. When this assumption is violated, the result is catastrophic failure: the classical command arrives, but the target qubit has already decohered. 
    Overcoming this ``latency-coherence gap'' is arguably the most severe networking challenge. It requires the classical networking community to shift away from purely centralized SDQN models and invent highly autonomous distributed control planes, predictive routing heuristics, and ultra-low-latency classical signaling channels that can outpace quantum decoherence.
\end{itemize}

\subsection{Control Plane Architectures}

To execute these time-sensitive tasks without violating the latency-coherence gap, a critical design dimension for any quantum network is the architecture of its classical control plane: who makes routing, scheduling, and allocation decisions, and how is the relevant network state information collected and distributed? Related work has begun to explore alternative control frameworks, from hierarchical multi-domain SD-QKD control to non-layered, resource-centric task-based control schemes for quantum networks~\cite{vicario2025hierarchicalsdqkd, horoschenkoff2025sdqkdncm}. In this tutorial, we group SDQN control planes into three broad categories~\cite{ornl2018sdqn_switch,aguado2020enablingQKDsdn,zhao2019sdqkdn,wang2023sliceQKDN}:

\begin{itemize}
    \item \textbf{Centralised Control:} A single controller---or a small cluster of controllers---holds a global view of the network topology and resource state, computes globally optimal routing decisions, and pushes instructions to individual nodes. The severe challenge in quantum networks is that this ``global view'' must include rapidly changing, probabilistic information (e.g., link existences, fidelities, and qubit wait times) that evolve on timescales set by coherence times rather than human-manageable update cycles.
    
    \item \textbf{Distributed Control:} Routing and allocation decisions are made locally at each node using a partial view of the network state, typically the states of its immediate entangled neighbours. This approach scales better and produces individually optimal decisions. It is particularly relevant in the quantum setting because it can react to probabilistic link outcomes immediately at the local node, completely avoiding the round-trip latency to a central controller.
    
    \item \textbf{Hybrid (Hierarchical) Control:} Certain decisions---such as long-term path computation, network utility optimization, and resource provisioning---are made centrally, while time-critical, event-driven decisions---such as triggering a swap when a particular entangled pair becomes available---are made locally. This separation mirrors the classical distinction between a slow control plane and a fast data plane, and is arguably the most practical model for near-term quantum networks.
\end{itemize}

Regardless of the specific control architecture chosen, the SDQN framework must execute three fundamental operational functions: Routing, Scheduling, and Resource Allocation. The following section details how these three control levers operate in practice.

\section{CONTROL PLANE FUNCTIONS: ROUTING, SCHEDULING, AND RESOURCE ALLOCATION}
\label{sec:control-functions}

As established in Section~\ref{sec:sdqn}, the SDQN controller acts as a centralized orchestrator, managing a complex, probabilistic, and fragile quantum data plane.  To effectively translate high-level application requests into physical realities, the controller relies on the aforementioned triad of fundamental control functions. Unlike classical networks, where routing, scheduling, and resource allocation are often loosely coupled, in a quantum network they are inextricably linked. A routing decision directly impacts the scheduling timeline, which in turn consumes the coherence budget allocated to a specific resource. For networking and communications researchers, this SDQN control triad is therefore the primary set of levers through which end-to-end quantum performance is shaped. Table~\ref{tab:control_triad} summarizes these three primary control levers.

\begin{table}[htbp]
\centering
\caption{The SDQN Control Triad vs. Quantum Metrics}
\label{tab:control_triad}
\renewcommand{\arraystretch}{1.4}
\begin{tabularx}{\columnwidth}{lXXX}
\hline
\textbf{Control Function} & \textbf{Primary Goal} & \textbf{Key Trade-Off Managed} & \textbf{Optimizes For} \\ \hline
\textbf{Routing} & End-to-end path selection & Hop distance vs. cumulative fidelity loss & Entanglement generation rate and path viability \\ \hline
\textbf{Scheduling} & Temporal orchestration of swaps and heralding & Throughput speed vs. memory decoherence time & Success probability within the coherence budget \\ \hline
\textbf{Allocation} & Division of physical network assets & Purification depth (quality) vs. yield (quantity) & Fair and efficient multi-flow resource sharing \\ \hline
\end{tabularx}
\end{table}

\subsection{Routing: Rate-Fidelity Trade-Off Selection}

In classical networks, routing algorithms (like Dijkstra's or Bellman-Ford) compute the shortest path based on additive link weights, such as latency or hop count. In quantum networks, however, classical additive routing is fundamentally inadequate because quantum path costs are non-additive and probabilistic. Specifically, as Caleffi mathematically proved~\cite{caleffi2017}, quantum routing metrics are generally not \textit{isotonic} due to decoherence and probabilistic swapping constraints; therefore, directly applying classical shortest-path algorithms can lead to sub-optimal routes or even routing inconsistencies. Despite this mathematical insight, comprehensive surveys show that a large segment of the networking and communications literature continues to default to Dijkstra variants by artificially simplifying link costs into additive metrics to maintain tractability~\cite{abane2025survey}.

Consequently, when an SDQN controller calculates a route, it is not merely connecting wires; it is actively selecting a chain of probabilistic entanglement swapping operations. The end-to-end performance is the product of the individual link performances, heavily penalized by the swapping operations themselves. For example, a 10-hop path where each link has a 90\% probability of successful entanglement generation yields an end-to-end success probability of roughly 35\% ($0.9^{10}$), completely ignoring the parallel degradation in fidelity. This simple calculation illustrates the exponential decay of path success probability with hop count even under optimistic per-link success rates.

Therefore, quantum routing metrics must be treated as projections of a multidimensional space, primarily balancing:
\begin{itemize}
\item \textbf{Hop Count (Distance):} Minimizing intermediate repeater nodes reduces the number of entanglement swaps required. Fewer swaps mean fewer probabilistic failure points and less algorithmic noise injected into the state, improving end-to-end fidelity.
\item \textbf{Expected Entanglement Rate ($R$):} Computed based on the probability of successful link generation and the speed of the hardware. A longer path utilizing highly efficient photonic links might offer a higher generation rate than a shorter path over lossy links.
\item \textbf{End-to-End Fidelity ($F$):} The controller must proactively estimate the expected fidelity degradation along the path, accounting for both swapping imperfections and the environmental noise of the chosen links. A path that maximizes the expected rate ($R$) can easily exploit noisy, fast links that degrade the end-to-end fidelity ($F$) below the application's required threshold.
\end{itemize}

Historically, pioneering quantum routing literature has often prioritized optimizing only one of these dimensions to maintain mathematical tractability. For example, the seminal multi-path routing framework by Pant et al.~\cite{Pant2019RoutingEntanglement} maximizes the \textit{Expected Entanglement Rate} by exploiting diverse network topologies to bypass probabilistic link failures and short coherence times. However, to achieve this rate-optimal scaling, the model treats entanglement generation largely as a binary success/fail event, abstracting away the continuous, non-additive degradation of \textit{Fidelity} across complex multi-hop paths. In a functional Quantum Internet, maximizing the raw generation rate offers limited utility if the resulting states arrive below the application's minimum fidelity threshold. Therefore, the SDQN controller cannot rely solely on throughput-maximizing algorithms; it must explicitly navigate the rate-fidelity trade-off.

Crucially, how the SDQN controller balances these competing metrics is not arbitrary; it must be dictated by the utility function of the application being served. A QKD application with a broad fidelity tolerance is well served by a throughput-maximizing metric, whereas a DQC application whose utility collapses below a high fidelity threshold is better served by a metric that strongly penalizes fidelity degradation, even at the cost of lower throughput.

Beyond selecting the appropriate optimization metric, the routing engine must also dictate the temporal execution strategy: specifically, choosing between \textit{proactive} and \textit{reactive} path computation. In proactive routing, the path and swapping strategy are determined before entanglement generation begins, using the expected physical-layer statistics. In reactive routing, path computation is deferred until a snapshot of currently available entangled links is known, computing paths dynamically on this instantaneous logical topology. Proactive routing offers more predictable quality but may waste attempts on failed links, while reactive routing exploits actual link availability but requires fast path computation within the strict coherence window.

\subsection{Scheduling: Coherence-Budget Management}

Once a route is established, the SDQN controller must orchestrate the precise timing of operations across the selected nodes. In a classical network, scheduling manages queueing delays to respect QoS bounds. In a quantum network, scheduling is instead a strict battle against physical decay: it is the management of a finite coherence budget. Every millisecond a qubit spends waiting in a quantum memory for a distant node to generate entanglement, or waiting for a classical heralding signal, its fidelity decays. If this wait time exceeds the memory's coherence time ($T_2$), the state collapses and the resource is destroyed.

This framing allows us to interpret the principal scheduling modes in terms of how they allocate the coherence budget:
\begin{itemize}
\item \textbf{Synchronous (Memoryless) Scheduling:} All nodes along a path attempt to generate their elementary entangled links simultaneously within a single time slot. Only if all links succeed is the swap performed; if any link fails, the entire attempt is discarded and the process restarts. This requires minimal memory coherence because qubits are never held waiting, but the end-to-end success probability is the product of all per-link probabilities, leading to very low throughput over many hops.
\item \textbf{Asynchronous (Memory-Assisted) Scheduling:} Nodes generate elementary links independently as soon as their memories are free, and a swap is triggered as soon as the two required adjacent links are simultaneously available at a repeater. This allows successful links to be stored in quantum memory while waiting for the remaining links to succeed, dramatically increasing the probability of completing the full path. The cost is that memories must hold entanglement for extended periods, heavily consuming the coherence budget. Centralized, batch-window schedulers that compute paths and edge capacities for a set of requests before each time window exemplify this style of asynchronous, memory-assisted operation~\cite{Li2021EffectiveRouting}.
\item \textbf{Hybrid and Virtual Scheduling:} Intermediate strategies such as pre-establishing multi-hop ``virtual links'' by performing swaps speculatively, or opportunistically forwarding entanglement hop-by-hop as links become available, seek to improve connectivity and reduce end-to-end latency without requiring all links to succeed simultaneously. These approaches are particularly valuable when the entanglement generation rate is high relative to the coherence time, allowing the logical entanglement graph to remain well connected and enabling reactive path computation to find paths quickly.
\end{itemize}

An important scheduling consideration that has no classical counterpart is the interaction between heralding latency and coherence time. Even in an asynchronous scheme, a swap cannot be performed immediately; the node must first receive the classical heralding signal confirming that the BSM succeeded, and the distant end node must receive the classical correction signal to apply the appropriate Pauli gate. This classical round-trip time is a hard lower bound on the time a memory qubit must wait, and must be rigorously budgeted against the coherence time. For network and protocol designers, this means scheduling decisions must jointly account for quantum operation timing and classical signaling delays when allocating the available coherence budget along a path.

\subsection{Resource Allocation: Managing a Perishable Commodity}

The final pillar of the control plane is Resource Allocation. While routing chooses the path and scheduling coordinates the time, resource allocation divides the physical assets---such as photon emission time-slots, classical bandwidth, and specific modes within a multimode quantum memory. Because entanglement is actively generated, immediately consumable, and highly perishable, allocating quantum resources is substantially more complex than allocating classical bandwidth.

The SDQN controller must make real-time decisions regarding:
\begin{itemize}
\item \textbf{Attempt Quotas:} How many times should a node attempt to generate a physical link before giving up and releasing the reserved memory modes for other tasks?

\item \textbf{Purification Depth:} If a generated link is too noisy, the controller must allocate resources for entanglement purification. Purification consumes multiple low-fidelity entangled pairs to distill a single higher-fidelity pair. The controller must decide exactly how many raw pairs to sacrifice to meet the routing threshold, representing a direct trade-off between network yield (quantity) and state quality.

\item \textbf{Memory Reservation:} Allocating specific, isolated memory registers to specific flows to prevent adjacent-channel crosstalk (noise interference between qubits stored in the same physical device).

\item \textbf{Network Sharing Models:} Deciding which classical sharing model to apply on bottleneck links. Options include reserving links exclusively for a single high-priority request (circuit-switching style), applying weighted fair sharing based on application priorities, or using statistical multiplexing that allocates links to whichever request's path can be completed first.
\end{itemize}

All of these decisions involve trade-offs driven by the application. A QKD application benefits from aggressive entanglement generation with minimal purification, maximizing rate. A DQC application benefits from deeper purification and more conservative path selection to guarantee fidelity. Meanwhile, a DQS application requires high-fidelity multipartite entanglement, demanding that resource allocation reserves sufficient memory capacity across many nodes simultaneously.

\subsection{Critical Assumptions and Network Control Challenges}

While the theoretical literature offers numerous algorithms for quantum routing, scheduling, and resource allocation, many of these proposals rely on idealized models of quantum hardware and classical control networks. For the networking and communications community, translating these theoretical algorithms into deployable network protocols requires dismantling several pervasive assumptions.

These assumptions represent specific, high-impact research opportunities for network engineers and computer scientists:

\begin{itemize}
\item \textbf{Deterministic State and Additive Costs (Routing):}
Many quantum routing protocols in the networking and communications literature assume that link metrics (e.g., entanglement generation probability or average fidelity) are known and stable during path computation, and that end-to-end performance can be optimized by applying classical shortest-path style strategies~\cite{Pant2019RoutingEntanglement,Shi2024ConcurrentEntanglementRouting,Sutcliffe2025FidelityAware}. For example, linear repeater baselines are frequently expressed in terms of the shortest-path length and per-link success probability~\cite{Pant2019RoutingEntanglement}, while concurrent entanglement routing models compute paths over a classical graph $G=\langle V,E,C\rangle$ using hop count and channel width as primary structural metrics before accounting for stochastic link states~\cite{Shi2024ConcurrentEntanglementRouting}. Even when fidelity is incorporated explicitly, it is often controlled via a single scalar parameter such as a memory cut-off time, trading off rate and fidelity along a restricted projection of the full cost space~\cite{Sutcliffe2025FidelityAware}. In reality, quantum channels fluctuate over time, and end-to-end path costs are fundamentally non-additive and often non-monotonic due to the multiplicative nature of swapping probabilities and the strict cut-offs imposed by fidelity thresholds~\cite{caleffi2017}. This gap between static, shortest-path-inspired models and dynamic, non-additive reality creates a need for dynamic, stochastic routing protocols that can operate on partial or outdated network telemetry. Moreover, because routing with non-additive fidelity constraints is mathematically NP-hard~\cite{caleffi2017}, there is a pressing requirement for fast, scalable routing heuristics that approximate high-quality paths within a fraction of a millisecond.

\item \textbf{The Single-Tenant Vacuum (Routing):}
Even when routing algorithms attempt to incorporate physical realism by modeling decoherence and probabilistic swapping, they frequently optimize for an isolated source-destination pair in a vacuum~\cite{caleffi2017}. This ``single-tenant'' assumption bypasses the reality of network-wide resource contention that occurs when concurrent applications share the same probabilistic hardware~\cite{abane2025survey}. Recent work has begun to address this challenge by jointly routing and allocating edge capacities for multiple simultaneous entanglement requests under finite memories and integrated purification, but still within a centrally scheduled, flow-allocation paradigm~\cite{Li2021EffectiveRouting}. In a functional Quantum Internet, routing cannot be decoupled from resource availability. This motivates integrated SDQN models in which path selection, scheduling, and multi-tenant resource allocation are coordinated jointly rather than treated as independent problems.

While the previous two assumptions concern how costs and contention are modeled in the routing problem, a separate class of assumptions concerns what information the controller can realistically obtain and act upon in real time.

\item \textbf{Instantaneous Classical Signaling and the ``God's-Eye'' View (Scheduling): }Theoretical scheduling models frequently abstract away the speed-of-light propagation delays of the classical control plane. For instance, recent proposals for optimizing global, satellite-based quantum networks rely on a centralized controller that gathers exact global telemetry and computes an optimal bipartite matching at the beginning of every time slot~\cite{panigrahy2022}. In practice, the classical signaling round-trip time required to gather this global state from orbiting satellites and transmit instructions back would consume hundreds of milliseconds. Because quantum memory coherence is finite, assuming instantaneous centralized control risks selecting schedules where target qubits decohere before the control instructions ever arrive. These latency constraints point toward highly optimized, distributed MAC-layer and link-layer signaling protocols that minimize classical round-trip time, as well as predictive scheduling algorithms that pipeline control signals and anticipate network states to preserve the coherence budget of quantum memories.

\item \textbf{Resource Fungibility and Perfect Isolation (Allocation):}
When dividing physical network assets, many quantum network resource allocation models treat quantum memories as classical buffer slots, modeling them as fungible, interchangeable servers in queueing systems or abstract FIFO buffers for entanglement requests~\cite{Gauthier2026EGS,bacciottini2025}. In these models, a memory qubit is effectively just a capacity unit that can be reserved, released, or dropped, and the internal structure of the device and potential interactions between stored qubits are not considered. Physically, however, multimode quantum memories and trapped-ion devices must be carefully engineered to suppress crosstalk between stored modes or neighboring ions, since addressing one mode or ion can otherwise disturb adjacent ones~\cite{Yang2018MultimodeMemory,Sotirova2024LowCrosstalk}. This hardware reality implies that the network control plane cannot treat quantum nodes as simple abstract queues. Instead, resource allocation algorithms must become ``hardware-aware'', merging classical queuing theory with quantum error propagation models, and carefully spacing out allocations not just in time, but also within the internal topology of the quantum memory to mitigate cross-tenant interference.

\item \textbf{Deterministic SLAs (Quality of Service): }
Classical resource allocation is built on providing deterministic guarantees, such as reserving a hard bandwidth of 1~Gbps for a specific tenant. Quantum links, by contrast, provide entanglement only probabilistically, with success rates and fidelities that fluctuate over time~\cite{abane2025survey,cacciapuotiQuantumInternetNetworking2020}. Emerging standardization efforts for QKD networks have begun to discuss QoS management and even QKD-specific service level agreements (SLAs)~\cite{ITUTY3809,ITUTY3829}, but these models remain largely tailored to key delivery rather than general entanglement services. Because entanglement generation is inherently probabilistic, classical-style deterministic SLAs are physically impossible to guarantee in general. This limitation suggests that new frameworks for \textit{stochastic QoS} and Quantum SLAs (QSLAs) are required, where guarantees are expressed in terms of confidence intervals (e.g., ``a 95\% probability of delivering 10 entangled pairs per second above 0.90 fidelity'') rather than fixed rates.

\item \textbf{Reactive Transport and Best-Effort Abstractions (Architecture): }
Because the classical Internet's best-effort, packet-switched architecture has been overwhelmingly successful, there is a natural instinct within the networking community to map these same transport principles onto the Quantum Internet. This includes exploring how classical TCP paradigms, Active Queue Management (AQM), and Explicit Congestion Notification (ECN) might be adapted to manage flows of entangled pairs~\cite{bacciottini2025}. While these classical analogies provide valuable conceptual stepping stones, relying on a fundamentally \textit{reactive} control paradigm can become a bottleneck for quantum operations. 

Classical transport protocols throttle traffic reactively---that is, only \textit{after} congestion is detected. In a quantum network, by the time an end node receives a classical congestion notification, intermediate switches have already committed time and resources to generate and buffer fragile entangled states. If queues fill up or buffering times exceed coherence limits, these expensive, probabilistically generated resources are irreversibly destroyed. This dynamic argues for a shift away from purely reactive, best-effort analogies toward strictly more \textit{proactive} control, where frameworks like SDQN and Q-NUM are used to globally schedule and allocate resources to guarantee path viability \textit{before} the physical generation of qubits begins.
\end{itemize}

To move beyond heuristic approximations and resolve these complex, multidimensional trade-offs systematically, the control plane requires a rigorous, application-aware optimization framework. As explored in the next section, this leads to the adoption of advanced network partitioning and the formulation of the Quantum Network Utility Maximization (Q-NUM) model.


\section{QUANTUM NETWORK VIRTUALIZATION AND PARTITIONING}
\label{sec:partitioning}

The control plane functions detailed in Section~\ref{sec:control-functions}---routing, scheduling, and resource allocation---provide the operational mechanics for establishing a single quantum connection. However, managing a large-scale, multi-tenant Quantum Internet requires a more granular strategy. Building on the virtualization concepts of classical SDN, the SDQN controller must divide and orchestrate the network's capabilities through two complementary strategies: Network Partitioning and Circuit Partitioning. Both approaches address the fundamental scalability challenges posed by limited qubit counts, decoherence, and noise in near-term quantum devices. For the network engineer, this introduces a radical paradigm shift: the network can no longer operate as a passive, application-blind data pipe. Because quantum states cannot be buffered, the architecture of the distributed quantum application (the circuit) directly dictates the real-time topology and coherence budget of the network. Optimizing these partitioning methods exposes profound gaps between theoretical algorithm design and the physical realities of quantum hardware.

\subsection{Quantum Network Partitioning (Slicing)}

Quantum Network Partitioning is a direct application of classical virtualization principles to the quantum domain. Analogous to 5G network slicing, it involves the logical division of a single physical network into multiple, isolated virtual networks (slices). This allows a provider to support diverse applications concurrently on shared hardware by allocating dedicated memory modes and entanglement links to each slice.

However, the fundamental difference lies in the nature of the resource being managed. Classical networks partition deterministic bandwidth; quantum networks must partition entanglement---a probabilistic, perishable commodity that is immediately consumed upon use. Consequently, managing a quantum slice involves orchestrating a continuous ``success by repetition'' lifecycle. The theoretical viability of quantum slicing currently rests on several idealized models that present major challenges for the classical networking community:
\begin{itemize}
    \item \textbf{The Flaw of Resource Fungibility:} Models often treat quantum resources (e.g., memory slots or entanglement links) as a uniform, perfectly divisible pool. This ignores the physical reality that each link and memory unit is a unique, non-interchangeable component with distinct noise characteristics and fidelities.
    
    \item \textbf{The Assumption of Perfect Isolation:} Partitioning presumes that logical software boundaries are sufficient to ensure physical isolation. In reality, executing operations in one slice can generate physical crosstalk, destroying fragile quantum states in adjacent qubits assigned to different slices.
    
    \item \textbf{Quantum Resource Fragmentation:} By permanently locking scarce physical resources into dedicated virtual slices, the network suffers from severe fragmentation. The network may fail to serve new requests that it could have otherwise fulfilled if resources were dynamically shared, a problem exacerbated by the difficulty of stitching together incompatible hardware platforms within a single slice.
    
    \item \textbf{The Multi-Tenancy Overhead:} Managing dozens of probabilistic slices simultaneously creates massive classical control traffic. The architecture assumes this control plane overhead will not become so large that classical commands arrive after the qubits in those slices have already decohered.
\end{itemize}

\subsection{Quantum Circuit Partitioning}

While Network Partitioning addresses how infrastructure is shared, Circuit Partitioning tackles how to execute a single quantum computation that is too large for any single quantum processor. It involves breaking down a quantum algorithm into smaller sub-circuits, which are then executed on separate, smaller quantum processors connected by the network. Conceptually, this is analogous to how a High-Performance Computing (HPC) cluster breaks a problem into parallel tasks. However, at the physical layer, classical HPC nodes exchange passive data, whereas quantum processors must generate and consume entanglement to perform remote gate operations and teleport qubits. Therefore, network engineers cannot leave circuit partitioning solely to quantum software compilers. The network is no longer a passive transport layer but an active component of the processor itself; it must deliver sufficiently high-fidelity resources and communicate results within the processor's time constraints. An entangled link with low fidelity does not just slow down the computation; it introduces errors into the quantum state that propagate through the entire algorithm. 

The circuit partitioning literature has grown rapidly in recent years. The partitioning task itself consists of two subproblems: how qubits are assigned to different processors, and how communication is achieved. These are interrelated, since the assignment of qubits determines which operations need to be covered by the network. Similarly, the choice of communication mechanism may change the assignment of qubits (i.e., if a qubit is teleported). The use of entanglement means that inter-QPU operations have a higher overhead than intra-QPU operations. Consequently, a typical optimization target for the partitioner is to minimize the inter-QPU entanglement requirements. On one hand, too much use of noisy entanglement will be detrimental to the output fidelity, while on the other, entanglement purification induces a delay which may approach or exceed the coherence times of qubits in the computation. Moreover, partitioning decisions directly affect how long quantum states must be stored while awaiting communication or synchronization, making memory coherence times a critical factor in the overall system performance.

Despite the breadth of work on entanglement-based circuit partitioning, many current strategies exhibit both conceptual limitations and rely on severe algorithmic assumptions:

\begin{itemize}
    \item \textbf{Limited Cost Model:} Many circuit partitioning methods restrict the cost model by either assuming a particular communication strategy and optimizing the assignment to minimize it~\cite{andres-martinezAutomatedDistributionQuantum2019}, or assume the assignment is given and optimize the choice of communication technique~\cite{wuEntanglementefficientBipartitedistributedQuantum2023,wuAutoCommFrameworkEnabling2022}. These are better treated under a unified cost model, since a locally optimal solution in one subproblem may be suboptimal in the global objective of entanglement consumption~\cite{davisDistributedQuantumComputing2023,burtGeneralisedCircuitPartitioning2024b, burtMultilevelFrameworkPartitioning2026}.

    \item \textbf{The ``Top-Down'' or ``Algorithm-First'' Fallacy:} Many partitioning algorithms focus purely on computational logic to find the optimal ``cut'' in a circuit, assuming the network is a subordinate provisioning service that can simply be configured on-demand to meet the algorithm's specific entanglement requirements. This disregards the fact that the cost of entanglement sharing at the network level depends on factors such as delays, rates and fidelities.
    
    \item \textbf{The ``God's-Eye View'' Controller:} Partitioning models assume the existence of a single classical entity with perfect, real-time, cross-layer knowledge of the entire distributed system. The controller is assumed to simultaneously know the application's computational needs, the precise fidelity of every physical link, and the exact gate error rates of every QPU. In practice, qubit error rates vary dramatically across a single device and drift over time~\cite{tannuNotAllQubits2019}, crosstalk between qubits cannot be continuously characterized without disturbing the computation~\cite{sarovarDetectingCrosstalkErrors2020}, and classical control plane latency may exceed qubit coherence times~\cite{cacciapuotiQuantumInternetNetworking2020}. The observer effect makes this perfect real-time holistic view physically unrealistic.

\end{itemize}
Several methods have begun to address these limitations individually. Optimizing the assignment of qubits alongside the choice of communication can reduce entanglement requirements~\cite{nikahdAutomatedWindowbasedPartitioning2021, sundaramDistributionQuantumCircuits2022,ferrariModularQuantumCompilation2023, kaurOptimizedQuantumCircuit2025}, particularly when done under a unified cost model~\cite{davisDistributedQuantumComputing2023,burtGeneralisedCircuitPartitioning2024b,burtMultilevelFrameworkPartitioning2026}. Beyond this, many algorithms are now designed to account directly for connectivity constraints in the network, though tend to either assume a particular network topology~\cite{ferrariCompilerDesignDistributed2021a, liuECDQCEfficientCompilation2025}, modify solutions with post-processing~\cite{andres-martinezDistributingCircuitsHeterogeneous2024, kaurOptimizedQuantumCircuit2025}, or sacrifice a unified cost model to achieve this~\cite{sundaramDistributionQuantumCircuits2022, sundaramDistributedQuantumComputation2024}. 

\subsection{Joint Network and Circuit Partitioning}

To be effective on general quantum circuits and networks, the SDQN paradigm must merge network and circuit partitioning into a unified, cross-layer optimization framework. In this joint model, the real-time, physically-limited state of the network acts as a hard constraint. The controller must simultaneously map sub-circuits to the best-suited processors while carving out a dedicated, on-demand virtual network slice to connect them. Within this slice of the network, the partitioner must also consider the local network topology alongside the circuit structure. Recent works have moved towards this by incorporating topology-dependent entanglement costs directly into the partitioning objective~\cite{sundaramDistributedQuantumComputation2024, burtEntanglementEfficientDistributionQuantum2025}. For large circuits and network slices, this can be done recursively, by partitioning over a clustered representation of the network and continuing to partition sub-circuits within each cluster, or ``slice'', of the network.

To understand this synergy, the closest classical analogy is Network Function Virtualization (NFV) Service Chaining. In NFV, an orchestrator co-optimizes the placement of virtual functions (like firewalls) onto physical servers and the allocation of network links to stitch them into a coherent service. Both NFV and joint quantum partitioning use the application's specific needs to dictate the topology and QoS of the virtual network being created. However, while a poorly configured classical NFV network merely makes a service slow, a low-fidelity quantum network slice can render the computation unreliable, as every network error propagates directly into the algorithm's final result.

While this co-design model ensures scarce resources are allocated efficiently, it creates a hyper-dimensional, NP-hard optimization problem. The SDQN controller must find the optimal solution within a massive search space encompassing all possible circuit partitions, processor mappings, and network resource allocations---objectives that are often in direct conflict.

Solving this complex optimization problem requires a ``unified, cross-layer cost model''---a universal mathematical abstraction that can normalize the ``cost'' of using a processor with lower gate fidelity against a network link with lower entanglement fidelity. Developing this universal abstraction is the primary motivation for adopting the Quantum Network Utility Maximization (Q-NUM) framework, which is detailed in the following section.

\section{QUANTUM NETWORK UTILITY MAXIMIZATION (Q-NUM)}
\label{sec:q-num}

As established in Sections~\ref{sec:control-functions} and~\ref{sec:partitioning}, attempting to jointly optimize network slicing, circuit partitioning, and routing creates an NP-hard, multi-objective problem. To prevent the SDQN controller from becoming paralyzed by these conflicting physical constraints, it requires a unified mathematical compass. In classical networks, this compass is provided by Network Utility Maximization (NUM)~\cite{Kelly1998RateControl}. In the quantum domain, we must adapt this framework into Quantum Network Utility Maximization (Q-NUM) to translate raw, incompatible physical metrics into a single, optimizable currency: application value.

Recent literature has begun to recognize this necessity of evaluating quantum networks not merely by their raw physical-layer metrics, but by the actual value they deliver to upper-layer applications. Foundational works have successfully mathematically mapped quantum properties---such as fidelity, negativity, and secret key rates---into distinct utility curves~\cite{Vardoyan2022QNUM}. Leveraging these mathematical formulations, researchers have developed powerful offline tools, utilizing Q-NUM frameworks to optimize pre-deployment network planning and link placement~\cite{Pouryousef2023QNUM}, as well as to benchmark the overarching economic and computational scaling laws of various topologies~\cite{lee2024QNUM}.

However, while the current literature heavily utilizes Quantum Network Utility for passive evaluation and offline architectural planning, a functional Quantum Internet requires transitioning this concept into an active, real-time control mechanism. To achieve this, we demonstrate how Q-NUM can be operationalized as the dynamic optimization engine for the SDQN controller. By defining specific utility functions for concurrent applications, we illustrate how a centralized control plane can use this mathematical lens to actively resolve multi-tenant resource contention, executing joint routing, scheduling, and allocation decisions that maximize the global value of the live network.

\subsection{From Classical to Quantum Utility}

In classical networks, NUM is a foundational framework used to allocate scarce resources, like bandwidth, efficiently and fairly among competing users~\cite{Kelly1998RateControl, Tychogiorgos2012NUM}. Utility is an abstraction of user satisfaction, typically represented as a non-decreasing, concave function of the data rate, $U_i(x_i)$, where $x_i$ is the rate for user $i$. The system-wide optimization objective is to maximize the total utility, $\max \sum_i U_i(x_i)$, subject to physical link capacities~\cite{Low1999OptFlowControl, Kelly1998RateControl}. Crucially, classical utility is generally \textit{monotonic}: more bandwidth is never worse for the application.

In quantum networks, the Q-NUM framework addresses a fundamentally different challenge: state synthesis and viability. The core task is not allocating robust bandwidth, but actively creating and maintaining a usable, high-quality resource---entanglement~\cite{lee2024QNUM, Vardoyan2022QNUM, Pouryousef2023QNUM}. Therefore, Q-NUM's objective is to maximize the rate of \textit{valuable, completed quantum tasks}.

Because quantum tasks are inherently fragile, quantum utility cannot be a function of rate alone; it must depend jointly on the expected entanglement generation rate ($R$) and the resulting end-to-end fidelity ($F$)~\cite{Vardoyan2022QNUM}. The Q-NUM framework maps every potential network operation into this two-dimensional $(R, F)$ capability space, where recent work demonstrates that different applications induce distinct utility shapes~\cite{Vardoyan2022QNUM}. Consequently, the SDQN controller's goal is not to maximize $R$ or $F$ blindly, but to maximize a specific Utility Function $U(R, F)$ defined by the end-user application. Table~\ref{tab:utility_functions} illustrates how drastically these utility shapes vary based on the application's unique sensitivity to errors versus speed~\cite{Vardoyan2022QNUM,lee2024QNUM,Gisin2002QCrypto,scarani2009security}.

\begin{table}[htbp]
\centering
\caption{Q-NUM Utility Functions by Application}
\label{tab:utility_functions}
\renewcommand{\arraystretch}{1.4}
\begin{tabularx}{\columnwidth}{lXXX}
\hline
\textbf{Application} & \textbf{Utility Function Shape} & \textbf{Fidelity Requirement} & \textbf{Rate Sensitivity} \\ \hline
\textbf{QKD \cite{Gisin2002QCrypto,scarani2009security}} & Near-linear with $R$ (after a threshold) & Low threshold ($\sim 0.85$) & High (Maximizes secret key rate) \\ \hline
\textbf{DQC \cite{Vardoyan2022QNUM,lee2024QNUM}} & Binary step-function & High threshold ($\sim 0.95$) & Low (Fidelity is paramount) \\ \hline
\textbf{DQS \cite{Vardoyan2022QNUM,lee2024QNUM}} & Exponential in $F$ & Very High ($\sim 0.99$) & Medium (Precision-first) \\ \hline
\end{tabularx}
\end{table}

Unlike classical applications that degrade gracefully under congestion, quantum tasks often exhibit a binary pass/fail outcome. If a DQC algorithm requires a fidelity of 0.95, delivering entanglement at a massive rate but with a fidelity of 0.94 has exactly zero utility. 

This creates a radical shift for the network engineer: quantum utility is \textit{non-monotonic} \cite{Vardoyan2022QNUM, lee2024QNUM}. Aggressively increasing the entanglement generation rate (e.g., by skipping purification steps) will inevitably degrade fidelity; if fidelity drops below the application's critical threshold, the utility instantly collapses to zero.

To mathematically model this sharp threshold effect, the utility function $U_j(R_j, F_j)$ for a given quantum application $j$ is formulated to explicitly balance this speed-quality trade-off. For an application with a strict minimum fidelity threshold, $F_{j,\text{th}}$, the utility can be approximated as~\cite{Vardoyan2022QNUM,Pouryousef2023QNUM}:
\begin{equation}
U_j(R_j, F_j) \approx R_j \cdot \phi_j(F_j)
\end{equation}
which follows the general Q-NUM philosophy of separating rate and quality contributions but adopts a simplified product form for clarity in this tutorial.
$\phi_j(F_j)$ is a quality factor that acts as a penalty function. It approaches 1 as fidelity approaches perfect purity, but crucially, it drops entirely to zero if $F_j < F_{j,\text{th}}$. The SDQN controller's ultimate optimization objective for the entire network is therefore to maximize the aggregate utility~\cite{Low1999OptFlowControl,Vardoyan2022QNUM}:
\begin{equation}
\max \sum_j U_j(R_j, F_j)
\end{equation}
subject to the constraints of the physical quantum memories and the probabilistic generation rates of the optical channels.

\subsection{Q-NUM as the Universal Cost Model}

This mathematical formulation provides the exact ``unified, cross-layer cost model'' required to solve the Joint Partitioning problem introduced in Section~\ref{sec:partitioning}. By framing routing, scheduling, and resource allocation as a unified Q-NUM optimization problem, the SDQN controller can provably maximize the actual value the network delivers to its users, moving beyond heuristic engineering to mathematically optimal quantum resource management.

Under the Q-NUM model, the SDQN control loop can be conceptualized as operating in three continuous steps:
\begin{enumerate}
    \item \textbf{Measure (Quantum Network Tomography):} To resolve the scalability bottleneck of brute-force QPT, the control plane requires efficient telemetry. The QNT techniques introduced in Section~\ref{sec:performance_metrics} could serve as the real-time sensory engine for this loop. While recent literature introduces QNT as a critical theoretical enabler for fidelity-aware routing and network management~\cite{deandrade2024_QNT}, embedding it explicitly as the first step of an SDQN control loop provides the concrete architectural mechanism to operationalize this telemetry. This integration would allow a centralized controller to continuously gather the realistic, composite noise models required to avoid the additive-cost fallacy.
    
    \item \textbf{Compute (Feasible Region Mapping):} Using the realistic link metrics provided by QNT, the controller could determine the achievable $(R, F)$ space for a given path. For instance, the controller might calculate that by invoking one round of entanglement purification over a newly characterized depolarizing channel, it can shift a link's performance from an initial state of $(R=100\text{Hz}, F=0.80)$ to a distilled state of $(R=30\text{Hz}, F=0.92)$.
    
    \item \textbf{Optimize (Utility Maximization):} Finally, the controller would apply the application's specific utility function to the calculated $(R, F)$ space to make the final routing and allocation decision. If the requesting application is QKD (which has a lower fidelity threshold), the controller could skip purification to maintain the 100Hz rate and maximize key generation. If the application is DQC, the controller would execute the purification to breach the $0.90$ fidelity threshold, willingly sacrificing rate to achieve the necessary computational quality.
\end{enumerate}

\subsection{Practical Operational Uses}

Beyond real-time routing and scheduling, the Q-NUM framework serves three vital roles in the maturation of the Quantum Internet:

\begin{itemize}
    \item \textbf{Multi-Tenant Admission Control:} By relying on Q-NUM, the network can perform intelligent admission control. If admitting a new QKD slice would cause the network-wide fidelity to drop, triggering a utility collapse for an existing DQC slice, the controller mathematically knows to reject the new request to preserve the total value of the network.
    \item \textbf{Architectural Benchmarking:} Q-NUM answers whether heterogeneous hybrid topologies, satellite downlinks, or pure fiber repeaters with deep purification deliver the most valuable completed tasks. It translates low-level hardware improvements (e.g., longer coherence times) directly into quantifiable end-user value.
    \item \textbf{Investment and Policy Strategy:} Linking physical metrics to aggregate utility provides a strategic compass. It allows network operators to prioritize capital investments that most expand the feasible task region, providing an application-aware benchmark that can be reported alongside traditional physical KPIs to prove the commercial viability of the infrastructure.
\end{itemize}


\section{DISTRIBUTED QUANTUM AI OVER IMPERFECT NETWORKS}
\label{sec:dqai}

The Q-NUM framework established in Section~\ref{sec:q-num} demonstrates how a quantum network must dynamically balance rate and fidelity to maximize application utility. Perhaps no application exposes the extreme limits of this trade-off more than \textit{Distributed Quantum Artificial Intelligence (DQAI)}. 

\subsection{Delineating the Scope: AI and Quantum Networking}
The intersection of \textit{Artificial Intelligence (AI)} and networking has expanded significantly with the advent of quantum technologies. To date, the majority of literature at this nexus is divided into two distinct paradigms:
\begin{enumerate}
    \item \textbf{Classical AI for Quantum Networks:} Utilizing classical machine learning (e.g., Deep Reinforcement Learning) as a control-plane mechanism to optimize quantum network operations, such as entanglement routing, swapping policies, and purification scheduling \cite{mahmud2025,Le2022DQRA}.
    \item \textbf{Quantum AI for Classical Networks:} Leveraging \textit{Quantum Machine Learning (QML)} algorithms to solve complex optimization problems in classical telecommunications, such as resource allocation in 6G and \textit{Space-Air-Ground Integrated Networks (SAGINs)} \cite{butt2025, Bouchmal2023_QAI_6G}.
\end{enumerate}

Readers interested in the algorithmic intricacies of these two paradigms are referred to existing comprehensive surveys on AI for quantum communications and quantum(-inspired) optimization for 6G networks~\cite{mahmud2025, Bouchmal2023_QAI_6G, butt2025}. In contrast, this section addresses a nascent, underexplored third frontier: \textit{the deployment of DQAI and Quantum Federated Learning (QFL) over physically constrained quantum networks.} 

While recent surveys on QFL focus on algorithmic speedups and mathematical convergence \cite{Chehimi2024QFLFoundations,Uddin2026QFLSurvey}, they overwhelmingly treat the underlying communication channels as idealized, noiseless, or entirely classical. Here, we abandon the assumption of an ideal network and analyze how the physical constraints of entanglement-assisted quantum networks fundamentally alter the design, training, and deployment of distributed quantum AI models.

\subsection{The Impact of Quantum Network Constraints on DQAI}

The prevailing paradigm for near-term quantum machine learning is largely based on \textit{Variational Quantum Algorithms (VQAs)}, where a classical optimizer iteratively trains a parameterized quantum circuit~\cite{cerezo2021variational}. While the trainability and on-device noise resilience of standalone VQAs have been extensively studied, scaling these architectures across multiple distinct quantum processing nodes introduces a fundamentally different class of physical limitations. As we transition from single-node QML to DQAI and QFL, the assumptions of pristine state geometry and deterministic local control break down.

In such distributed frameworks, multiple geographically separated nodes collaboratively train a global quantum model, such as a \textit{Quantum Neural Network (QNN)} or a \textit{Variational Quantum Eigensolver (VQE)}~\cite{peruzzo2014vqe, mcclean2016theoryVQE}. While recent landmark experiments have successfully demonstrated deterministic DQC over dedicated, point-to-point optical links~\cite{Main2025DQCNature}, scaling these operations to a shared, multi-tenant Quantum Internet presents a severe cross-layer challenge. Theoretical models, such as the distributed gate architecture proposed by Gyongyosi and Imre~\cite{Gyongyosi2021ScalableDQC}, mathematically prove that the objective functions of these distributed computations are strictly bottlenecked by network decoherence and unitary gate delays. Consequently, the transmission of quantum states to evaluate global cost functions or teleport quantum gradients cannot occur in an idealized vacuum; the physical network constraints previously discussed in this tutorial introduce severe machine learning bottlenecks:

\begin{itemize}

    \item \textbf{Probabilistic Entanglement and the ``Quantum Straggler'' Problem:} In classical distributed ML, heterogeneous compute times or network latencies manifest as familiar straggler effects, but synchronous global aggregation with deterministic, reliable communication is still a reasonable default. Recent quantum federated learning frameworks adopt this same paradigm, assuming that each round a central server can synchronously collect local quantum model updates (e.g., update unitaries) from a selected set of nodes and aggregate them into a new global model~\cite{xia2021quantumfed}. In a future deployment over a physical quantum network, however, conveying quantum information or entanglement between nodes and server will rely on probabilistic operations such as entanglement generation and swapping, and on finite-coherence quantum memories. As the number of participating nodes grows, coordinating simultaneous, high-fidelity entanglement across all required links in a fixed time window becomes increasingly unlikely, while qubits that wait for slower links accumulate decoherence. This ``quantum straggler'' effect suggests that strictly synchronous QFL protocols designed for classical networks may suffer severe latency and reliability penalties once quantum networking constraints are taken into account.

    \item \textbf{Fidelity Degradation and ``Quantum Gradient Noise'':} In QML, gradients are typically estimated via the \emph{parameter-shift rule}, which requires repeated circuit evaluations and accurate measurement of phase-dependent expectation values. State-of-the-art QFL optimization algorithms, including geometrically-aware methods like the Quantum Natural Gradient that navigate the quantum state space via the Fubini–Study metric tensor~\cite{stokes2020quantum}, typically assume that clients can reliably estimate such local gradients on their quantum hardware and that the resulting classical gradient (or update) vectors can then be aggregated at a central server without further distortion~\cite{qi2022federated}. In practice, however, decoherence and gate noise perturb the measurement statistics used in these estimators, inducing a form of \emph{quantum gradient noise} whose structure is dictated by the underlying quantum dynamics rather than by simple additive perturbations. At the same time, some recent QFL works model randomness or privacy by adding i.i.d. Gaussian- or Laplace-distributed noise directly to classical gradient vectors~\cite{li2021quantum}, which is convenient for analysis but does not capture the highly state- and device-dependent way in which physical decoherence erodes phase coherence and biases expectation values. When quantum states must be stored for extended times or transmitted over imperfect links, this decoherence can drive the effective signal-to-noise ratio of parameter-shift estimates below a usable regime, so that beyond some fidelity threshold gradients become dominated by physical noise and distributed training may fail to converge in practice.

    \item \textbf{No-Cloning and Checkpointing Failures:} Classical distributed ML relies heavily on data redundancy and fine-grained checkpointing, where intermediate activations and gradients can be replicated and resent across workers. Recent quantum learning systems adopt a similar manager–worker abstraction, scheduling quantum learning \emph{subtasks} (circuits) to multiple workers and aggregating results on a classical host, in direct analogy with classical cloud-compute clusters~\cite{donofrio2023distributed}. However, because the no-cloning theorem forbids copying unknown quantum states, classical-style redundancy and state-level checkpointing cannot be implemented for quantum data: an in-flight quantum state lost due to a device or link failure cannot be recovered or “resent,” and must instead be re-prepared by re-executing the corresponding circuits from classical descriptions of the data and parameters. Consequently, while classical parameters can still be checkpointed, any failure that destroys quantum state forces recomputation of the affected quantum subtasks, limiting the applicability of classical retransmission and checkpointing strategies in distributed quantum learning.

    \item \textbf{The Entanglement Scarcity vs. Purification Trade-off:} DQAI systems present a perfect Q-NUM optimization problem. To mitigate gradient noise, networks utilize entanglement purification. However, because purification consumes multiple pairs, DQAI systems face a strict resource trade-off: allocating the coherence budget to transmit a larger volume of model parameters with high noise, versus consuming those pairs for purification, resulting in fewer parameter transmissions but higher accuracy.
\end{itemize}

\subsection{Open Research Challenges and Opportunities}
To realize Distributed Quantum AI over practical quantum networks, the networking and machine learning communities must collaboratively address several open challenges:

\subsubsection{Entanglement-Aware Quantum Model Parallelism}
To overcome the qubit limitations of near-term hardware, researches have increasingly turned to distributed paradigms, notably separating workloads into data-distributed federated learning and circuit-distributed model parallelism~\cite{wu2023distributed}. However, to bypass the severe fragility of in-flight quantum states and the difficulty of high-fidelity inter-processor entanglement distribution, many current model-parallel architectures adopt circuit-partitioning schemes with purely classical inter-node communication, where sub-circuits are executed locally and only classical measurement statistics or feedforward signals are exchanged between processors~\cite{hwang2024distributed}. For instance,~\cite{FelixKin2025} recently demonstrated a distributed photonic QNN that completely bypasses inter-node entanglement by exchanging only classical measurement statistics to reconstruct neural weights via tensor networks, proving highly resilient to near-term hardware noise. This entirely avoids the entanglement-distribution bottleneck at the network layer but, by construction, restricts inter-node correlations to classical ones, forfeiting genuine nonlocal entanglement across the distributed model even though each node may still host strongly entangled subcircuits. At the same time, recent evidence shows that such classical-communication schemes can match the expressiveness and task performance of entangling, quantum-communication schemes at shallow circuit depths for certain workloads, underscoring that the loss of nonlocal entanglement need not immediately translate into a loss of accuracy in all regimes~\cite{hwang2024distributed}. 

To preserve these correlations across a true Quantum Internet, classical bandwidth-based splitting must be replaced by \textit{entanglement-aware model splitting}. Building on the circuit partitioning concepts in Section~\ref{sec:partitioning}, future research must explore how to dynamically partition a deep QNN across multiple nodes based on real-time link fidelities. Highly entangled operations must be localized within nodes possessing high-quality internal Quantum Processing Units (QPUs), while the network is utilized only for transmitting states that are mathematically robust to the specific fidelity degradation of the available optical paths.

\subsubsection{Asynchronous and Noise-Resilient QFL Aggregation}
Because heterogeneous delays from probabilistic quantum operations and limited quantum resources can create severe ``quantum straggler'' effects, strictly synchronous QFL becomes impractical over near-term quantum communication networks. Recent foundational work explicitly identifies QFL de-synchronization and increased training latency, induced by imperfect quantum operations, decoherence in quantum memories, and classical control delays, as major challenges for deploying QFL over physical quantum networks~\cite{chehimi2023foundations}. To address this, there is a clear need for \textit{asynchronous Quantum Federated Learning} algorithms that can gracefully tolerate delayed or effectively lost contributions when learning parameters are carried by noisy quantum channels. Moreover, aggregation must be co-designed with the quantum network control plane: global model updates should re-weight or discount client contributions based on the estimated fidelities of the quantum states or paths used to transmit their parameters, using QCN-level fidelity and routing metrics as direct proxies for update reliability~\cite{chehimi2023foundations}. Pioneering architectures are already adopting this paradigm; for example, Chen et al.~\cite{FelixKin2025} utilize asynchronous consensus protocols for gradient aggregation across decentralized photonic nodes, effectively eliminating the quantum straggler bottleneck entirely.

\subsubsection{Distributed Quantum Data Loading (QRAM)}
A major, largely unaddressed bottleneck in DQAI is the loading of classical training data into distributed quantum states. \textit{Quantum Random Access Memory (QRAM)} architectures were originally proposed to embed classical data into coherent superpositions of memory cells, enabling superposition queries over large datasets~\cite{giovannetti2008qram}. More recent work has begun to explore QRAM-based quantum data centers that explicitly integrate QRAM with quantum networking infrastructure~\cite{jiang2023quantumdatacenter}. However, synchronizing and distributing a coherent QRAM state across a noisy quantum network to feed geographically separated QNNs remains a massive open challenge, requiring novel, network-aware data-loading protocols.

\subsubsection{Blind Quantum Machine Learning (BQML)}
A profound opportunity unique to quantum networks is the realization of unconditionally secure distributed AI. \textit{Blind Quantum Machine Learning (BQML)} builds on blind quantum computation protocols, allowing a resource-constrained client to delegate the training or inference of a QNN to a powerful, remote quantum server while keeping its inputs, outputs, and (in some settings) model structure hidden from the server~\cite{fitzsimons2017private,broadbent2009ubqc}. Recent work has shown how blind quantum computing can be combined with variational quantum classifiers and federated learning, enabling multiple clients to collaboratively train a shared quantum model with strong information-theoretic privacy guarantees~\cite{li2021qflbqc,li2024bqml}. Through the exchange of (possibly entangled) qubits and carefully randomized measurement patterns, the client can leverage the server's computational power without relying on computational hardness assumptions, in contrast to classical homomorphic encryption schemes. Designing entanglement-routing and scheduling strategies that explicitly optimize for low-latency, high-fidelity BQML deployments over quantum networks remains an open and highly promising research direction.

As quantum networks transition from theoretical constructs to physical deployments, the assumption of idealized communication in Distributed Quantum AI must be discarded. The ultimate success of QFL and DQAI will hinge not only on advances in variational algorithms and quantum optimization, but on their deliberate co-design with entanglement routing, purification scheduling, QRAM-based data loading, and Q-NUM-driven memory and resource management across the network stack.


\section{Future Directions and Open Challenges}
\label{sec:future-directions}

The transition from isolated, physics-driven quantum experiments to a globally accessible, scalable Quantum Internet requires resolving a multitude of engineering and architectural bottlenecks. While the frameworks discussed in this paper—such as SDQN, the synthesized protocol stack, and the Q-NUM optimization model—provide the necessary conceptual foundation, their practical realization demands significant ongoing research. The future trajectory of quantum networking is defined by the necessity to close the ``simulation-reality gap'' and build robust classical control layers capable of mastering the fundamental unpredictability of quantum physics.

\subsection{Cross-Layer, Heterogeneity-Aware Orchestration}
The paramount focus for the next phase of SDQN development is the creation of a truly heterogeneity-aware control plane. Future controllers cannot rely on simple, static extensions of classical protocols like OpenFlow. They must evolve into sophisticated optimization engines capable of managing a radically diverse hardware ecosystem.

This involves orchestrating protocols across entirely incompatible physical platforms—for example, coordinating an entanglement swap between a fast, low-fidelity photonic system and a slow, high-fidelity trapped-ion memory. The control plane must seamlessly handle the complex transduction layers required to interface these systems, dynamically balancing conflicting physical requirements across the network in real-time. Moving beyond heuristic algorithms to provably optimal, AI-driven resource orchestration that can react within the microsecond constraints of the latency-coherence gap is a monumental open challenge.

\subsection{Multi-Domain Federation and the ``Network of Networks''}
\label{subsec:multi_domain_federation}
As the Quantum Internet scales, it will mirror classical network evolution, shifting from isolated intranets to a globally federated ``network of networks.'' These domains will be owned and operated by disparate administrative entities, including universities, government laboratories, and commercial telecom providers. 

Connecting these heterogeneous domains requires the invention of standardized protocols for quantum inter-domain routing (a quantum equivalent to BGP). To facilitate cross-domain resource sharing while maintaining data sovereignty, quantum networks must implement robust, standardized QoS metrics and QSLAs. When an application in Domain A requests a quantum link to Domain B, the inter-domain gateway must negotiate complex, multi-dimensional parameters: the required entanglement generation rate (pairs per second), latency bounds for classical heralding, and the minimum acceptable state fidelity. 

\subsection{Security Models Beyond Trusted Relays}
To truly realize the transformative security potential of quantum communication, the network architecture must advance beyond the limitations of the current generation of hardware. Today, many long-distance quantum networks (particularly QKD deployments) rely on the ``trusted relay'' model, where quantum states are converted back to classical bits at intermediate nodes to be re-encrypted and re-transmitted. This fundamentally breaks end-to-end quantum security, introducing classical vulnerabilities at every repeater node.

Future network architectures must be designed to uphold end-to-end security based strictly on physical principles. This requires the development of decentralized, device-independent key management systems and the successful, scalable deployment of true quantum repeaters capable of blind entanglement swapping without observing the payload.

\subsection{Standardization of the Quantum-Classical Interface}
Finally, the establishment of robust abstraction layers purpose-built for quantum resources is urgently needed. Current classical standards are ill-equipped to describe a resource that is probabilistically generated, immediately consumable, and highly perishable. 

To enable a vibrant, multi-vendor ecosystem, standardization bodies (such as the IETF, IEEE, and ITU-T) must formalize the QNOS APIs and the classical-quantum metadata formats. These standards must effectively capture the unique properties of entanglement, allowing the SDQN controller to function universally as a complex resource lifecycle manager. Standardizing this interface is the critical final step in transitioning quantum networking from a bespoke physics experiment into a foundational pillar of global Information and Communication Technology.

\section{Conclusion}\label{sec:conclusion}
The development of the Quantum Internet is rapidly transitioning from foundational physics verification to a complex, multi-scale network engineering challenge. This tutorial has sought to serve as a definitive, network-centric guide, facilitating this transition by bridging the persistent knowledge and technical language gap between the physics and engineering communities.

We have established that realizing a pragmatic and sustainable Quantum Internet requires a massive paradigm shift away from traditional classical networking principles, which rely on deterministic signal amplification and packet retransmission. Because the No-Cloning and Information-Disturbance theorems physically forbid these operations, we have argued that quantum networks must adopt a fundamentally symbiotic, dual-plane architecture. We systematically established that the physical quantum data plane, characterized by fragile, non-readable payloads, is functionally dependent on a high-throughput, low-latency classical control plane for essential telemetry and synchronization.

Furthermore, we provided a logical progression of concepts to operationalize this architecture and close the pervasive ``simulation-reality gap.'' We conceptualized the QNOS as the essential middleware required to abstract the severe heterogeneity of physical quantum hardware. Building upon this, we synthesized a unifying structure for the SDQN control plane, emphasizing that joint routing, scheduling, and resource allocation are not classical graph-theory abstractions, but an inseparably linked physical control triad governed by strict latency-coherence budgets. 

Crucially, to guide pragmatic network design and multi-tenant slice virtualization, we adapted the Q-NUM framework. This formulation provides a foundational cost model for the control plane to dynamically negotiate the inherent, physically dictated trade-offs between entanglement generation rate and end-to-end fidelity, entirely bypassing the classical additive-cost fallacy.

Finally, we extended this framework to the emerging frontier of DQAI over physically imperfect networks. This analysis provides insight into how higher-layer applications like federated learning cannot be designed in an idealized theoretical vacuum, as phenomena like the ``quantum straggler'' problem and decoherence-induced gradient noise fundamentally reshape learning model viability and convergence.

Ultimately, this tutorial has equipped the network engineer with the pragmatic operational mindset required to architect the Quantum Internet. For the field to advance, the classical networking community must now move beyond idealized simulators and actively embrace the standardization of hardware-aware abstractions, ensuring the Quantum Internet matures into a programmable, multi-tenant utility for future advanced services.

\section*{ACKNOWLEDGMENT}
The authors would like to gratefully acknowledge Michael R. Hanks for his valuable insights, detailed explanations, and fruitful discussions regarding the quantum mechanical concepts presented in this tutorial. Gemini AI (Google) was used for editing and grammar enhancement.

\bibliographystyle{IEEEtran}
\bibliography{references} 

\end{document}